\documentclass[11pt, a4paper]{article}
\pdfoutput=1

\usepackage{url}

\usepackage{amsmath}
\usepackage{amsfonts}
\usepackage{amssymb}
\usepackage{bm}
\usepackage{bbm}
\usepackage{verbatim}
\usepackage{dsfont}
\usepackage{booktabs}
\usepackage{slashed}
\usepackage{nicefrac}
\usepackage{mathtools}

\usepackage{epsfig}
\usepackage{color}
\usepackage[table]{xcolor}
\usepackage{graphicx}
\usepackage[font=small]{caption}
\usepackage[listofformat=empty,subrefformat=empty]{subfig}

\usepackage{float}
\usepackage{overpic}
\usepackage{tikz}
\usepackage{tikz-cd} 

\usepackage{enumerate}
\usepackage{hhline}
\usepackage{multirow}

\usepackage[nosort]{cite}
\usepackage{xspace}
\usepackage{setspace}

 \usepackage[pdftitle={},
   pdfauthor={Dewi Gould, Ling Lin, Evyatar Sabag},
   pdfsubject={Swampland, supergravity, symmetries},
   bookmarksopen, bookmarksnumbered, bookmarksopenlevel=2, colorlinks=false, linkcolor=blue, citecolor=blue, urlcolor=blue]{hyperref}

\usepackage[nameinlink,capitalize]{cleveref}

\makeatletter{}
\g@addto@macro\bfseries{\boldmath}
\makeatother

\newcommand*{\bbZ}{\ensuremath{\mathbb{Z}}}
\newcommand*{\bbR}{\ensuremath{\mathbb{R}}}

\newcommand*{\cN}{\ensuremath{\mathcal{N}}}

\usepackage[normalem]{ulem}

\newcommand{\be}{\begin{equation}}
\newcommand{\ee}{\end{equation}}
\newcommand{\ba}{\begin{aligned}}
\newcommand{\ea}{\end{aligned}}

\newcommand{\cA}{\mathcal{A}}

\newcommand{\cD}{\mathcal{D}}

\newcommand{\cF}{\mathcal{F}}

\newcommand{\cL}{\mathcal{L}}

\newcommand{\cS}{\mathcal{S}}

\def\half{{\frac{1}{2}}}

\def\Z{{\mathbb{Z}}}

\begin{document}

\baselineskip=18pt  
\numberwithin{equation}{section}  
\allowdisplaybreaks  


\thispagestyle{empty}


\vspace*{0.6cm}
\begin{center}
{\huge{\textbf{Swampland Constraints on}}} \\
\bigskip
{\huge{\textbf{the SymTFT of Supergravity}}}\\
 \vspace*{1.5cm}
Dewi S.W. Gould$^{1}$, Ling Lin$^{2,3}$, Evyatar Sabag$^{1}$\\

 \vspace*{1.0cm}
{\it $^1$Mathematical Institute, University of Oxford\\
Andrew-Wiles Building, Woodstock Road, Oxford, OX2 6GG, UK}\\
\vspace*{.3cm}
{\it $^2$Dipartimento di Fisica e Astronomia, Universit\`{a} di Bologna, via Irnerio 46, Bologna, Italy}\\
\vspace*{.1cm}
{\it $^3$INFN, Sezione di Bologna, viale Berti Pichat 6/2, Bologna, Italy}

\vspace*{0.8cm}
\end{center}
\vspace*{.5cm}

\noindent
We consider string/M-theory reductions on a compact space $X=X^\text{loc} \cup X^\circ$, where $X^\text{loc}$ contains the singular locus, and $X^\circ$ its complement.
For the resulting supergravity theories, we construct a suitable Symmetry Topological Field Theory (SymTFT) associated with the boundary $\partial X^\text{loc} \coprod \partial X^\circ$.
We propose that boundary conditions for different BPS branes wrapping the same boundary cycles must be correlated for the SymTFT to yield an absolute theory consistent with quantum gravity.
Using heterotic/M-theory duality, this constraint can be translated into a field theoretic statement, which restricts the global structure of $d\geq 7$, $\cN=1$ supergravity theories to reproduce precisely the landscape of untwisted toroidal heterotic compactifications.
Furthermore, for 6d $(2,0)$ theories, we utilize a subtle interplay between gauged 0-, 2-, and 4-form symmetries to provide a bottom-up explanation of the correlated boundary conditions in K3 compactifications of type IIB.

\newpage
\setcounter{tocdepth}{2}
\tableofcontents



\section{Introduction}
The existence of global symmetries in quantum field theory (QFT) has allowed for various insights, including restrictions on correlations functions and the computation of 't Hooft anomalies which constrain RG-flows and can give insights to strongly coupled regimes of theories. Rephrasing global symmetries in the language of topological operators \cite{Gaiotto:2014kfa} opened the way to finding many new forms of symmetries that were not considered before, referred to as \emph{generalized symmetries}. Many of these newly considered symmetries are discrete. One compact way to contain information on such symmetries and their anomalies is to encode them in a Symmetry Topological Field Theory (SymTFT) in one dimension higher than the QFT \cite{Gaiotto:2020iye, Apruzzi:2021nmk, Freed:2022qnc}. We provide a detailed introduction to this in Section \ref{sec:symtftintro}.

When coupling such QFTs to gravity to obtain a quantum gravity it is conjectured all such global symmetries should be gauged or broken \cite{Banks:1988yz,Banks:2010zn}, including generalized symmetries \cite{Harlow:2018tng,Heidenreich:2021xpr}. Part of the Swampland program \cite{Vafa:2005ui,Arkani-Hamed:2006emk,Ooguri:2006in} (see also \cite{Brennan:2017rbf,Palti:2019pca,vanBeest:2021lhn}) is focused on gaining insight about quantum gravity by understanding how such global symmetries from the effective field theory description are broken or gauged when the theory is coupled to gravity. On the other hand, one can also consider the top down approach of compactifying a string theory to build an effective field theory with a known UV completion, and follow the process of decoupling gravity. In this process one can expect various global symmetries to emerge including generalized ones. 

In this paper we will mostly focus on supergravity models in $d \geq 6$ with 16 supercharges in various dimensions that can be engineered via compactifications of string or M-theory on a compact internal manifold $X$ \cite{Eguchi:1980jx}.
These have an effective field theory description (or, in the case of 6d $(2,0)$ theories, a tensor gauge theory) as a super-Yang--Mills theory with gauge algebra $\mathfrak{g}$ \cite{Aspinwall:1994rg,Narain:1985jj} weakly coupled to gravity, whose global structure (i.e., the gauge group $G$ whose associated algebra is $\mathfrak{g}$) is determined by the spectrum of massive non-local operators.
In the limit of decoupling gravity, these become the Wilson and 't Hooft operators (or suitable analogues in the 6d $(2,0)$ case) \cite{Kapustin:2014gua,Gaiotto:2014kfa}.

From a bottom-up perspective, by which we mean the low-energy description of the (light) local operator spectrum, such global structures are choices that are not constrained by usual effective field theory (EFT) conditions.
The same holds for string theory constructions of non-gravitational theories.
In such cases, the compactification space $X^\text{loc}$ is non-compact, and different global structures are determined by different boundary conditions at infinity ($\partial X^\text{loc}$) \cite{Witten:1998wy,GarciaEtxebarria:2019caf,Morrison:2020ool,Albertini:2020mdx,Apruzzi:2021nmk}. Specifically, one needs to assign Dirichlet or Neumann boundary conditions to background gauge fields associated with discrete global symmetries. It was recently understood that these symmetries are generated by specific BPS-branes wrapping torsional boundary cycles \cite{Apruzzi:2022rei,GarciaEtxebarria:2022vzq,Heckman:2022muc}. These boundary conditions are constrained by field theory considerations requiring a maximal set of mutually local operators \cite{Aharony:2013hda} and no gauge anomalies. The former constraint ensures the field theory is an \textit{absolute} quantum field theory. Once we couple gravity back, due to the global constraints of compact geometries, many such choices cannot be realized in top-down constructions of gravitational theories from string theory, without an apparent field theoretic explanation of why they are inconsistent.

In the philosophy of the Swampland program, this discrepancy between possible string realizations and apparently consistent low-energy theories warrants further investigation, in order to clarify if there are yet-to-be discovered string constructions, or if there is a deeper, quantum-gravitational reason why these field theories are inconsistent.
Of course, there is a third option, that there exists effective field theories coupled to gravity with a consistent UV-completion that do not originate from string theory.
However, recent works, in particular for effective supergravity theories in large numbers of dimensions \cite{Kumar:2009us,Adams:2010zy,Kim:2019vuc,Kim:2019ths,Cvetic:2020kuw,Montero:2020icj,Hamada:2021bbz,Bedroya:2021fbu}, provide strong support for ``String Universality'' for $d \geq 6$ theories with 16 supercharges:
namely, that every gravitational EFT with consistent UV-completion has a string theory realization.

Regardless of one's preference, delineating between these scenarios first requires a precise characterization of the EFTs with known string constructions from those without, in terms of data that is accessible from a bottom-up perspective.
Given that the global structure of the EFT is encoded in some SymTFT, we propose in this work a characterization of the supergravity theories in the ``String Landscape'' in terms of boundary conditions of the SymTFT.

To this end, we first construct a $(d+1)$-dimensional SymTFT of a $d$-dimensional effective field theory, engineered from string/M-theory on a compact $X$, that will contain contributions from both the local model $X^\text{loc}$ that contains the singular locus of the compact space $X$ and its complement in $X$ which we denote $X^\circ$, such that $X=X^\text{loc} \cup X^\circ$. More precisely, the EFT associated to $X$ can be thought as coupling the EFTs on the local patches $X^\text{loc}$ and $X^\circ$ via massive states from strings/branes wrapping cycles $\Sigma \subset X$ that stretch across $X^\text{loc} \cap X^\circ$ \cite{Cvetic:2023pgm,Baume:2023kkf}.
The cycles $\Sigma$ can be then thought of as a ``gluing'' $\Sigma = \Sigma^\text{loc} \cup \Sigma^\circ$ of relative cycles $\Sigma^\text{loc},\Sigma^\circ$ on each patch.
The SymTFT of the supergravity theory in question is constructed by reduction of the underlying String/M-theory on $\partial X^\text{loc} \coprod \partial X^\circ$, which has a natural interpretation as the tensor product SymTFT for the gauge sectors localized on $X^\text{loc}$ and $X^\circ$. 

The above is a SymTFT of the supergravity model in the following sense. The usual local data which defines the supergravity theory is fully obtained from the disjoint union $X^{\text{loc}} \coprod X^\circ$; e.g., the gauge algebra in $d \geq 7$ supergravity theories is specified by the gauge theories on these local geometries.
Then, the spectrum of non-local operators compatible with this local data are a priori encoded in the reduction on the boundary $\partial X^{\text{loc}} \coprod \partial X^{\circ}$ of the disjoint union. 
Finally, a polarization of this SymTFT is required to specify an \textit{absolute} supergravity theory or equivalently to fix the global structure.

Field theoretically, we may choose various different boundary conditions of this SymTFT, so long as they give a mutually local set of charged defects.
Roughly, this amounts to specifying which cycles in the local patches are wrapped by certain BPS $p$-branes.
However, in the gravitational theory defined on $X$, there is no such choice; instead, every compact cycle $\Sigma \simeq \Sigma^\text{loc} \cup \Sigma^\circ$ must be wrapped by all possible $p$-branes (of suitable dimension and compatibility with supersymmetry requirements).
In the product SymTFT, this translates into \emph{identical} boundary condition choices for different branes associated to the relative cycles $\Sigma^\text{loc}$ and $\Sigma^\circ$.
This means that for the global structure of the gravitational EFT, higher form symmetries of different degrees, generated by wrapping BPS-branes on the same boundary cycle will have \emph{correlated boundary conditions}.
Requiring this correlation, which can be phrased purely in terms of the local data that enters the definition of the supergravity EFT, thus provides a bottom-up characterization of the global structures realized in string constructions.
We will elaborate in detail on this in two classes of models.

\paragraph{Gauge group topology of ${\cal N}=1$, $d \geq 7$ supergravity theories.}
In the case of $\cN=1, d\geq 7$, the supergravity EFT is specified by the gauge algebra $\mathfrak{g} \oplus \mathfrak{u}(1)^b$.
Then, the correlation between the electric 1-form and magnetic $(d-3)$-form center symmetries, which form the defect group $D = {\cal Z}^{(1)} \oplus {\cal Z}^{(d-3)}$, of the gauge sector imposes Dirichlet/Neumann boundary conditions for the \emph{same} subgroup ${\cal C} \subset {\cal Z}$.
The global structure is thus specified, without any reference to a concrete top-down construction, in terms of a ``maximally mixed'' polarization,
\begin{align}
    \Lambda = {\cal C}^{(1)} \oplus {\cal C}^{(d-3)} \subset D \, ,
\end{align}
which leads to the gauge group $[\widetilde{G} \times U(1)^b]/{\cal C}$, where $\widetilde{G}$ is the simply-connected group with algebra $\mathfrak{g}$.

While we derive this polarization for 7d theories engineered from M-theory on compact K3 surfaces by adapting the results of \cite{Cvetic:2023pgm} to the supergravity SymTFT, it turns out that the very same structure also arises in untwisted toroidal compactifications of the heterotic string,\footnote{One key feature of these models relevant to us is that in $d$ dimensions, the total gauge rank including $\mathfrak{u}(1)$ factors is $26-d$.} whose global structures in $d \geq 7$ have been recently computed systematically \cite{Font:2020rsk,Cvetic:2021sjm,Cvetic:2022uuu}.
In fact, as we will explain, this criterion is intimately tied to a mixed-anomaly-argument involving the 1-form center symmetry which rules out certain non-Abelian gauge group topologies in the supergravity landscape \cite{Apruzzi:2020zot,Cvetic:2020kuw} (see also \cite{Montero:2020icj} for alternative arguments based on other Swampland principles), though it is crucial now to include the center anomalies of Abelian gauge factors.

Combining the requirement for a maximally mixed polarization with the absence of 1-form symmetry anomalies, we then obtain a constraint that rules out supergravity theories without a stringy realization.
E.g., the majority of gauge groups that correspond to a ``purely electric'' polarization, i.e., of the form $\widetilde{G} \times U(1)^b$, which cannot be ruled out by the anomaly argument of \cite{Apruzzi:2020zot,Cvetic:2020kuw} alone, are now incompatible with the maximally mixed polarization.

One important caveat is that we do not have a bottom-up argument \emph{why} the polarization has to be maximally mixed.
Furthermore, we are only restricting ourselves to models on the branch of the supergravity moduli space that have gauge rank $26-d$.\footnote{String compactifications to $d\geq 7$ with ${\cal N}=1$ supersymmetry can only realize gauge algebras $\mathfrak{g} \oplus \mathfrak{u}(1)^b$ with certain values of their rank. As argued in \cite{Montero:2020icj}, Swampland arguments reproduce (largely) these restrictions.}
String realizations of the other branches involve either ``frozen'' singularities in M-theory \cite{Witten:1997bs,deBoer:2001wca,Tachikawa:2015wka,Bhardwaj:2018jgp}, twisted heterotic compactifications \cite{Chaudhuri:1995fk,Mikhailov:1998si,Font:2021uyw,Fraiman:2021soq}, or other exotic brane setups \cite{Aharony:2007du}; it would be important to find characterizations of their global structures in terms of the SymTFT.

\paragraph{The global structure of 6d $(2,0)$ supergravity.}
Another important case demonstrating the above constraints on SymTFT boundary conditions is the 6d $(2,0)$ theories. The 2-form defect group symmetry for this case is special \cite{Gukov:2020btk} as there is no distinction between electric and magnetic symmetries. Thus, a consistent choice of polarization will be equivalent to the geometric constraints imposed on the string charges of the 2-form symmetry in IIB compactifications on K3s. Moreover, this case has an additional part of the defect group coming from 0- and 4-form symmetries generating an electric/magnetic pair. Carefully analysing the anomalies of the theory shows one can write a mixed gauge gravity anomaly relating the 0- and 2-form symmetries. For this anomaly to vanish in the geometric reduction from type IIB, one precisely requires the 0- and 2-form symmetries SymTFT boundary conditions to be correlated. Thus, in this case one can show that the boundary conditions of the SymTFT for the effective theory can be completely determined by polarization and mixed gauge gravity anomalies constraints. 

Drawing inspiration from this case, it would be interesting to see if also in the class of $d>6$ theories mentioned above, there may more subtle terms in the SymTFT that can provide a bottom-up justification for requiring a maximally mixed polarization.

\paragraph{Plan of the paper.} The paper is organized as follows: In Section \ref{sec:symtftintro} we review some important aspects of discrete symmetries, SymTFTs, and polarizations, where first we explain some basic notions in general and later we show how these notions arise in the context of geometric engineering. In Section \ref{S:MMP-K3} we thoroughly analyze the case of 7d $\cN=1$ theories engineered by K3-compactifications of M-theory. Specifically, we examine how the homology structure of the K3 and its division to two parts is related to the lattice perspective and how it affects the SymTFT and the required boundary conditions. In Section \ref{S:MMP_d>7} we consider $\cN=1,\,d>7$ supergravity theories resulting from untwisted heterotic compactifications on tori. In these cases we show how geometric properties can be associated with lattice embeddings and how these in turn necessitates the \emph{maximally mixed polarization}. In Section \ref{S:CBCreq} we generalize the results of the former two sections to a geometric reduction of string/M-theory and show the results qualitatively remain the same. This leads to the \emph{correlated boundary conditions} requirement and the \emph{maximally mixed polarization} requirement which we state explicitly summarizing our main results. We conclude this section with examples that one can study in the future to corroborate our constraints, and also give the explicit example of 6d $(2,0)$ theories. We also provide two appendices where we list some lattice facts and discuss instanton fractionalization in the lattice description.

\section{Global structures and the SymTFT}
\label{sec:symtftintro}

In this section we review various aspects of discrete symmetries and how they are encoded using a SymTFT as a $(d+1)$-dimensional bulk attached to a $d$-dimensional spacetime. In particular we focus on the defect group, the SymTFT, and how these arise in string theory realizations.

\subsection{The defect group, SymTFT and polarizations}
\label{sec:defectgroupsymtftandpol}

The SymTFT \cite{Gaiotto:2020iye, Apruzzi:2021nmk, Freed:2022qnc} is a topological field theory in one greater dimension than that of the QFT of interest. Its purpose is to lift the symmetry properties of a QFT away from the dynamics, providing a framework in which symmetries, their anomalies and generalized charges can be studied in isolation. It has seen many applications \cite{Kaidi:2022cpf, Kaidi:2023maf, Cordova:2023bja, Zhang:2023wlu,Bhardwaj:2023idu,Sun:2023xxv} including in string theory set-ups \cite{vanBeest:2022fss,Apruzzi:2022rei,Chen:2023qnv, Apruzzi:2023uma,Antinucci:2022vyk,Apruzzi:2022dlm, Lawrie:2023tdz,Bah:2023ymy, Baume:2023kkf,Yu:2023nyn}.

Related, and particularly relevant in the context of string-/M-theory realizations, is the concept of the \textit{defect group} \cite{DelZotto:2015isa,Morrison:2020ool,Albertini:2020mdx} which encodes information about the possible global symmetries of a QFT.

In this section we review both concepts and their relationship to one another. 

\paragraph{The SymTFT. } The $(d+1)$-dimensional bulk TQFT has two boundaries, denoted $\mathfrak{B}^{\rm phys}$ and $\mathfrak{B}^{\rm sym}$, the former is not gapped generically whilst the latter is gapped. In order to get back a $d$-dimensional QFT including the theory dynamics and symmetries one can compactify the interval between the SymTFT two boundaries. The resulting QFT is a relative theory as in general it will hold mutually non-local operators. This can be prevented by imposing boundary conditions for the TQFT, that will leave a maximal set of mutually local operators, meaning the resulting QFT is an absolute theory. In a more mathematical language, the topological defects of the SymTFT corresponding to a symmetry category $\cS$ make up the Drinfeld center of $\cS$ \cite{Bhardwaj:2023ayw}.

The topological defects of the bulk TQFT encode both the topological defects of the absolute QFT, as well as the (generically non-topological) charged objects under these symmetry defects. The final role played by these defects is determined entirely by their boundary conditions.
 
\paragraph{Finite, abelian symmetries. } Restricting our attention to quantum field theories with finite, abelian higher-form symmetries only, vastly simplifies this SymTFT structure. 

In particular, in this simplification the SymTFT has a simple action formulation
\be\label{eq:SymTFTabelianfinite}
S_{\rm SymTFT} = \frac{1}{2\pi}\sum_{p,i,j} \int_{M_{d+1}}  \kappa_{ij}^{(p)} B_p^i dC_{d-p}^j + \cA(\{B_p^i\}) \,,
\ee
where $\kappa_{ij}^{(p)} \in \Z$, $B$ and $C$ are $U(1)$-valued background gauge fields for finite, abelian global symmetries with degree given by their subscript, and $\cA$ denotes a set of possible (mixed) 't Hooft anomaly terms.

Let's focus on the simplest case of \eqref{eq:SymTFTabelianfinite} for concreteness:
\be\label{eq:simpleSymTFT}
S_{\rm SymTFT} = \frac{N}{2\pi} \int_{M_{d+1}}  B_p dC_{d-p} \,.
\ee
This action is invariant under gauge transformations
\be\ba
B_p &\to B_p + d\lambda_{p-1} \,, \\
C_{d-p} &\to C_{d-p} + d\lambda_{d-p-1}\,.
\ea\ee
This ``BF''-theory has $(d-p)$- and $p$-form global symmetries generated by
\be
U(M_p) = e^{ i \int_{M_p} B_p} \,, \quad V(M_{d-p}) = e^{ i \int_{M_{d-p}} C_{d-p}}\,,
\ee
respectively. They shift the fundamental fields by
\be
C_{d-p} \to C_{d-p} + \frac{1}{N} \xi_{d-p}\,, \quad B_p \to B_p + \frac{1}{N} \xi_{p} \,,
\ee
respectively, with $\xi$ properly normalized flat gauge fields. The BF term in \eqref{eq:simpleSymTFT} forces $U$ and $V$ to not commute
\be\label{eq:linking}
U(M_p) V(M_{d-p}) = \text{exp}\left( \frac{2\pi i L(M_p, M_{d-p})}{N} \right) V(M_{d-p}) U(M_p) \,,
\ee
where $L(\cdot, \cdot)$ denotes the linking number of the two manifolds in the bulk spacetime.

\paragraph{Boundary conditions. } To land on an absolute QFT, we must pick a set of boundary conditions consistent with the topological terms \eqref{eq:SymTFTabelianfinite}. 
Concretely, for \eqref{eq:simpleSymTFT} we can demonstrate this effect in the above example. If both $B_p$ and $C_{d-p}$ are fixed on the boundary (given Dirichlet boundary conditions), the operators $U$ and $V$ will descend to extended operators of the boundary QFT. Their linking \eqref{eq:linking}  implies that these extended objects are not mutually local. A consistent \textit{absolute} QFT requires that a maximal, mutually local subset of such extended objects is chosen \cite{Aharony:2013hda}. 

One particular choice would be to trivialize the $U$ operators on the boundary\footnote{Trivialize here means $U=1$ on the boundary.} ($B_p$ Dirichlet), whilst $V$ operators remain non-trivial on the boundary ($C_{d-p}$ Neumann). In this instance the $V$ operators remain topological, and now link with the $U$ operators in a manner which recreates the standard symmetry action via linking. Specifically, the $V$ operators in this case will generate an $(p-1)$-form $\Z_N$ symmetry in spacetime with the boundary of the $U$ operators spanning between the two boundaries resulting in charged objects under this $(p-1)$-form symmetry, with charge $L(M_p, M_{d-p})/N$, in spacetime.

\paragraph{Non-genuine operators. }
As was explained above, boundary conditions will leave some of the operators in the bulk topological in spacetime and others not. In addition, one can also expect to have operators spanning between the two boundaries of the bulk SymTFT that will result in non-genuine operators in spacetime. This will happen in the above case for if we take the $V$ operators to span between the two boundaries, with Neumann boundary conditions for $C_{d-p}$. These would correspond to spacetime $(d-p-1)$ objects attached to $(d-p)$ topological objects.

\begin{figure}[h!]
$$
\begin{tikzpicture}[x=1cm,y=1cm] 
\draw[line width=0.3mm] (-5,-3) -- (-5,-7);
\draw[line width=0.3mm] (-5,-3)--(-3,-1); 
\draw[line width=0.3mm] (-5,-7)--(-3,-5); 
\draw[line width=0.3mm] (-3,-5) -- (-3,-1);
\draw [blue, thick, fill=red,opacity=0.1] 
(-5,-6) -- (0,-6) -- (2,-4) -- (-3,-4) -- (-5,-6);
\draw [blue, thick] 
(-5,-6) -- (-3,-4) -- (-5,-6);
\draw [blue, thick, fill=red,opacity=0.1] 
(-5,-6) -- (-3,-4) -- (-3,-1) -- (-5,-3) -- (-5,-6);
\node[above, color=red] at (0.5,-5) {\large{$U(\Sigma)$}};
\node[above, color=blue] at (-4,-4.5) {\large{$L(\gamma)$}};
\node[below] at (-7,-6) {\large{boundary $QFT$}};
\end{tikzpicture}
$$

\caption{Consider a bulk topological operator denoted $U(\Sigma)$, which we have already given \textit{Neumann} boundary conditions. If we try to place the operator perpendicular to the boundary, as demonstrated here, it ``bends'' to become parallel. In other words, the line it ends on in the boundary, $L(\gamma)$, is in-fact \textit{non-genuine} since it comes with a $U$ operator attachment in the boundary.
\label{fig:nongenuineoperator}}
\end{figure}
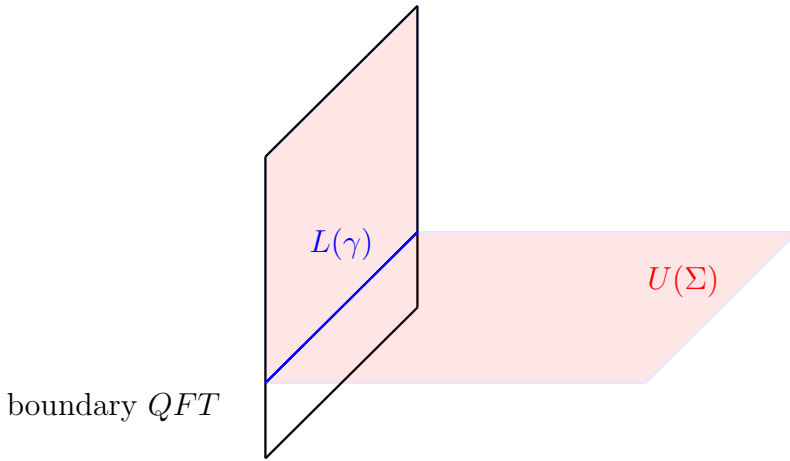

Here we briefly review the appearance of non-genuine operators in the simplest construction. A \emph{non-genuine operator} is defined as a $q$-dimensional operator that is attached to a collection of $p$-dimensional operators, with $p>q$. Let us begin with some SymTFT operator $U(\Sigma)$. 
Suppose we pick Neumann boundary conditions for this operator, but then try to force this operator to lie perpendicular to the boundary. This situation is depicted in Figure \ref{fig:nongenuineoperator}. We are naturally lead to the existence of non-genuine operators\cite{ Gaiotto:2014kfa,Aharony:2013hda,Bergman:2022otk}.\footnote{Note that in recent work \cite{Baume:2023kkf} a different but related notion of genuine/ non-genuine operators is discussed with relation to multi-sector QFTs.}

\paragraph{The defect group. }
The defect group $D$ is a group encoding information about the possible global symmetries of a QFT. In particular it specifies the failure of defining a partition function for the QFT: when $D$ is non-trivial, one obtains a partition vector encoding the various choices of ``global form'', i.e. global symmetries, of the QFT.

Labelling the set of non-dynamical operators of dimension $p$ as $\{\mathfrak{D}_p \}$, and the set of dynamical operators of dimension $p$ as $\{ \cD_p \}$, $D$ is defined as 
\be
D = \bigoplus_{n=1}^{d-1} D^{(n)}\,,
\ee
where $D^{(n)}$ is the group of equivalence classes formed by operators $\{\mathfrak{D}_p \}$ modulo screening by operators $\{ \cD_{p} \}$.

Generically there is a non-degenerate pairing on the defect group\footnote{The phase representing the non-commuting of operators in \eqref{eq:linking} is an example of this pairing.}
\be\label{eq:defect_group-pairing}
\langle \cdot, \cdot \rangle: D \times D \to U(1) \,.
\ee
Non-trivial values of this pairing between two defects describe their mutual non-locality. We can land on consistent QFTs by picking a subgroup
\be\label{eq:polarizationgrading}
\Lambda = \bigoplus_{n=1}^{d-1} \Lambda_n \,, \quad \Lambda_n \subseteq D^{(n)} \,.
\ee
Consistency forces this subgroup to be a \textit{polarization}, $\Lambda$ of $D$. Concretely, a \textbf{polarization} is defined as a subgroup of $D$ which is maximal and isotropic, i.e., on which the pairing trivializes. 
Note, ``maximal'' here refers to $\Lambda \subset D$ as a subgroup;\footnote{Note that this notion of maximality differs from the ``maximal'' in \textit{maximally mixed polarization}. In the latter case, ``maximal'' refers to fact that the amount of ``electric'' and ``magnetic'' symmetries given Dirichlet conditions is as large as it can be.} together with the fact it is isotropic with respect to a non-degenerate pairing implies the condition $\vert \Lambda \vert^2 = \vert D \vert$ on the order of the group.
Each choice $\Lambda$ describes a different global form for the QFT. Below we describe how, after picking a polarization, topological operators generating the defect group can correspond to symmetry generators and charged objects. The \textit{charge} of the charged objects under some generator is exactly given by computing the pairing $\langle \cdot, \cdot \rangle$ between the charged object and the generator.

\paragraph{Relation to SymTFT. } The first set of terms in \eqref{eq:SymTFTabelianfinite} are often referred to as ``BF-terms''. They are intricately related to the defect group above: $B_p$ and $C_{d-p}$ are background gauge fields for finite $(p-1)$- and $(d-p-1)$-form ``defect group'' symmetries respectively. 

Denoting
\be
\text{SNF}(\kappa^{(p)}) = \text{diag}\{l_1^{(p)}, \dots, l_i^{(p)}, \dots \} \,,
\ee
where SNF is the Smith normal form, we can write
\be
D = \bigoplus_{p,i} \left( \Z_{l_i^{(p)}}^{(p-1)} \oplus \Z_{l_i^{(p)}}^{(d-p-1)} \right) \,.
\ee
For example, in the trivial case where $\kappa$ is a diagonal matrix, the defect group is given by
\be
D = \bigoplus_{p,i} \left( \Z_{\kappa_{ii}}^{(p-1)} \oplus \Z_{\kappa_{ii}}^{(d-p-1)} \right) \,.
\ee
The defect group is generated by a set of topological operators. These are the same set of topological operators which make up the SymTFT. The choice of polarization precisely correspond to the choice of boundary conditions for the background gauge fields in the SymTFT that leads to an absolute theory. Thus, the choice of polarization, will leave some of these defect group operators topological, i.e. they will correspond to symmetry generators, while other operators that link with the former operators in spacetime will become the charged under these symmetries. The charge of the operators will correspond to the phase generated by commuting these operators with the topological operators.

\subsection{String/M-theory origin of defects and symmetries}
\label{S:StringSyms}
In this section we will review the string/M-theory origin of the defect group, polarization, and the resulting higher-form symmetries of QFTs realized in string theory constructions. This will include a summary of some of the results appearing in \cite{DelZotto:2015isa, GarciaEtxebarria:2019caf, Albertini:2020mdx, Morrison:2020ool, Apruzzi:2021nmk}

We start by considering a $D$ dimensional string/M-theory placed on a purely geometric background
\be
\mathbb{R}^{1,d-1} \times X_{D-d}\,,
\ee
where $X_{D-d}$ is a non-compact singular Ricci flat special holonomy $(D-d)$ dimensional space, with localized singularities that are not necessarily isolated. Since $X$ is non compact, gravity is decoupled. Compactifying the original string/M-theory on such a geometry will result in a supersymmetric field theory living on the $\mathbb{R}^{1,d-1}$ spacetime.

\paragraph{The internal space geometry.}
Since we are focusing on purely geometric backgrounds, the information about the defect group is encoded geometrically in the long exact sequence in relative homology.
For the cases we are interested in, we will assume $H_{m+1}(\partial X) = H^{\widetilde{d}-(m+2)}(\partial X) = 0 = H_m(X,\partial X) = H^{\widetilde{d}-m}(X)$, and $H_{m+1}$ is free.\footnote{For example when $D-d=4$ and $m=p+n=1$, we will consider spaces where the boundary topology is $\partial X \cong \coprod_i S^3/\Gamma_i$, with $\Gamma_i \subset SU(2)$ a finite group of ADE type. These spaces have $H_2(\partial X^\circ) = 0$, and $H_1(\partial X^\circ) = \prod_i \text{Ab}[\Gamma_i]$ pure torsion.}
The sequence then truncates to
\be \label{E:rel_hol-Xo}
    0 \rightarrow H_{m+1}(X) \xrightarrow{J_{m+1}} H_{m+1}(X, \partial X) \xrightarrow{D_{m+1}} H_{m}(\partial X) \xrightarrow{I_{m}} H_{m}( X) \rightarrow 0 \, .
\ee
There is a dual sequence in cohomology,
\be
    0 \rightarrow H^{\widetilde{d}-(m+1)}(X, \partial X) \xrightarrow{{J}^*_{m+1}} H^{\widetilde{d}-(m+1)}(X) \xrightarrow{{D}^*_{m+1}} H^{\widetilde{d}-(m+1)}(\partial X) \xrightarrow{{I}^*_{m}} H^{\widetilde{d}-m}(X, \partial X) \rightarrow 0 \, ,
\ee
where $\widetilde{d}=D-d$, and each term is isomorphic to the corresponding term in \eqref{E:rel_hol-Xo} by Lefschetz-duality.

Note that the gauge fields of the $(d+1)$-dimensional SymTFT are obtained from Kaluza--Klein reductions of the $D$-dimensional string/M-theory fields on $H^{\widetilde{d}-(m+1)}(\partial X)$.
In contrast, dynamical gauge fields in the $d$-spacetime dimensions are obtained from reducing the string/M-theory fields on $H^{\widetilde{d}-(m+1)}(X)$.
When $H_m(X) = 0$ one can identify $H^{\widetilde{d}-(m+1)}(\partial X) = H^{\widetilde{d}-(m+1)}(X)/\text{im}({J}^*_{m+1})$ as the discrete group that remains after the dynamical $U(1)$s --- arising from the free group $H^{\widetilde{d}-(m+1)}(X)$ --- are screened by dynamical charges labelled by $H^{\widetilde{d}-(m+1)}(X, \partial X) \cong H_{m+1}(X)$.

In general, however, $H^{\widetilde{d}-(m+1)}(X)$ can contain a torsion subgroup.
That is, by the Universal Coefficient Theorem, we have
\begin{align}
    H^{\widetilde{d}-(m+1)}(X) = \text{Hom}(H_{m+1}(X), \Z) \oplus \text{Ext}(H_m(X), \Z) \, ,
\end{align}
where $\text{Ext}(H_m(X),\mathbb{Z}) =: H_m(X)^\vee \cong \text{Tor}(H_m(X))$.
Because $H_{m+1}(X) \cong H^{\widetilde{d}-(m+1)}(X, \partial X)$ is free, its image under the injective map $J_{m+1} \simeq {J}^*_{m+1}$ must lie in $\text{Hom}(H_{m+1}(X), \mathbb{Z})$.

This allows us to rewrite the relative (co-)homology sequences above as short exact sequences,
\begin{equation}\label{eq:comm-diag_of_SES_rel-hom}
	\begin{tikzcd}[row sep = normal, column sep= normal  ]
		0 \arrow[r] & \frac{\text{Hom}(H_{m+1}(X), \mathbb{Z})}{J_{m+1}(H_{m+1}(X))} \oplus H_{m}(X)^\vee \arrow{r}{}{\widetilde{D}_{m+1}^*} \arrow{d}{}{\cong} & H^{\widetilde{d}-(m+1)}(\partial X) \arrow{r}{}{I_m^*} \arrow{d}{}{\cong} & H^{\widetilde{d}-m}(X, \partial X) \arrow{d}{}{\cong} \arrow[r] & 0 \\
		0 \arrow[r] & \frac{H_{m+1}(X,\partial X)}{J_{m+1}(H_{m+1}(X))} \arrow{r}{}{\widetilde{D}_{m+1}} \oplus H_m(X)^\vee & H_{m}(\partial X) \arrow{r}{}{I_m} & H_m(X) \arrow[r] & 0
	\end{tikzcd}
\end{equation}
where we now identify 
\be\label{E:cohomSeqRes}
    \widetilde{D}_{m+1}^*\left( \frac{\text{Hom}(H_{m+1}(X), \mathbb{Z})}{J_{m+1}(H_{m+1}(X))} \right) = \frac{\text{Hom}(H_{m+1}(X), \mathbb{Z})}{J_{m+1}(H_{m+1}(X))} \subset H^{\widetilde{d}-(m+1)}(\partial X)    
\ee
as the background fields that arise from screening dynamical $U(1)$ gauge fields.

Let us emphasize that $H^{\widetilde{d}-(m+1)}(\partial X)$ contains in general also the torsional subgroup $\widetilde{D}_{m+1}^*(H_m(X)^\vee) \cong H_m(X)^\vee \subset H^{\widetilde{d}-(m+1)}(X)$.
This means that the SymTFT field arising from KK-reducing along a $(\widetilde{d}-(m+1))$-cocycle $\breve{\gamma}_{\widetilde{d}-(m+1)} \in H^{\widetilde{d}-(m+1)}(\partial X)$ is not a background field for the $d$-dimensional boundary QFT, but rather itself a dynamical field arising from KK-reducing along $\breve{\gamma}_{\widetilde{d}-(m+1)} \in H^{\widetilde{d}-(m+1)}(X)$.
We will come back to this phenomenon in Section \ref{sec:internalspacewithinherentchoiceofbc}.

\paragraph{Defect group. } By wrapping BPS branes on non-compact cycles of $X$ extending from the singularity to the boundary one can obtain supersymmetric non-dynamical defects. Similarly by wrapping BPS branes on compact cycles of $X$ one can generate dynamical objects that can screen some of the former defects. This results in the so called defect group \cite{DelZotto:2015isa, Albertini:2020mdx, Morrison:2020ool} which is the set of all unscreened defects
\be
\label{E:Hom2defect}
D=\bigoplus_{n=1}^{d-1} D^{(n)}\,,\qquad D^{(n)} = \bigoplus_{p} \frac{H_{p-n+1}(X,\partial X)}{J_{p-n+1}(H_{p-n+1}(X))} \cong \bigoplus_{p} \frac{H_{p-n+1}(X,\partial X)}{H_{p-n+1}(X)}\,,
\ee
where $n$ is the dimension of the defect, and $p$ is the spatial dimension of supersymmetric p-branes in the chosen string/M-theory. These defects of dimension $n$ are the charged objects under the possible $n$-form global symmetry.

\paragraph{Polarization and boundary conditions. } As explained in Section \ref{sec:defectgroupsymtftandpol} the defect group has a natural pairing. This has been given various equivalent string theory explanations including in terms of flux non-commutativity \cite{Witten:1998wy,GarciaEtxebarria:2019caf} and linking configurations of branes \cite{Apruzzi:2023uma, Heckman:2022muc}.

In the geometric setting, this pairing arises from the intersection product on $X$, or the linking product on $\partial X$.
Namely, $n$-dimensional objects from $p$-branes wrapped on $(p-n+1)$-dimensional relative cycles $\Sigma_{p-n+1}$, and $m$-dimensional objects from $(q=D-p-4)$-branes\footnote{Recall that in string/M-theory in $D$ dimensions, for $p$- and $q$-branes electromagnetically dual to each other, one generally needs $p+q = D-4$.} wrapped on $(q-m+1)$-dimensional cycles $\Sigma_{q-m+1}$ will also be electric-magnetic duals to each other in $d$-dimensional spacetime, if $n+m = d-2$, which is equivalent to $(p-n+1) + (q-m+1) = D-d = \widetilde{d}$.
The defect group pairing \eqref{eq:defect_group-pairing} is then determined by the geometric linking pairing 
\begin{align}
    \ell: H_k (\partial X) \times H_{\widetilde{d}-1 -k}(\partial X) \rightarrow \mathbb{Q}/\Z
\end{align}
as
\begin{align}
    \langle \Sigma_{p-n+1} , \Sigma_{q-m+1} \rangle = \ell (D_{p-n+1} (\Sigma_{p-n+1}) , D_{q-m+1}(\Sigma_{q-m+1}) ) \, .
\end{align}

In general, in order to get an absolute $d$ dimensional field theory one would need to choose a polarization that will reduce the defect group to the global symmetry of the field theory. This is to avoid a situation where there are extended operators in the field theory which have mutually non-local correlators \cite{Aharony:2013hda}.
Concretely, one must only allow $p$- and $(q=D-p-4)$-branes to wrap respective cycles such that $\langle \Sigma_{p-n+1} , \Sigma_{q-m+1} \rangle =0$.
Focusing on the $n$-dimensional objects, this is achieved in the SymTFT framework by giving $p$-branes on the boundary cycles $\sigma_{p-n} = D_{p-n+1}(\Sigma_{p-n+1})$ Dirichlet boundary conditions, forcing the corresponding SymTFT defects to stretch between the boundary and the SQFT interface.
This procedure gives rise to extended $n$-dimensional objects in $d$-dimensional theory charged under a higher-form global symmetry.
The symmetry generators are precisely the $q$-branes wrapping boundary cycles $\widetilde\sigma_{q-m}$ which link \emph{non}-trivially with $\sigma_{p-n}$; in the SymTFT they arise by imposing Neumann boundary conditions for $q$-branes on these $\widetilde\sigma_{q-m}$, with the charge given by $\ell(\sigma_{p-n}, \widetilde\sigma_{q-m})$.
Analogously, the $m$-dimensional charged objects from $q$-branes with Dirichlet boundary conditions on $\sigma_{q-m} = D_{q-m+1}(\Sigma_{q-m+1})$ are charged under symmetry generators which are $p$-branes on $\widetilde\sigma_{p-n}$ with Neumann boundary conditions.

Recently it has been understood that the string theory origin of the topological generators of generalized symmetries are branes wrapped on boundary cycles ``at infinity'' \cite{Apruzzi:2022rei,Heckman:2022muc,GarciaEtxebarria:2022vzq}. The above branes given Neumann boundary conditions are exactly these generators: they will correspond to topological defects generating global symmetries in the absolute theory. Finally note that the non-torsional factors of the defect group $D$ will correspond to the continuous symmetries, while the torsional factors will correspond to the discrete symmetries. In this work we will focus on the latter.

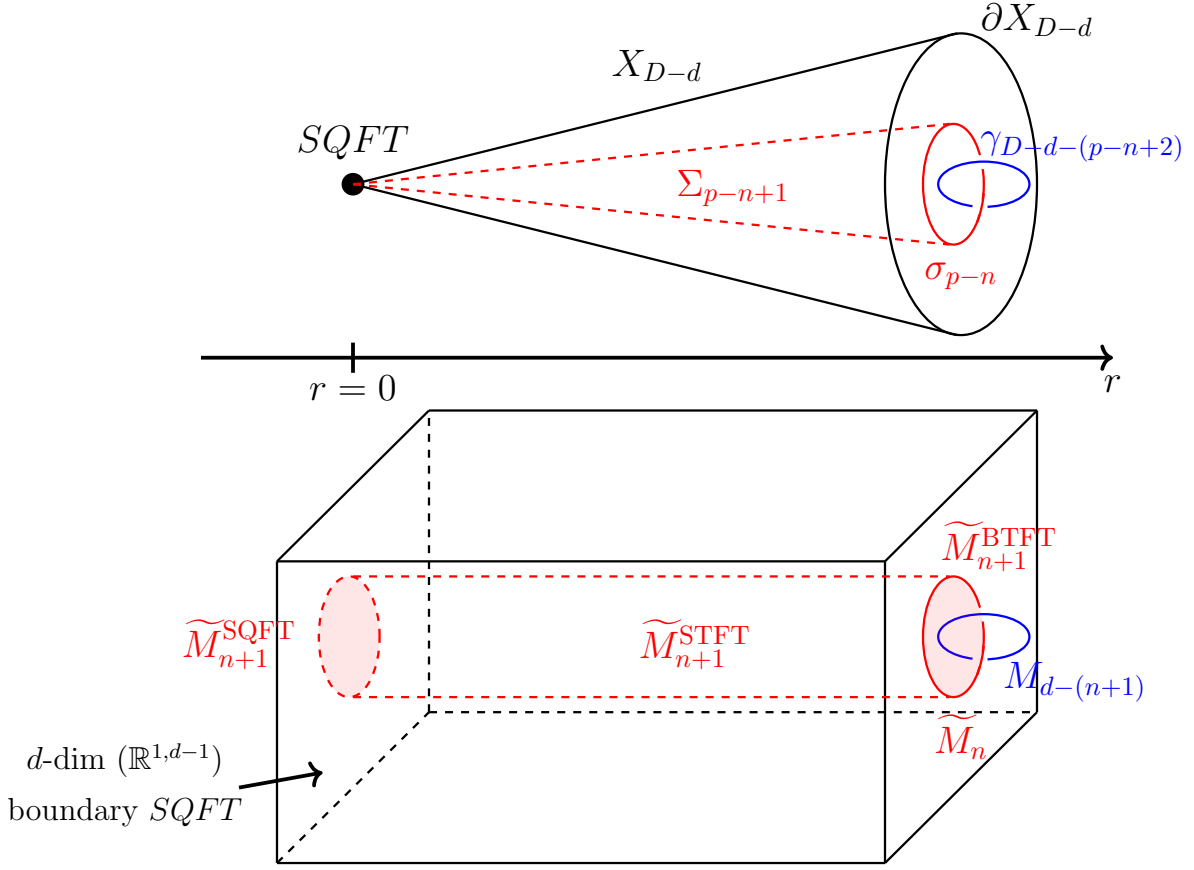
\begin{figure}[p]
$$
\begin{tikzpicture}[x=1cm,y=1cm] 

\begin{scope}[shift={(0,-0.7)}] 

\fill[black] (-4,3) circle (0.15);
\draw[line width=0.3mm] (4,3) ellipse (1 and 2);
\draw[line width=0.3mm] (-4,3)--(3.95,1); 
\draw[line width=0.3mm] (-4,3)--(3.95,5); 

\node[above] at (5,4.8) {\Large{$\partial X_{D-d}$}};
\node[above] at (0,4.2) {\Large{$X_{D-d}$}};
\node[above] at (-4,3.2) {\Large{$SQFT$}};

\draw[line width=0.3mm,red] (3.9,3) ellipse (0.4 and 0.8);
\draw[line width=0.3mm,red,dashed] (-4,3)--(3.85,2.2); 
\draw[line width=0.3mm,red,dashed] (-4,3)--(3.85,3.8); 
\node[red] at (1,2.95) {\Large{$\Sigma_{p-n+1}$}};
\node[below,red] at (4,2.1) {\Large{$\sigma_{p-n}$}};

\fill[white] (4.3,3.3) circle (0.1);
\draw[line width=0.3mm,blue] (4.3,3) ellipse (0.6 and 0.3);
\fill[white] (4.25,2.7) circle (0.1);
\draw[line width=0.3mm,red] (4.215,2.5) arc (340:355:1.5);
\node[blue] at (5.6,3.5) {\Large{$\gamma_{D-d-(p-n+2)}$}};

\end{scope}

\begin{scope}[shift={(0,0.3)}] 

\draw[line width=0.3mm] (-5,-3)--(3,-3); 
\draw[line width=0.3mm] (-5,-7)--(3,-7); 
\draw[line width=0.3mm] (-5,-3) -- (-5,-7);
\draw[line width=0.3mm] (3,-3) -- (3,-7); 

\draw[line width=0.3mm] (-5,-3)--(-3,-1); 
\draw[line width=0.3mm] (3,-3)--(5,-1); 
\draw[line width=0.3mm] (-3,-1) -- (5,-1);

\draw[line width=0.3mm] (3,-7)--(5,-5); 
\draw[line width=0.3mm] (5,-5) -- (5,-1);

\draw[line width=0.3mm,dashed] (-5,-7)--(-3,-5); 
\draw[line width=0.3mm,dashed] (-3,-5) -- (-3,-1);
\draw[line width=0.3mm,dashed] (-3,-5) -- (5,-5);

\draw[line width=0.3mm,red] (3.9,3-7) ellipse (0.4 and 0.8);
\draw[line width=0.3mm,red,dashed] (-4.05,3-7) ellipse (0.4 and 0.8);
\draw[line width=0.3mm,red,dashed] (-4,2.2-7)--(3.85,2.2-7); 
\draw[line width=0.3mm,red,dashed] (-4,3.8-7)--(3.85,3.8-7); 
\node[red] at (0.5,2.9-7) {\Large{$\widetilde{M}_{n+1}^{\rm{STFT}}$}};
\node[below,red] at (4,2.1-7) {\Large{$\widetilde{M}_{n}$}};

\fill[white] (4.3,3.3-7) circle (0.1);
\draw[line width=0.3mm,blue] (4.3,3-7) ellipse (0.6 and 0.3);
\fill[white] (4.25,2.7-7) circle (0.1);
\draw[line width=0.3mm,red] (4.215,2.5-7) arc (340:355:1.5);
\node[blue] at (5.5,2.4-7) {\Large{$M_{d-(n+1)}$}};

\fill[red,opacity=0.1] (3.9,3-7) ellipse (0.4 and 0.8);
\fill[red,opacity=0.1] (-4.05,3-7) ellipse (0.4 and 0.8);
\node[red] at (-5.5,2.9-7) {\Large{$\widetilde{M}_{n+1}^{\rm{SQFT}}$}};
\node[red] at (4.5,4.2-7) {\Large{$\widetilde{M}_{n+1}^{\rm{BTFT}}$}};

\node[above] at (-7,-6) {\large{$d$-dim ($\mathbb{R}^{1,d-1}$)}};
\node[below] at (-7,-6) {\large{boundary $SQFT$}};
\draw[->, line width=0.5mm] (-5.5,-6)--(-4.4,-5.8); 

\end{scope}

\draw[->, line width=0.5mm] (-6,0)--(6,0); 
\node[below] at (6,-0.1) {\Large{$r$}};
\draw[line width=0.5mm] (-4,-0.2)--(-4,0.2); 
\node[below] at (-4,-0.1) {\Large{$r=0$}};

\end{tikzpicture}
$$

\caption{The geometry of the space $\mathbb{R}^{1,d-1} \times X_{D-d}$ where the upper part displays the non-compact space $X$ while the lower part displays the $\mathbb{R}^{1,d-1}$ and the radial direction of $X$.
The lower part effectively shows the SymTFT with the SQFT boundary on the left and the topological boundary on the right. On the SQFT boundary we display a $(n+1)$-dimensional surface $\widetilde{M}_{n+1}^{\rm{SQFT}}$ as a pinkish filled ellipse. Similarly on the topological boundary we display a $(n+1)$-dimensional surface $\widetilde{M}_{n+1}^{\rm{BTFT}}$ as a pinkish filled ellipse, with $\partial \widetilde{M}_{n+1}^{\rm{BTFT}}=\widetilde{M}_{n}$. The unification $\widetilde{M}_{n+1} = \widetilde{M}_{n+1}^{\rm{SQFT}} \cup \widetilde{M}_{n+1}^{\rm{STFT}} \cup \widetilde{M}_{n+1}^{\rm{BTFT}}$ creates a closed $(n+1)$-dimensional surface spanning between the two boundaries of the SymTFT, drawn as a capped cylinder. An absolute theory is found by enforcing boundary conditions on BPS-branes wrapping the topological boundary cycles. Let us consider wrapping a $p$-brane on $\sigma_{p-n}\times \widetilde{M}_{n+1}$. When $\sigma_{p-n}$ is trivialized in $X$, meaning $\sigma_{p-n}\in \text{ker}(I_{p-n})$, the operator supported on $\widetilde{M}_{n+1}^{\rm{SQFT}}$ is trivialized, while the topological operator on $\widetilde{M}_{n+1}^{\rm{BTFT}}$ is trivialized when we choose Dirichlet boundary conditions for the $p$-brane wrapping the boundary cycle $\sigma_{p-n}$. This means the corresponding $n$-dimensional operator in spacetime will be a genuine charged operator under a global $n$-form symmetry when we choose Dirichlet boundary conditions. If we choose Neumann boundary conditions this $n$-dimensional operator will be non-genuine as it will be the boundary of a non-trivial operator supported on $\widetilde{M}_{n+1}^{\rm{BTFT}}$. The genuine operator will be charged under a global $n$-form symmetry generated by a $(d-n-1)$-dimensional operator created by wrapping a magnetic dual $(D-p-4)$-brane on $\gamma_{D-d-(p-n+2)} \times M_{d-(n+1)}$ if we set Neumann boundary condition for it on the topological boundary.
Note that both the $\sigma_{p-n}$ and $\gamma_{D-d-(p-n+2)}$ cycles, and the $\widetilde{M}_{n}$ and $M_{d-(n+1)}$ cycles link in $\partial X$ and $\mathbb{R}^{1,d-1}$ spaces, respectively. This is required in order for the resulting charged objects to be charged under the symmetry generated by their respective topological operators.
\label{F:Geometry&Defects}}
\end{figure}

In Figure \ref{F:Geometry&Defects} we summarize the above story, whilst also explaining the action of the topological operators on the extended defects from a brane perspective.

\paragraph{The SymTFT and supergravity. } In \cite{Apruzzi:2021nmk} the SymTFT was first derived in string theory constructions in a geometric engineering context on non-compact space $X$ in M-theory. In this case, the TFT was derived by a differential cohomology reduction on $\partial X$. This generically results in a TQFT action of the form \eqref{eq:SymTFTabelianfinite} including BF terms and Chern-Simons (CS) terms depending on homology of $X$ and the string/M-theory BPS brane dimensions. A similar procedure is valid in holographic contexts \cite{Witten:1998wy, Belov:2006xj}.
Recent applications include both geometric engineering \cite{Apruzzi:2021nmk, Apruzzi:2022dlm} and holographic contexts \cite{vanBeest:2022fss,  Antinucci:2022vyk, Apruzzi:2022rei, Bah:2023ymy}.

\paragraph{Example. } Consider a theory engineered by placing M-theory on an internal geometry $X^\text{loc}$ created by the disjoint union of all the singular local neighborhoods of the space $T^4/\Z_2$, which corresponds to $16$ $A_1$ singularities, or $\mathbb{C}^2/\mathbb{Z}_2$.
The exact sequence of \eqref{E:rel_hol-Xo} is translated for $p=2$ and $n=1$ to
\be
    0 \rightarrow H_2(X^\text{loc}) \xrightarrow{J_2} H_2(X^\text{loc}, \partial X^\text{loc}) \xrightarrow{D_2} H_1(\partial X^\text{loc}) \xrightarrow{I_1} H_1( X^\text{loc}) \rightarrow 0 \, .
\ee
which for this example gives the exact sequence of groups
\be
    0 \rightarrow \Z^{16} \xrightarrow{J_2} \Z_2^{16} \xrightarrow{D_2} \Z_2^{16} \xrightarrow{I_1} 0 \rightarrow 0 \, .
\ee
The set of all unscreened defects given in \eqref{E:Hom2defect} indicates the defect group in this case is
\be
D = \left(\Z_2^{(1)} \times \Z_2^{(4)}\right)^{16}\,.
\ee
Here $\Z_2^{(n)}$ is a putative $n$-form symmetry in the relative QFT.
Because $H_1(X^\text{loc})=0$, the linking of boundary 1-cycles can be computed, using $H_1(\partial X^\text{loc}) = H_2(X^\text{loc}, \partial X^\text{loc})/H_2(X^\text{loc}))$, using the intersection product on $H_2(X^\text{loc}, \partial X^\text{loc}) = ({\bf A}_1^*)^{\oplus 16}$.
The dual of the ${\bf A}_1$ lattice has one generator $e^*$ with self-pairing $e^* \cdot e^* = \tfrac12$, and is orthogonal to the other 15 generators that span $H_2(X^\text{loc}, \partial X^\text{loc})$.
Thus, for the boundary cycles associated with each $\{0, \sigma_1^{i}\}= \Z_2^i\subset \Z_2^{16} = H_1(\partial X^\text{loc})$, the pairing is $\langle \sigma_1^i \, , \, \sigma_1^j \rangle = \tfrac{\delta{ij}}{2}$.
This means that M2- and M5-branes link non-trivially if and only if they wrap the same cycle $\sigma_1^i$.
\begin{table}[h]
\renewcommand{\arraystretch}{1.2}
    \centering
    \begin{tabular}{|c|c|c|}
    \hline
         & $\Z_2^{(1),i}$ & $\Z_2^{(4),i}$  \\
         \hline
         \hline
         charged objects & M2 on $\sigma_1^i \times \mathbb{R}^+$ & M5 on $\sigma_1^i \times \mathbb{R}^+$ \\
         generators & M5 on $\sigma_1^i$ & M2 on $\sigma_1^i$ \\
         \hline
    \end{tabular}
    \caption{Generators and charged objects in the \textit{defect group}. We stress that these are ``defect group symmetries'' --- a consistent polarization must be chosen amongst these to obtain true symmetries of an absolute QFT. $\mathbb{R}^+$ denotes the radial direction of the geometric construction.}
    \label{tab:wrappedbranechargeandgenlabels}
\end{table}

Let us analyze the $\Z_2^{(n),1}$ symmetries for concreteness. Because $\sigma_1^1$ is self linked, we are not free to place both M5 and M2 branes on it --- this will lead to non-local correlators in the boundary theory. We must pick consistent boundary conditions. As explained above, this is equivalent to a choice of which operator trivializes on the boundary.

One choice of boundary condition would be to maintain the $\Z_2^{(1),1}$ symmetry on the boundary theory: meaning $\Z_2^{(4),1}$ is gauged. Equivalently, at the level of a SymTFT --- the background field for the former is given Dirichlet boundary conditions, while the latter has Neumann. Alternatively, we could keep $\Z_2^{(4),1}$: this amounts to reversing everything. These two choices correspond to keeping either column 1 or column 2 of Table \ref{tab:wrappedbranechargeandgenlabels} - they are mutually exclusive at the level of the absolute QFT (neglecting for now the option of mixed polarizations).

For completeness, we have the same flexibility for the symmetries with all other $i$ superscripts. This system can be equivalently described by $16$ BF terms
\be
S_{\rm SymTFT} = \int_{M_{8}} 2 \sum_{i=1}^{16} B_{2}^{(i)} dC_{5}^{(i)} \,,
\ee
where $B_{2}^{(i)}$ are background gauge fields for $\Z_2^{(1),i}$ symmetries, and similarly $C_{5}^{(i)}$ are background gauge fields for $\Z_2^{(4),i}$.

\subsection{Internal space with inherent choice of boundary conditions.} 
\label{sec:internalspacewithinherentchoiceofbc}

In the 7d example above, $H_{1}(X) = 0$ simplified the discussion of the defect group considerably, because it agrees with the symmetry group $H_1(\partial X)^{(1)} \oplus H_1(\partial X)^{(4)}$ of the SymTFT obtained from dimensional reduction on $\partial X$.
In this section we will consider the more general case where we place the $D$ dimensional string/M-theory on a background
\be
\mathbb{R}^{1,d-1}\times X_{D-d}^\circ\,,
\ee
where $X^\circ_{D-d} \equiv X^\circ$ is a non-compact Ricci flat special holonomy space with $H_{p-n}(X^\circ) \neq 0$.
For our physical considerations, such a space arises as the complement of the singular loci inside a compact manifold.\footnote{For example when $D-d=4$ it would correspond to a K3 space with $\mathbb{C}^2/\Gamma$ sub-spaces removed.} In other words, gluing this subspace to the singular geometries discussed before would correspond to coupling the spacetime field theory to gravity. 
We will assume that $X^\circ$ upholds the other assumptions around \eqref{E:rel_hol-Xo}.

We would like to argue that for compactifications on such $X^\circ$, there is  only one choice of boundary conditions allowed for BPS-branes wrapping some cycles in the internal geometry, as stated after \eqref{E:cohomSeqRes}. 
This is because the strata (summands labelled by $p$ in \eqref{E:Hom2defect}) of the defect group $D^{(n)}$ is no longer the same group seen by the boundary reduction,
\be
\frac{H_{p-n+1}(X^\circ,\partial X^\circ)}{H_{p-n+1}(X^\circ)} \ne H_{p-n}(\partial X^\circ) \, ,
\ee
due to $H_{p-n}(X^\circ)\ne 0$. 
This means that we will have several $(p-n)$-cycles $\sigma$ on $\partial X^\circ$ that will not be ``trivialized'' in $X^\circ$, i.e., lie in the kernel of $I_{p-n}$ in \eqref{eq:comm-diag_of_SES_rel-hom} which is induced by the inclusion $\partial X^\circ \hookrightarrow X^\circ$.
Therefore, there exists a $(p-n+1)$-chain in $X^\circ$ which ``interpolates'' between $\sigma$ on the asymptotic boundary $\partial X^\circ$, and $I_{p-n}(\sigma)$ in the interior of $X^\circ$.
Because the $(p-n+1)$-chain exists in the interior of $X^\circ$, it must be wrapped.\footnote{This is similar to $(p-n+1)$-cycles in $X^\circ$, we are always wrapped, and give rise to, e.g., W-bosons for $(p-n+1)=2$ in M-theory.} Now imagine giving $p$-branes on $\sigma \notin \text{ker}(I_{p-n})$ Dirichlet boundary conditions, and wrap $p$-branes on the $(p-n+1)$-chains in $X^\circ$ which has $\sigma$ as part of its boundary.
Then, the resulting $n$-dimensional objects in spacetime must be attached to $(n+1)$-dimensional topological objects corresponding to the $p$-brane wrapping the $(p-n)$-cycles $I_{p-n}(\sigma) \subset X^\circ$.
Therefore, the operator constructed above is actually a non-genuine operator and not a charged object.
In addition, we can consider operators constructed by wrapping $(D-p-4)$-branes on $(D-d-(p-n+2))$ boundary cycles $\gamma$ linked with $\sigma$ in $\partial X^\circ$. These operators will be paired to the former operators in a manner requiring us to give them Neumann boundary conditions \cite{Heckman:2022muc}. This means that wrapping relative cycles with $\gamma$ as their boundary will also generate non-genuine operators. So, in total we will not get a maximal set of mutually local operators, as required for a polarization. 
This indicates that in such a geometry we can only choose Neumann boundary conditions for $p$-branes wrapping such $(p-n)$-cycles $\sigma\notin \text{ker}(I_{p-n})$.

Let us compare the current case of a geometry with $H_{p-n}(X^\circ)\ne 0$ to the former case where we had $H_{p-n}(X)=0$. 
Assuming that the boundaries agree, $\partial X = \partial X^\circ$, then we would obtain in both cases topological $(n+1)$-dimensional objects in the $(d+1)$-dimensional bulk SymTFT that are free to move to the $d$-dimensional boundary field theory, if we impose Neumann boundary conditions on $(p-n)$-cycles in $\partial X = \partial X^\circ$.
The difference between the cases appears when we try to impose Dirichlet boundary conditions on $p$-branes wrapping such $(p-n)$-cycles $\sigma$. 
If $\sigma$ is trivialized in $X^\circ$ or $X$, meaning $\sigma \in \text{ker}(I_{p-n}) = \text{Im}(D_{p-n+1})$, the operators generated will be $n$-dimensional charged objects in spacetime with the topological $(n+1)$-dimensional operators attached to them being trivialized. 
This can be understood by considering operators generated by wrapping $p$-branes on an internal space $(p-n+1)$ relative cycle $\Sigma_{p-n+1}$ with $D_{p-n+1}\Sigma_{p-n+1} = \sigma$. 
In the SymTFT $(d+1)$-dimensional space this results in an operator with worldvolume $\widetilde{M}_{n+1}$, as shown in the lower part of Figure \ref{F:Geometry&Defects}, as the union $\widetilde{M}_{n+1} = \widetilde{M}_{n+1}^{\rm{SQFT}} \cup \widetilde{M}_{n+1}^{\rm{STFT}} \cup \widetilde{M}_{n+1}^{\rm{BTFT}}$. Due to the Dirichlet boundary conditions the operator supported on $\widetilde{M}_{n+1}^{\rm{BTFT}}$ is trivialized, and the operator supported on $\widetilde{M}_{n+1}^{\rm{SQFT}}$ is trivialized since the boundary $(p-n)$-cycles lie in $\text{ker}(I_{p-n})$. 
For the space $X^\circ$ there is a possibility for boundary $(p-n)$-cycles not to lie in $\text{ker}(I_{p-n})$. 
In this case the operator supported on $\widetilde{M}_{n+1}^{\rm{SQFT}}$ is not trivialized giving a $(n+1)$ topological operator attached to a $n$-dimensional operator. This will correspond to a non-genuine operator instead of an expected charged operator, meaning this Dirichlet boundary condition is inconsistent with the bulk of $X^\circ$ and we must choose Neumann boundary conditions.

Note that the inherent choice of Neumann boundary conditions for a subset of boundary $(p-n)$-cycles also means that the Pontryagin-dual subset of $(\widetilde{d}-p+n-2)$-cycles, which link non-trivially with those $(p-n)$-cycles, must receive Dirichlet boundary conditions.
The latter gives rise to charged defects in the absolute theory, which then must be mutually local with respect to each other \cite{Heckman:2022muc}.

\paragraph{Example. } 
To illustrate the above discussion, consider a theory engineered by placing M-theory with on an internal geometry $X^\circ_4$ generated by removing all the singular points of the space $T^4/\Z_2$. 
As such, the boundary homology will be the same as for $\partial X^\text{loc}$ above, namely, $H_1(\partial X^\circ) = \Z_2^{16}$.
The exact sequence of \eqref{E:rel_hol-Xo} is translated for $p=2$ and $n=1$ to
\be
    0 \rightarrow H_2(X^\circ) \xrightarrow{J_2} H_2(X^\circ, \partial X^\circ) \xrightarrow{D_2} H_1(\partial X^\circ) \xrightarrow{I_1} H_1( X^\circ) \rightarrow 0 \, ,
\ee
which for this example gives the exact sequence of groups
\be
    0 \rightarrow \Z^6 \xrightarrow{J_2} \Z^6 \oplus \Z_2^5 \xrightarrow{D_2} \Z_2^{16} \xrightarrow{I_1} \Z_2^5 \rightarrow 0 \, .
\ee
From $H_2(X^\circ) = \mathbb{Z}^6$, the local gauge theory is $\mathfrak{u}(1)^6$.
Here, the map $J_2$ is multiplication by 2 into the $\mathbb{Z}^6$ factors in $H_2(X^\circ, \partial X^\circ)$ \cite{Cvetic:2023pgm} (see also \cite{Taimanov_2018,10951846-5fb6-3a7d-9f1d-aabe9b3c1767}).

This means that the set of all unscreened defects given in \eqref{E:Hom2defect}, i.e., the defect group in this case, is
\be
D = \left(\Z_2^{(1)} \times \Z_2^{(4)}\right)^{11}\,.
\ee
On the other hand if we consider the SymTFT of this theory by deriving it from the homological structure of $\partial X^\circ$ it seems the defect group is actually
\be
D_\text{SymTFT} = \left(\Z_2^{(1)} \times \Z_2^{(4)}\right)^{16}\,.
\ee
The SymTFT can be described by BF terms as follows
\be
S_{\rm SymTFT} = \int_{M_{8}} 2 \sum_{i=1}^{16} B_{2}^{(i)} dC_{5}^{(i)} \,,
\ee
where $B_{2}^{(i)}$($C_{5}^{(i)}$) are the background gauge field for $\Z_2^{(1),i}$($\Z_2^{(4),i}$). Note this is exactly the same SymTFT as in the former example of Subsection \ref{S:StringSyms}. Before we address this alleged contradiction let us only consider the parts of the defect group which does not arise from the free part of $H_2(X^\circ,\partial X^\circ)$ that are the same for both defect groups. This will correspond to removing a $\left(\Z_2^{(1)}\times \Z_2^{(4)}\right)^6$ factor from both defect groups.
This leaves us with a residual SymTFT defect group $\left(\Z_2^{(1)} \times \Z_2^{(4)}\right)^{10}$ which still does not match the residual of the defect group $D$ $\left(\Z_2^{(1)} \times \Z_2^{(4)}\right)^5$. This is resolved by understanding that attaching the $X^\circ$ bulk inherently implements Neumann boundary conditions to background gauge fields of a subgroup $\left(\Z_2^{(1)} \times \Z_2^{(4)}\right)^5$, associated with cycles $\sigma_i \notin \text{ker}(I_1)$, on the SymTFT, as explained after \eqref{E:cohomSeqRes}.

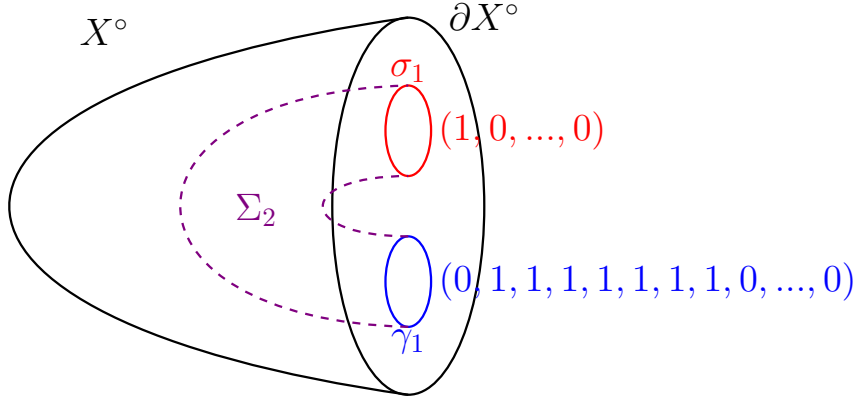
\begin{figure}[t!]
$$
\begin{tikzpicture}[x=1cm,y=1cm] 

\draw[line width=0.3mm] (4,3) ellipse (1 and 2.5);
\draw[line width=0.3mm] (4,0.5) .. controls (-3,1.5) and (-3,4.5) .. (4,5.5);

\node[above] at (5,5.2) {\Large{$\partial X^\circ$}};
\node[above] at (0,5) {\Large{$X^\circ$}};

\draw[line width=0.3mm,red] (4,4) ellipse (0.3 and 0.6);
\draw[line width=0.3mm,blue] (4,2) ellipse (0.3 and 0.6);

\node[red] at (4,4.8) {\Large{$\sigma_{1}$}};
\node[blue] at (4,1.2) {\Large{$\gamma_{1}$}};
\node[red] at (5.5,4) {\Large{$(1,0,...,0)$}};
\node[blue] at (7.15,2) {\Large{$(0,1,1,1,1,1,1,1,0,...,0)$}};

\draw[line width=0.3mm,violet,dashed] (4,2.6) .. controls (2.5,2.6) and (2.5,3.4) .. (4,3.4);
\draw[line width=0.3mm,violet,dashed] (4,1.4) .. controls (0,1.5) and (0,4.5) .. (4,4.6);
\node[violet] at (2,2.95) {\Large{$\Sigma_{2}$}};

\end{tikzpicture}
$$

\caption{The relative 2-cycle $\Sigma_2$ with $\sigma_1$ as one of his boundaries in the internal geometry of $X^\circ$. One can see that the second boundary of $\Sigma_2$ is $\gamma_1$ where $\sigma_1= 
\gamma_1 + (1,1,1,1,1,1,1,1,0,...,0)$, with $(1,1,1,1,1,1,1,1,0,...,0)\in \text{ker}(I_1)$.
\label{F:BCgeometry}}
\end{figure}

Let us also show the imposed boundary conditions from a more physical perspective. For this end, we consider a 1-cycle $\sigma_1\in H_{1}(\partial X^\circ)=\Z_2^{16}$ and also $\sigma_1\notin \text{ker}(I_1)$, generated by the element $(1,0,...,0)$. If we try to impose Dirichlet boundary conditions to M2-branes wrapping this cycle we find that wrapping M2-branes on the relative 2-cycle that $\sigma_1$ is its boundary will have another boundary with a boundary 1-cycle $\gamma_1$ such that $[\sigma_1]=[\gamma_1]$ in $H_{1}(X^\circ)$, see Figure \ref{F:BCgeometry}. In the SymTFT picture this will result in what is shown in the lower part of Figure \ref{F:Geometry&Defects} ($n=1$), with an operator supported on the two dimensional surface $\widetilde{M}_2=\widetilde{M}_{2}^{\rm{SQFT}} \cup \widetilde{M}_{2}^{\rm{STFT}} \cup \widetilde{M}_{2}^{\rm{BTFT}}$ where the part supported on $\widetilde{M}_{2}^{\rm{SQFT}}$ not being trivial. This will result in a non-genuine operator in space time as if we gave Neumann boundary conditions. Thus, we must choose Neumann boundary conditions for such a cycle. In our example there are five such $\Z_2$ cycles that must be given Neumann boundary conditions, and due to their pairing with the additional five $\Z_2$ cycles the latter must receive Dirichlet boundary conditions. Note that when we state boundary conditions for cycles we mean that any BPS-brane wrapping these cycles will get the same boundary condition. This completely determines the boundary conditions for all the cycles unrelated to the gravitational $\mathfrak{u}(1)$'s.

\subsection{Some specifics of SymTFTs on complex surfaces}\label{sec:surface-specifics}

As we will be dealing mostly with $X$ being complex surfaces (i.e., real four-dimensional mani\-folds), it is worth summarizing some aspects related to the fact that 2-cycles on $X$ intersect with themselves.
Likewise, their boundary 1-cycles will link with themselves in the three-dimensional boundary $\partial X$.
The intersection product on $H_2(X)$, which under our assumptions is a finitely generated free Abelian group, is a non-degenerate symmetric pairing
\begin{align}
    H_2(X) \times H_2(X) \rightarrow \mathbb{Z} \, , \quad (\Sigma, \widetilde{\Sigma}) \mapsto \Sigma \cdot \widetilde\Sigma \, ,
\end{align}
which gives $H_2(X)$ a lattice structure.
In this setting, we can identify the free part of the relative cycles, $\text{Hom}(H_2(X),\mathbb{Z}) \subset H_2(X, \partial X)$, with the \emph{dual lattice},
\begin{align}
    \text{Hom}(H_2(X),\mathbb{Z}) = H_2(X)^* := \left\{ \sum_i \lambda_i \, \Sigma_i \, , \, \lambda_i \in \mathbb{Q} \, \middle| \, \forall \widetilde\Sigma \in H_2(X) : \, \,  \sum_i \lambda_i \Sigma_i \cdot \widetilde\Sigma_i \in \mathbb{Z}  \right\} \, ,
\end{align}
which has the same rank as $H_2(X)$.
The intersection product extends naturally to the dual lattice, and generally gives fractional intersection numbers. 
This gives rise to a $\mathbb{Q}/\mathbb{Z}$-valued product on the \emph{discriminant group} $H_2(X)^*/H_2(X)$; the latter is clearly a subgroup of $H_1(\partial X)$, and the linking pairing on this subgroup agrees with the product induced by the intersection product,
\begin{align}
    \ell(D_2\Sigma^\text{rel}_2, D_2 \widetilde\Sigma^\text{rel}_2) = \Sigma_2^\text{rel} \cdot \widetilde\Sigma_2^\text{rel} \mod \mathbb{Z} \quad \text{for} \quad \Sigma^\text{rel} , \widetilde\Sigma^\text{rel} \in \text{Hom}(H_2(X), \mathbb{Z})\, .
\end{align}
We refer to Appendix \ref{sec:latticeappendix} for more background material on lattices.

\subsubsection*{Linking pairing, bulk torsion, and mutual locality}

As we have seen above, the boundary cohomology $H_{1}(\partial X^\circ)$ is richer than the discriminant group when $H_1(X^\circ) \neq 0$.
This has an important consequence for the linking properties of boundary 1-cycles and the appearance of torsion in $H_1(X^\circ)$.
Namely, given that we have argued above that any $p$-brane on $H_1(\partial X^\circ) \ni \sigma_1 \notin \text{ker}(I_1)$ must have Neumann boundary conditions, any $(q=D-p-4)$-brane wrapping 1-cycles $\widetilde\sigma_1$ that link non-trivially with these $\sigma_1$ must have Dirichlet boundary conditions, i.e., wrap the relative cycles $\widetilde\Sigma_2$ with $D_2\widetilde\Sigma_2 = \widetilde\sigma_1$.
But then, mutual locality for $p$- and $q$-branes wrapping these cycles requires that $\ell(\widetilde\sigma_1, \widetilde\sigma_1) = \widetilde\Sigma_2 \cdot \widetilde\Sigma_2 \mod \mathbb{Z} = 0$.\footnote{E.g., in M-theory compactified on $X^\circ$ to 7d, M2- and M5-branes wrapping relative 2-cycles with boundary $\widetilde\sigma_1$ give Wilson lines and 't Hooft 4-surfaces in spacetime, which must be mutually local to each other.}

Going back to the sequence \eqref{eq:comm-diag_of_SES_rel-hom}, which we reproduce here for convenience,
\begin{align}
    0 \rightarrow \underbrace{\frac{H_2(X^\circ)^*}{H_2(X^\circ)} \oplus H_1(X^\circ)^\vee}_{\cong H_2(X^\circ,\partial X^\circ)/H_2(X^\circ)} \xrightarrow{\widetilde{D}_2} H_1(\partial X^\circ) \xrightarrow{I_1} H_1(X^\circ) \rightarrow 0 \, ,
\end{align}
the above considerations imply the following.
First, the 1-cycles that are forced to have Dirichlet boundary conditions are in the subgroup $\widetilde{D}_2(H_1(X^\circ)^\vee) \cong H_1(X^\circ)^\vee \subset H_1(\partial X^\circ)$, and $H_1(\partial X^\circ)/\text{im}(\widetilde{D}_2) \cong H_1(X^\circ)$ is given Neumann boundary conditions.
The linking on $H_1(\partial X^\circ)$ then induces a non-degenerate linking pairing 
\begin{align}\label{eq:H1_pairing-Xo}
    H_1(X^\circ)^\vee \times H_1(X^\circ) \rightarrow \mathbb{Q}/\mathbb{Z} \, ,    
\end{align}
associated with the BF-terms in the SymTFT derived from $\partial X^\circ$.
Then, mutual locality for M2- and M5-brane states on these cycles require,
\begin{align}
    \ell(\widetilde\sigma_1, \widetilde\sigma'_1) = 0 \quad \text{for} \quad \widetilde\sigma_1, \widetilde\sigma_1' \in H_1(X^\circ)^\vee \subset H_1(\partial X^\circ) \, .
\end{align}
Furthermore, the linking pairing on $\widetilde{D}_2 [H_2(X^\circ)^*/H_2(X^\circ)]$ is, as usual, inherited from the (fractional) intersection pairing on $H_2(X^\circ)$, which is non-degenerate.
Since the pairing $H_1(\partial X^\circ)$ is non-degenerate, it is then natural to expect that $\ell(v_1, \widetilde\sigma_1) = 0$ for $v_1 \in \widetilde{D}_2[H_2(X^\circ)^* / H_2(X^\circ)]$ and $\widetilde\sigma_1 \in \widetilde{D}_2(H_1(X^\circ)^\vee)$.
Physically, this means that the inherent choice of Dirichlet boundary conditions for $H_1(X^\circ)^\vee$ does not impose any restrictions on boundary condition choices for cycles in $\widetilde{D}_2[H_2(X^\circ)^* / H_2(X^\circ)]$.
While we do not have a proper mathematical proof of these assertions for general $X^\circ$, we will later on argue for these properties in the concrete setting where $X^\circ$ is the complement of ADE singularities inside of a K3 surface.
These will be the relevant setting for the physical models that appear in this paper.

In summary, we have shown that the effective field theory from M-theory on $X^\circ$ with $H_1(X^\circ) \neq 0$ comes with an inherent global symmetry $H_1(X^\circ)^{(1)} \oplus H_1(X^\circ)^{(4)}$, and has a ``reduced'' defect group $[H_2(X^\circ)^*/H_2(X^\circ)]^{(1)} \oplus [H_2(X^\circ)^*/H_2(X^\circ)]^{(4)}$ (compared to the naive group $H_1(\partial X^\circ)^{(1)} \times H_1(\partial X^\circ)^{(4)}$); the latter has a SymTFT description in terms of M-theory on $\partial X^\circ$, but with the inherent boundary condition choices.

\section{Maximally mixed polarization in K3-compactifications of M-theory}
\label{S:MMP-K3}

When we place string/M-theory on a compact internal space $X_{D-d}$, the resulting effective theory has dynamical gravity, and hence should have no global symmetries.
From the SymTFT perspective discussed above, this is obvious, because there is no boundary, $\partial X_{D-d} = \emptyset$.
However, from a low-energy EFT perspective, we have a supergravity theory with some massless matter charged under some gauge sector, which in general has a non-trivial defect group for which a polarization needs to be specified.
In the No-Global-Symmetries context, the subgroup with Neumann boundary conditions for its background gauge fields corresponds to gauged symmetries, and the subgroup with Dirichlet boundary conditions will be screened, i.e., broken.

To see which polarization describes the theory on the compact manifold, it is instructive to think of $X_{D-d}$ as gluing together two non-compact regions, $X^\circ$ and $X^\text{loc}$.
With the geometric details spelled out in \cite{Cvetic:2023pgm}, we provide a field theoretic extension, by interpreting this geometric operation as identifying the boundary conditions for the two SymTFTs on $X^\circ$ and $X^\text{loc}$, respectively.
This, in turn, allows us to give a characterization of the global structure of the low-energy EFT in terms of its SymTFT.

Let us look at the particular case of M-theory compactified on $X_4 \equiv X$ a K3 surface, giving rise to a 7d ${\cal N}=1$ supergravity theory.
The non-Abelian gauge sector comes from local ADE singularities of type $\mathfrak{g}_i$, modelled on $X_i \cong \mathbb{C}^2 / \Gamma_i$.
Throughout this section, we will work with (the homology of) the blow-up resolution $\widehat{X}_i$.
The root lattice $L_\mathfrak{g}$ of the ADE algebra $\mathfrak{g} = \bigoplus_i \mathfrak{g}_i$ is given by the homology of the ``local'' part $X^\text{loc} = \bigcup_i \widehat{X}_i$, $L_\mathfrak{g} = H_2(X^\text{loc})$.
The compact K3 is then the union $X = X^\text{loc} \cup X^\circ$ with another open patch $X^\circ$, which has no singularities and only contributes $b = \text{rank}(H_2(X^\circ))$ Abelian gauge factors.
The low-energy effective supergravity action is now completely determined by this gauge algebra specification.

Field theoretically, the global structure of the gauge group would now correspond to a choice of polarization for the ``product'' SymTFT, defined as M-theory on the disjoint union $\partial X^\text{loc} \coprod \partial X^\circ$.
From the top-down perspective, this choice is determined geometrically:
The intersection $X^\text{loc} \cap X^\circ$ deformation-retracts to $\partial X^\text{loc} = \partial X^\circ$, along which relative cycles are glued into compact cycles of $X$.
In what follows, we will explain how this process can be interpreted as a choice of boundary condition for the SymTFT, and present a succinct characterization of the resulting polarization, i.e., the global structure of the supergravity theory.

\paragraph{Maximally mixed polarization. } Concretely, we will demonstrate that gluing procedure enforces a \textit{maximally mixed polarization} on the 7d theory. Focusing in particular on the 1-form and 4-form symmetry contributions to the defect group of the 7d theory, we have
\be
D = D^{(1)} \oplus D^{(4)} \,.
\ee
As explained in Section \ref{sec:symtftintro}, a \textit{polarization} is a choice of subgroup $\Lambda \subset D$ which is maximal and isotropic. A \textit{maximally mixed} polarization in this instance is defined as one where
\be
\Lambda_{\text{maximally mixed}} = \Lambda_1 \oplus \Lambda_4 \,, \quad \Lambda_1 \cong \Lambda_4
\ee
in the notation of \eqref{eq:polarizationgrading} ($\Lambda_1 \subset D^{(1)}$ and $\Lambda_4 \subset D^{(4)}$), where $\cong$ is an isomorphism at the level of groups. In other words, there is an ``equal amount'' of 1-form and 4-form symmetry specified in the polarization.

Note, in Section \ref{S:MMP_d>7} we will discuss spacetime $d \geq 7$ in which this definition is generalized to:
\be
\Lambda_{\text{maximally mixed}} = \Lambda_1 \oplus \Lambda_{d-3}\,, \quad \Lambda_1 \cong \Lambda_{d-3} \,.
\ee

\subsection{The gluing process}
\label{SS: Gluing}

Let us review the underlying geometry, following the discussion in \cite{Cvetic:2023pgm}.
Associated with the covering $X = X^\text{loc} \cup X^\circ$, in which we assume all singularities have been resolved through blow-up, the Mayer--Vietoris sequence truncates to
\begin{align}\label{eq:les_MV}
    0 \to H_2(X^\circ) \oplus H_2(X^\text{loc}) \xrightarrow{\jmath_{2}} \underbrace{H_{2}(X)}_{= {\rm II}_{3,19}} \xrightarrow{\partial_{2}} H_{1}( \underbrace{\partial X^\text{loc}}_{=\partial X^{\circ}} ) \xrightarrow{\imath_{1}} H_{1}(X^{\circ}) \to 0 \, ,
\end{align}
where $H_1(X^\text{loc} \cap X^\circ) = H_1(\partial X^\text{loc}) = H_1(\partial X^\circ)$ is due to the deformation retraction.

The map $\jmath_2$ has two important properties.
First, it is injective, as evident from the sequence, and so the free groups $H_2(X^\circ)$ and $H_2(X^\text{loc}) = L_\mathfrak{g}$ can be viewed as sublattices (with pairing given by the intersection product) of $H_2(X)$.
Second, since $H_1(\partial X^\text{loc})$ is pure torsion, $\text{im}(\jmath_2) \cong H_2(X^\circ) \oplus H_2(X^\text{loc})$ must be of the same rank as $H_2(X)$.

Lastly, an important feature of the gluing is the existence of a \emph{primitive} sublattice $M \subset H_2(X)$ (meaning $H_2(X)/M$ is torsion free, see Appendix \ref{sec:latticeappendix}), which is of the same rank as $H_2(X^\text{loc})$ and also contains $\jmath_2(H_2(X^\text{loc})) \cong H_2(X^\text{loc})$.
Viewing both $M$ and $H_2(X^\text{loc})$ as sublattices of $H_2(X)$, which has an integer intersection pairing (i.e., elements in $M$ and $H_2(X^\text{loc})$ intersect integrally as well), allows us to identify $M$ also as a sublattice of the dual lattice $H_2(X^\text{loc})^* = H_2(X^\text{loc}, \partial X^\text{loc})$.
This implies that the quotient
\begin{align}
    \frac{M}{\jmath_2(H_2(X^\text{loc}))} = \frac{M}{H_2(X^\text{loc})} \subset \frac{H_2(X^\text{loc})^*}{H_2(X^\text{loc})} = {\cal Z}(\widetilde{G})\,,
\end{align}
is a subgroup of the center of simply-connected group $\widetilde{G}$ whose ADE-algebra is the gauge algebra of the local theory on $X^\text{loc}$.
By excising the singularities from $X$, one can argue that there is a canonical pairing \cite{Cvetic:2023pgm}
\begin{align}
    \frac{M}{H_2(X^\text{loc})} \times H_1(X^\circ) \rightarrow \mathbb{Q}/\mathbb{R} \, ,
\end{align}
thus identifying the lattice quotient $M/H_2(X^\text{loc}) \cong H_1(X^\circ)^\vee$ as the Pontryagin-dual to $H_1(X^\circ)$.

This pairing turns out to be the same as the pairing \eqref{eq:H1_pairing-Xo} for torsion homology on $X^\circ$.
To see this, we take the cokernel of $\jmath_2$ in \eqref{eq:les_MV}, which yields a short exact sequence,
\begin{align}
    0 \rightarrow \text{coker}(\jmath_2) \xrightarrow{\widetilde{\partial}_2} H_1(\partial X^\circ) \xrightarrow{\imath_1} H_1(X^\circ) \rightarrow 0 \, .
\end{align}
Since the last map $\imath_1$ is induced by the inclusion $\partial X^\circ \hookrightarrow X^\circ$, it is identical to the map $I_1$ that appears in the relative homology sequence \eqref{eq:comm-diag_of_SES_rel-hom} for $X^\circ$, which we reproduce here for convenience (and where we have identified $\text{Hom}(H_2(X^\circ),\mathbb{Z}) = H_2(X^\circ)^*$): 
\begin{align}\label{eq:les-to-ses-relative_Xo}
\begin{split}
    0 \rightarrow H_2(X^\circ) \xrightarrow{J_2^\circ} H_2(X^\circ,\partial X^\circ) \cong H_2(X^\circ)^* \oplus H_1(X^\circ)^\vee \xrightarrow{D_2^\circ} H_1(\partial X^\circ) \xrightarrow{I_1} H_1(X^\circ) \rightarrow 0 \, , \\
    \Rightarrow \qquad 0 \rightarrow \frac{H_2(X^\circ)^*}{H_2(X^\circ)} \oplus H_1(X^\circ)^\vee \xrightarrow{\widetilde{D}_2^\circ} H_1(\partial X^\circ) \xrightarrow{I_1} H_1(X^\circ) \rightarrow 0 \, .
\end{split}
\end{align}
This means that the two short exact sequences are actually the same, allowing us to identify
\begin{align}
\begin{split}
    \text{coker}(\jmath_2) = \frac{H_2(X)}{H_2(X^\circ) \oplus H_2(X^\text{loc})} \cong \frac{H_2(X^\circ)^*}{H_2(X^\circ)} \oplus H_1(X^\circ)^\vee \cong \frac{H_2(X^\circ)^*}{H_2(X^\circ)} \oplus \frac{M}{H_2(X^\text{loc})} \, .
\end{split}
\end{align}
This identification can be understood as gluing relative 2-cycles $\Sigma_2^\circ \in H_2(X^\circ, \partial X^\circ)$ and $\Sigma_2^\text{loc} \in H_2(X^\text{loc}, \partial X^\text{loc})$ along $\sigma_1 \in \text{ker}(\imath_1) = \text{im}(\widetilde\partial_2) \cong \text{coker}(\jmath_2)$ into compact 2-cycles $\Sigma_2 \in H_2(X)$, up to 2-cycles in $\text{im}(\jmath_2) = \text{ker}(\partial_2) = H_2(X^\circ) \oplus H_2(X^\text{loc})$.
All compact 2-cycles in $X$ must arise this way (including compact cycles in $H_2(X^\circ)$ and $H_2(X^\text{loc})$, which can be thought of as glued to the trivial cycle on the other side).

\begin{figure}[t!]
$$
\begin{tikzpicture}[x=1cm,y=1cm] 

\begin{scope}[shift={(0,0)}] 
\fill[black] (-6,3) circle (0.15);
\draw[line width=0.3mm] (0,3) ellipse (1 and 2.5);
\draw[line width=0.3mm] (-6,3)--(-0.05,0.5); 
\draw[line width=0.3mm] (-6,3)--(-0.05,5.5); 

\node[above] at (1,5.2) {\Large{$\partial X_{4}^\text{loc}$}};
\node[above] at (-3,4.5) {\Large{$X_{4}^\text{loc}$}};
\node[above] at (-5.5,3.4) {\Large{$SQFT$}};

\draw[line width=0.3mm,red] (0,2) ellipse (0.3 and 0.6);
\draw[line width=0.3mm,red,dashed] (-6,3)--(-0.05,1.4); 
\draw[line width=0.3mm,red,dashed] (-6,3)--(-0.05,2.6); 

\draw[line width=0.3mm,blue] (0,4) ellipse (0.3 and 0.6);
\draw[line width=0.3mm,blue,dashed] (-6,3)--(-0.05,3.4); 
\draw[line width=0.3mm,blue,dashed] (-6,3)--(-0.05,4.6); 

\node[blue] at (0,4.8) {\Large{$\sigma_{1}$}};
\node[red] at (0,1.1) {\Large{$\widetilde{\sigma}_{1}$}};

\node[blue] at (-2,3.6) {\Large{$\Sigma_{2}^\text{loc}$}};
\node[red] at (-2,2.4) {\Large{$\widetilde{\Sigma}_{2}^\text{loc}$}};
\end{scope}

\begin{scope}[shift={(3.5,3)}] 

\draw[line width=0.3mm] (0,3) ellipse (1 and 2.5);
\draw[line width=0.3mm] (0.05,0.5) .. controls (7,-0.5) and (7,6.5) .. (0.05,5.5);
\draw[line width=0.3mm] (3,2.25-1) .. controls (2.5,2.5-1) and (2.5,3.5-1) .. (3,3.75-1);
\draw[line width=0.3mm] (3-0.2,2.5-1) .. controls (3.5-0.2,2.75-1) and (3.5-0.2,3.25-1) .. (3-0.2,3.5-1);

\draw[line width=0.3mm] (3,2.25+1) .. controls (2.5,2.5+1) and (2.5,3.5+1) .. (3,3.75+1);
\draw[line width=0.3mm] (3-0.2,2.5+1) .. controls (3.5-0.2,2.75+1) and (3.5-0.2,3.25+1) .. (3-0.2,3.5+1);

\draw[line width=0.3mm,red] (0,2) ellipse (0.3 and 0.6);
\draw[line width=0.3mm,red,dashed] (0,1.4) .. controls (5.5,0) and (5.5,4) .. (0,2.6);

\draw[line width=0.3mm,blue] (0,4) ellipse (0.3 and 0.6);
\draw[line width=0.3mm,blue,dashed] (-0.05,3.4) .. controls (2.6,3.3) and (2.6,4.7) .. (-0.05,4.6);

\node[blue] at (0,4.8) {\Large{$\sigma_{1}$}};
\node[red] at (0,1.1) {\Large{$\widetilde{\sigma}_{1}$}};

\node[blue] at (1.2,4) {\Large{$\Sigma_{2}^\circ$}};
\node[red] at (1.7,2) {\Large{$\widetilde{\Sigma}_{2}^\circ$}};

\end{scope}

\begin{scope}[shift={(3.5,-3)}] 

\draw[line width=0.3mm] (0,3) ellipse (1 and 2.5);
\draw[line width=0.3mm] (0.05,0.5) .. controls (7,-0.5) and (7,6.5) .. (0.05,5.5);
\draw[line width=0.3mm] (3,2.25-1) .. controls (2.5,2.5-1) and (2.5,3.5-1) .. (3,3.75-1);
\draw[line width=0.3mm] (3-0.2,2.5-1) .. controls (3.5-0.2,2.75-1) and (3.5-0.2,3.25-1) .. (3-0.2,3.5-1);

\draw[line width=0.3mm] (3,2.25+1) .. controls (2.5,2.5+1) and (2.5,3.5+1) .. (3,3.75+1);
\draw[line width=0.3mm] (3-0.2,2.5+1) .. controls (3.5-0.2,2.75+1) and (3.5-0.2,3.25+1) .. (3-0.2,3.5+1);

\node[above] at (3,5.5) {\Large{$X^\circ_{4}$}};

\draw[line width=0.3mm,red] (0,2) ellipse (0.3 and 0.6);
\draw[line width=0.3mm,blue] (0,4) ellipse (0.3 and 0.6);
\draw[line width=0.3mm,violet,dashed] (0,2.6) .. controls (1.5,2.6) and (1.5,3.4) .. (0,3.4);
\draw[line width=0.3mm,violet,dashed] (-0.05,1.4) .. controls (3,1.5) and (3,4.5) .. (-0.05,4.6);

\node[blue] at (0,4.8) {\Large{$\sigma_{1}$}};
\node[red] at (0,1.1) {\Large{$\widetilde{\sigma}_{1}$}};
\node[violet] at (1.7,3) {\Large{$\Sigma_{2}^\circ$}};

\end{scope}

\end{tikzpicture}
$$

\caption{The geometry of the local internal space $X_{4}^\text{loc}$ appears on the left, and possible completions to a compact space appear on the right. The upper right picture describes two possible completions of cycles: Both cycles $\Sigma_{2}^\text{loc}$ and $\widetilde{\Sigma}_{2}^\text{loc}$ are similarly completed by $\Sigma_{2}^\circ$ and $\widetilde{\Sigma}_{2}^\circ$ to a closed cycle $\Sigma_{2}\simeq \Sigma_{2}^\text{loc} \cup_{\sigma_1} \Sigma_{2}^\circ$ and $\widetilde{\Sigma}_{2}\simeq \widetilde{\Sigma}_{2}^\text{loc} \cup_{\widetilde\sigma_1} \widetilde{\Sigma}_{2}^\circ$, respectively, where in both cases $\sigma_{1},\widetilde{\sigma}_{1}\in \text{ker}(\imath_{1})$.
In the SymTFT for the supergravity EFT, we associate Dirichlet boundary conditions with $\sigma_{1}$ and $\widetilde{\sigma}_{1}$ in $X^\text{loc}$. 
The cases differ by the fact that $\Sigma_{2}\in \frac{M}{H_2(X^\text{loc})} \subset H_{2}(X)$, while $\widetilde\Sigma_{2}\notin \frac{M}{H_2(X^\text{loc})}$, meaning the latter is associated with the $U(1)$ sector from $H_2(X^\circ)$ while the former is not. The lower right picture describes the case when $\sigma_{1},\widetilde{\sigma}_{1}\notin \text{ker}(\imath_{1})$, but $[\sigma_{1}] + [\widetilde{\sigma}_{1}] = 0 \in H_{1}(X^\circ)$,
In the supergravity SymTFT, we associate Neumann boundary condition with $\sigma_{1}$ and $\widetilde{\sigma}_{1}$.
\label{F:CycleCompK3}}
\end{figure}
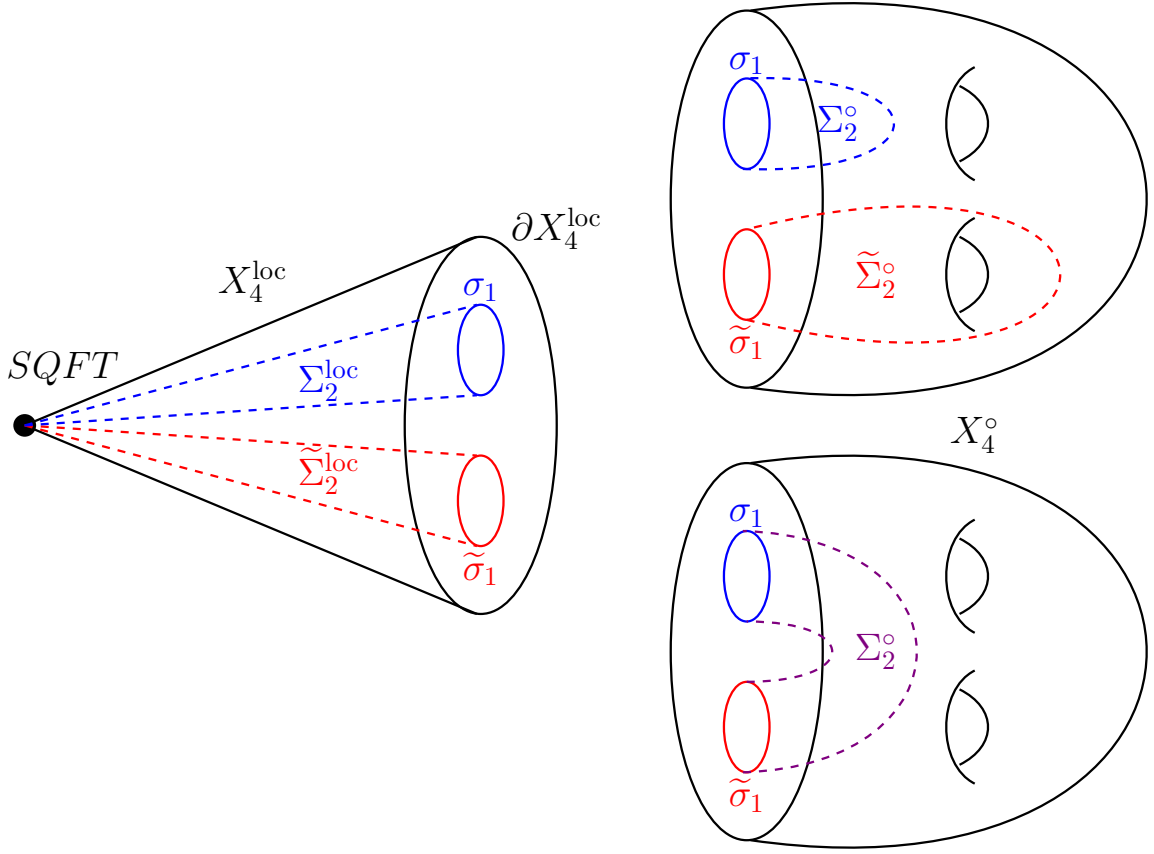

Since every $\sigma_1 \in H_1(\partial X^\text{loc})$ is the boundary of a relative 2-cycle $\Sigma_2^\text{loc}$ in $X^\text{loc}$, one may wonder how these are completed in the compact geometry.
From the above discussion one can find there are three ways to do so, depending on $D_2^\text{loc}(\Sigma_2^\text{loc})$, which we will explain now (see also Figure \ref{F:CycleCompK3} for an illustration).

\begin{itemize}

    \item For $\sigma_1 = D^\text{loc}_{2}(\Sigma_2^\text{loc}) \in H_1(X^\circ)^\vee \subset \text{coker}(\jmath_2)$, there is, up to compact 2-cycles in $X^\circ$, a corresponding relative 2-cycle $\Sigma_2^\circ \in \text{Ext}(\text{Tor}(H_1(X^\circ),\Z)) \cong H_1(X^\circ)^\vee \subset H_2(X^\circ, \partial X^\circ)$ with $D_2^\circ(\Sigma_2^\circ) = \sigma_1$.
    These two glue along $\sigma_1$ into a compact 2-cycle $\Sigma_2 \in H_2(X)$ with $\partial_2 \Sigma_2 = \sigma_1$.
    We denote this relation by $\Sigma_2 \simeq \Sigma_2^\text{loc} \cup_{\sigma_1} \Sigma_2^\circ$.
    As we will explain momentarily, $\Sigma_2^\circ$ in this case does \emph{not} carry any defect charge under the Abelian gauge factors.
    To illustrate this, we depicted them in blue in Figure \ref{F:CycleCompK3}, as cycles that on the $X^\circ$ side can be ``contracted'' to the boundary, in the sense that they do not have a proper intersection with the compact 2-cycles on $X^\circ$.

    \item For $\widetilde\sigma_1 = D^\text{loc}_{2}(\widetilde\Sigma_2^\text{loc}) \in \frac{H_2(X^\circ)^*}{H_2(X^\circ)} \subset \text{coker}(\jmath_2)$, there is (modulo $H_2(X^\circ)$), a relative 2-cycle $\widetilde\Sigma_2^\circ \in H_2(X^\circ)^* = \text{Hom}(H_2(X^\circ),\Z) \subset H_2(X^\circ,\partial X^\circ)$ with $D_2^{\circ}(\widetilde\Sigma_2^\circ) = \widetilde\sigma_1$.
    These glue into $\widetilde\Sigma_2 \simeq \widetilde\Sigma_2^\text{loc} \cup_{\widetilde\sigma_1} \widetilde\Sigma_2^\circ$.
    In this case, $\widetilde\Sigma_2^\circ$ intersects (fractionally) the compact 2-cycles in $X^\circ$, so we depicted them in red as ``incontractible'' chains on $X^\circ$ in Figure \ref{F:CycleCompK3}.

    \item The more interesting case arises when $0 \neq \sigma_1 = D_2^\text{loc}(\Sigma_2^\text{loc}) \notin \text{coker}(\jmath_2) = \text{ker}(\imath_1) = \text{ker}(I_1)$.
    Because of the sequences \eqref{eq:les-to-ses-relative_Xo}, $[\sigma_1 \mod \text{ker}(I_1)]$ defines a non-trivial class in $H_1(X^\circ) = \text{Tor}(H_1(X^\circ))$.
    But then, there must exists another non-trivial class $0 \neq [\widetilde\sigma_1 \mod \text{ker}(I_1)] \in H_1(X^\circ)$ such that $[\sigma_1] + [\widetilde\sigma_1] = 0 \in H_1(X^\circ)$.
    This means that there is a relative 2-cycle $\Sigma_2^\circ$ such that $D_2^\circ(\Sigma_2^\circ) = \sigma_1 + \widetilde\sigma_1 \in \text{ker}(I_1) = \text{Im}(D_2^\circ)$.
    Therefore, we see that there is a compact 2-cycle $\Sigma_2$ which is the gluing of $\Sigma_2^\text{loc}, \widetilde\Sigma_2^\text{loc}$ on $X^\text{loc}$ and $\Sigma_2^\circ$ on $X^\circ$ along the boundary $\sigma_1 + \widetilde\sigma_1$.
    This is depicted in the lower part of Figure \ref{F:CycleCompK3}.
    In fact, the relative cycle $\Sigma_2^\circ$ must be in $H_1(X^\circ)^\vee$, i.e., ``contractible''.
    
\end{itemize}

\subsection{The global structure from gluing}

One can readily read off the global gauge group structure $[(\widetilde{G}/{\cal C}_\text{loc}) \times U(1)^{b}]/{\cal C}_\text{extra}$ \cite{Cvetic:2023pgm}:\footnote{Notice in particular that the sequence $0 \rightarrow {\cal C}_\text{loc} \rightarrow {\cal C}_\text{full} := \text{coker}(\jmath_2) \rightarrow {\cal C}_\text{extra} \rightarrow 0$ from \cite{Cvetic:2023pgm} always splits in M-theory compactified on K3s, due to the Universal Coefficient Theorem applied to $H_2(X^\circ, \partial X^\circ)$.}
\begin{align}\label{eq:global_gauge_group_geometry}
    {\cal C}_\text{loc} \cong H_1(X^\circ)^\vee = \frac{M}{H_2(X^\text{loc})} \, , \qquad {\cal C}_\text{extra} \cong \frac{H_2(X^\circ)^*}{H_2(X^\circ)} = \frac{\text{coker}(\jmath_2)}{H_1(X^\circ)^\vee} \, .
\end{align}
From the effective field theory perspective, this may be interpreted as having a gauged ${\cal C}_\text{loc} \oplus {\cal C}_\text{extra}$ 1-form symmetry.
The new insight we extract from the geometry is that this automatically implies that there is also a gauged ${\cal C}_\text{loc} \oplus {\cal C}_\text{extra}$ 4-form symmetry.
From the point of view of the SymTFT on $\partial X^\text{loc} \coprod \partial X^\circ$, this amounts to specifying a maximal isotropic subgroup of both the 1-form and the 4-form symmetry to have Dirichlet boundary conditions.
This is what we refer to as a ``maximally mixed'' polarization.

\subsubsection*{Deconstructing the intersection product on $X$}

To see this in detail, we make use of the fact that the map $\jmath_2$ is also a lattice embedding with respect to the intersection pairing.
That is, the free groups $H_2(X^\circ)$ and $H_2(X^\text{loc})$, with their individual intersection pairing, are orthogonal sublattices of the even self-dual lattice $H_2(X) = {\rm II}_{3,19}$.
Moreover, since $\text{rank}(H_2(X)) = \text{rank}(H_2(X^\circ))+\text{rank}(H_2(X^\text{loc}))$, we have $H_2(X) \otimes \mathbb{Q} = (H_2(X^\circ) \otimes \mathbb{Q}) \oplus (H_2(X^\text{loc}) \otimes \mathbb{Q})$.
Because $H_2(X)$ is also free, any $\Sigma_2 \in H_2(X)$ can therefore be orthogonally decomposed into fractional linear combinations of elements in $H_2(X^\circ)$ and $H_2(X^\text{loc})$:
\begin{align}
    \Sigma_2 = S^\text{loc} + S^\circ \, , \qquad S^\text{loc} \in H_2(X^\text{loc}) \otimes \mathbb{Q} \, , \quad S^\circ \in H_2(X^\circ) \otimes \mathbb{Q} \, ,
\end{align}
where the sum is addition of lattice vectors.
The intersection product in the compact space can be then equivalently computed in the local patches, via
\begin{align}
    \Sigma_2 \cdot \widetilde\Sigma_2 = S^\text{loc} \cdot \widetilde{S}^\text{loc} + S^\circ \cdot \widetilde{S}^\circ \, .
\end{align}
This is closely related to the gluing procedure $\Sigma_2 \simeq \Sigma_2^\text{loc} \cup_{\sigma_1} \Sigma_2^\circ$ with relative 2-cycles.
Namely, relative 2-cycles $\Sigma_2^\spadesuit \in H_2(X^\spadesuit)^* = \text{Hom}(H_2(X^\spadesuit),\mathbb{Z}) \subset H_2(X^\spadesuit ,\partial X^\spadesuit)$ can also be expressed as fractional linear combinations of compact 2-cycles on the local patch $X^\spadesuit$.
For $X^\spadesuit = X^\text{loc}$, all relative 2-cycles are of this form, so $\Sigma_2^\text{loc}$ directly represents a vector in $H_2(X^\text{loc}, \partial X^\text{loc}) = H_2(X^\text{loc})^*$, so $\Sigma_2^\text{loc} = S^\text{loc} \in H_2(X^\text{loc}, \partial X^\text{loc}) \subset H_2(X^\text{loc}) \otimes \mathbb{Q}$.
On the other hand, only $\Sigma_2^\circ \in \text{Hom}(H_2(X^\circ), \mathbb{Z}) \subsetneq H_2(X^\circ, \partial X^\circ)$ with $D_2^\circ \Sigma_2^\circ \in {\cal C}_\text{extra} \subset H_1(\partial X^\circ) = H_1(\partial X^\text{loc})$ can be expressed in this way.

This means that for $\Sigma_2 \simeq \Sigma_2^\text{loc} \cup_{\sigma_1} \Sigma_2^\circ$ with $\partial_2 \Sigma_2 =\sigma_1 \in {\cal C}_\text{extra} \subset \text{coker}(\jmath_2)$, the orthogonal decomposition $\Sigma_2 = S^\text{loc} + S^\circ$ agrees with the gluing, and $S^\text{loc} = \Sigma_2^\text{loc} \in H_2(X^\text{loc}, \partial X^\text{loc}) = H_2(X^\text{loc})^*$, $S^\circ = \Sigma_2^\circ \in \text{Hom}(H_2(X^\circ),\mathbb{Z}) = H_2(X^\circ)^* \subset H_2(X^\circ, \partial X^\circ)$ can be regarded as vectors in the respective dual lattices.
Meanwhile, for $\Sigma_2 \simeq \Sigma_2^\text{loc} \cup_{\sigma_1} \Sigma_2^\circ$ with $\partial_2 \Sigma_2 \in {\cal C}_\text{loc}$, we still have $S^\text{loc} = \Sigma_2^\text{loc} \in H_2(X^\text{loc})^*$, but the relative cycle $\Sigma_2^\circ$ must correspond to the zero vector $S^\circ = 0 \in H_2(X^\circ)^*$, so that the intersection of $\Sigma_2$ with any other compact cycle $\widetilde\Sigma_2$ is identical to the intersection of $\Sigma_2^\text{loc}$ with $\widetilde\Sigma_2^\text{loc}$.

\subsubsection*{Linking properties on $X^\circ$ from intersections on $X^\text{loc}$}

It is important to stress that for $\partial_2 \Sigma_2 \in {\cal C}_\text{loc}$, $\Sigma_2^\circ$ is not a trivial relative homology class in $H_2(X^\circ,\partial X^\circ)$, even though it is ``invisible'' for the intersection product on $H_2(X)$.
Instead, it captures the fact that $\text{coker}(\partial_2) \cong \text{im}(\imath_1) \neq 0$ in \eqref{eq:les_MV}.
In turn, this is related to some linking properties we asserted in Section \ref{sec:surface-specifics} for the the relative homology (see \eqref{eq:les-to-ses-relative_Xo}) on $X^\circ$.
That is, the subgroups $\widetilde{D}_2^\circ[H_2(X^\circ)^*/H_2(X^\circ)] \cong {\cal C}_\text{extra}$ and $\widetilde{D}_2^\circ(H_1(X^\circ)^\vee) \cong {\cal C}_\text{loc}$ of $H_1(\partial X^\circ)$ are orthogonal to each other with respect to the linking pairing $\ell$ on $H_1(\partial X^\circ)$.
Instead, $\widetilde{D}_2^\circ(H_1(X^\circ)^\vee) \subset H_1(\partial X^\circ)$ links with non-trivial representatives of $H_1(\partial X^\circ)/\text{ker}(I_1) \cong H_1(X^\circ)$, as indicated by Pontryagin duality.
Furthermore, we have also argued, based on the mutual locality in the gauge theory on $X^\circ$, that $\widetilde{D}_2^\circ (H_1(X^\circ))$ has trivial linking amongst itself.

While we did not have a rigorous argument based purely on the geometry of $X^\circ$ to support these claims, we can use the properties of $X$ and $X^\text{loc}$ to confirm them in case $X^\circ$ can be glued to a (union of) ADE patch into a K3, which is anyways the scenario we are interested in.
Namely, because $X^\circ$ and $X^\text{loc}$ have the same boundary $\partial X^\circ = \partial X^\text{loc}$, the linking properties on the boundary must agree.
On $X^\text{loc}$, we can use $H_1(\partial X^\text{loc}) = H_2(X^\text{loc})^* / H_2(X^\text{loc})$, with the linking given by intersection modulo integers, which in turn is related to the intersection in $X$, which we know must be integer.

Explicitly, as explained above, the gluing identifies $\widetilde\partial_2[M/\jmath_2(H_2(X^\text{loc}))] \subset H_1(\partial X^\text{loc})$ with $\widetilde{D}_2^\circ[H_1(X^\circ)^\vee] \subset H_1(\partial X^\circ)$, so that there is a $\Sigma_2 \in M \subset H_2(X)$ with $\partial_2 \Sigma_2 = \sigma_1 \in \widetilde{D}_2^\circ[H_1(\partial X^\circ)^\vee] = {\cal C}_\text{loc}$.
Since $M$ is an overlattice of $\jmath_2(H_2(X^\text{loc})) \cong H_2(X^\text{loc})$ of the same rank, which must have integer pairing with any element of $H_2(X^\text{loc})$, we may also view it as a sublattice of $H_2(X^\text{loc})^* = H_2(X^\text{loc}, \partial X^\text{loc})$, so we can compute the linking of $\sigma_1 \in \widetilde{D}_2^\circ[H_1(X^\circ)^\vee] = {\cal C}_\text{loc}$ with any other boundary cycle $\widetilde\sigma_1$ as $\ell(\sigma_1, \widetilde\sigma_1) = \Sigma_2^\text{loc} \cdot \widetilde\Sigma_2^\text{loc} \mod \mathbb{Z}$, where $\Sigma_2 = \Sigma_2^\text{loc} \in M \subset H_2(X^\text{loc})^*$ and $\widetilde\Sigma_2^\text{loc} \in H_2(X^\text{loc})^*$ with $D_2^\text{loc}(\Sigma_2^\text{loc}) = \sigma_1$ and $D_2^\text{loc}(\widetilde\Sigma_2^\text{loc}) = \widetilde\sigma_1$.
Crucially, being a sublattice of $H_2(X)$, $M$ must also be an \emph{integer} lattice, so that $\ell(\sigma_1, \widetilde\sigma_1) = \Sigma_2^\text{loc} \cdot \widetilde\Sigma_2^\text{loc} = 0 \mod \mathbb{Z}$ if $\widetilde\sigma_1 \in {\cal C}_\text{loc} \Leftrightarrow \widetilde\Sigma_2^\text{loc} \in M$.
Meanwhile, for $\widetilde\sigma_1 = \partial_2 \widetilde\Sigma_2 \in {\cal C}_\text{extra}$, we use the orthogonal decomposition $\widetilde\Sigma_2 = \widetilde\Sigma_2^\text{loc} + \widetilde\Sigma_2^\circ$ and the fact that $M$ is orthogonal to $H_2(X^\circ)$ in $H_2(X)$ to find
\begin{align}
    \ell(\sigma_1,\widetilde\sigma_1) = \Sigma_2^\text{loc} \cdot \widetilde\Sigma_2^\text{loc} \mod \mathbb{Z} = \Sigma_2^\text{loc} \cdot \widetilde\Sigma_2 \mod \mathbb{Z} = \Sigma_2 \cdot \widetilde\Sigma_2 \mod \mathbb{Z} = 0 \mod \mathbb{Z} \, ,
\end{align}
because intersections in $H_2(X)$ are integer.

\subsubsection*{Boundary conditions from gluing}

With this, we can argue for the gauge group structure from the SymTFT of the theory on $X^\text{loc} \coprod X^\circ$ as follows.
For any compact 2-cycle, the orthogonal decomposition $\Sigma_2 = \Sigma_2^\text{loc} + \Sigma_2^\circ$ corresponds to the fact that $\Sigma_2$ specifies a bi-charged state $(\Sigma_2^\text{loc}, \Sigma_2^\circ)$ for the defect group of the product theories on $X^\text{loc}$ and $X^\circ$.
Because in the compact geometry, compact 2-cycles are wrapped indiscriminately by both M2- and M5-branes, we can interpret this as the compact geometry specifying Dirichlet boundary conditions for both M2- and M5-branes wrapping $\Sigma_2^\text{loc}$ in $X^\text{loc}$ and $\Sigma_2^\circ$ in $X^\circ$.

The symmetry generators that are trivialized with this boundary condition, and which can be interpreted as ``being gauged'', correspond to the boundary cycles in the ``diagonal'' $\text{coker}(\jmath_2) \cong \text{im}(\partial_2)$ subgroup:
\begin{align}
    (\sigma_1^\text{loc}, \sigma_1^\circ) = (D_2^\text{loc} \Sigma_2^\text{loc}, D_2^\circ \Sigma_2^\circ) = (\partial_2 \Sigma_2, \partial_2 \Sigma_2) = (\sigma_1, \sigma_1) \in H_1(\partial X^\text{loc}) \times H_1(\partial X^\circ) \, .
\end{align}
Since both M2- and M5-branes on these are given Dirichlet boundary conditions, the resulting electrically and magnetically charged states must be mutually local with respect to each other.
Geometrically, this is guaranteed by the integer intersection product between compact 2-cycles in $X$:
\begin{align}
\begin{split}
    \langle (\Sigma_2^\text{loc}, \Sigma_2^\circ)_{\text{M2}} \, , \, (\widetilde\Sigma_2^\text{loc}, \widetilde\Sigma_2^\circ)_{\text{M5}} \rangle & = \ell(\sigma_1^\text{loc}, \widetilde\sigma_1^\text{loc}) + \ell(\sigma^\circ_1, \widetilde\sigma_1^\circ) \\
    & = \Sigma_2^\text{loc} \cdot \widetilde\Sigma_2^\text{loc} + \Sigma_2^\circ \cdot \widetilde\Sigma_2^\circ \mod \Z = \Sigma_2 \cdot \widetilde\Sigma_2 \mod \Z = 0 \, .
\end{split}
\end{align}
Notice that this equation also be read as the charged defects from M2-(M5-)branes on $(\Sigma_2^\text{loc}, \Sigma_2^\circ)$ to be uncharged under the symmetry generators from M5-(M2-)branes on $(\widetilde\sigma_1^\text{loc}, \widetilde\sigma_1^\circ)$, which is consistent with the interpretation that these symmetries have been gauged.

Furthermore, because of the feature of the local theory on $X^\circ$ surrounding the non-triviality of $H_1(X^\circ)$, the subgroup $(0,\sigma_1) \in H_1(\partial X^\text{loc}) \oplus H_1(\partial X^\circ)$ with $\sigma_1 \in H_1(X^\circ)^\vee$ is also given Dirichlet boundary condition for both M2- and M5-branes.
Because of the orthogonality properties on $H_1(\partial X^\circ)$ discussed above, the linking pairing on $\text{coker}(\jmath_2) \oplus \{(0, \sigma_1) \, | \, \sigma_1 \in H_1(X^\circ)^\vee\} \subset H_1(\partial X^\text{loc}) \oplus H_1(\partial X^\circ)$ is completely trivial.
Since 
\begin{align}
    |\text{coker}(\jmath_2)| \times |H_1(X^\circ)^\vee| = |H_1(\partial X^\circ)| = |H_1(\partial X^\text{loc})| = \sqrt{|H_1(\partial X^\text{loc}) \oplus H_1(\partial X^\circ)|}
\end{align}
from the short exact sequence \eqref{eq:les-to-ses-relative_Xo}, this specifies a polarization $\Lambda$ for the product SymTFT:
\begin{align}
    \begin{split}
    \Lambda_{\text{maximally mixed}} & =  [\text{coker}(\jmath_2) \oplus H_1(X^\circ)^\vee]^{(1)} \oplus [\text{coker}(\jmath_2) \oplus H_1(X^\circ)^\vee]^{(4)} \\
    & \subset D = [H_1(\partial X^\text{loc}) \oplus H_1(\partial X^\circ)]^{(1)} \oplus [H_1(\partial X^\text{loc}) \oplus H_1(\partial X^\circ)]^{(4)}\,.
    \end{split}
\end{align}
By giving this subgroup Dirichlet boundary conditions, we obtain the charged defects which are identified with the dynamical objects in the gravitational theory from M2-/M5-branes wrapping compact 2-cycles in $X$.
They screen/break the global symmetry of the absolute field theory, whose generators come from M5-/M2-branes on $[H_1(\partial X^\text{loc}) \oplus H_1(\partial X^\circ)]/\Lambda$ which receive Neumann boundary conditions.

Finally, to explain the particular structure $[(\widetilde{G}/{\cal C}_\text{loc}) \times U(1)^b]/{\cal C}_\text{extra}$, notice that, due to orthogonality of $M$ and $H_2(X^\circ)^* \subset H_2(X^\circ) \otimes \mathbb{Q}$, the linking of any charged defect $(\Sigma_2^\text{loc}, \Sigma_2^\circ)$ (not necessarily coming from a compact 2-cycle in $X$) with $\sigma_1 \in {\cal C}_\text{loc}  = \{\partial_2 \Sigma_2 \, | \, \Sigma_2 \in M\}$ does not depend on the charge under the Abelian factors (which determined by $\Sigma_2^\circ \in H_2(X^\circ)^*$).
Hence, ${\cal C}_\text{loc}$ embeds trivially into the Abelian factors and acts only on the non-Abelian symmetry $\widetilde{G}$.

\subsubsection*{Maximally mixed polarization}

The important feature that we want to highlight here is that the geometrically specified boundary conditions for the various cycles applies simultaneously to M2- and M5-branes.
Loosely speaking, it means that there is an ``equal amount'', namely $\text{coker}(\jmath_2)$, of electric and magnetic states that are given Dirichlet (and Neumann) boundary conditions, i.e., the polarization is ``maximally mixed''.
This is to be contrasted with other field-theoretically viable polarizations, e.g., the ``purely electric'' polarization where only electric charges are present, and where the gauge group is simply-connected.
Analogously, there is the ``purely magnetic'' polarization, where the fundamental group of the gauge group is the largest.
Such polarizations would be admissible in the M-theory geometry that is truly the disjoint union $X^\text{loc} \coprod X^\circ$, in which case M2-branes (M5-branes) are given Dirichlet boundary conditions on both factors (modulo the subtlety surrounding $H_1(X^\circ) \neq 0$ which we will come back to momentarily).
In general, there are many other choices of consistent polarizations in the disjoint-union setting, all of which would specify a global structure for the supergravity theory that naively would be consistent.
These could have various amounts of electric/magnetic defects with Dirichlet boundary conditions, i.e., non-maximally mixed polarizations.
None of these correspond to the global structure of M-theory on a compact K3.
Hence, we may interpret the compact geometry as placing the majority of such global structures (for a given gauge algebra) into the Swampland, and pointing towards a deeper reason why only maximally mixed polarizations are allowed in the Landscape.

\subsubsection*{Example: $T^4/\Z_2$}

Let us return to the example of $X = T^4 / \mathbb{Z}_2$, for which we have already analyzed the local patches $X^\text{loc} \cong \coprod_{i=1}^{16} \mathbb{C}^2/\mathbb{Z}_2^{(i)}$ and $X^\circ = X \setminus X^\text{loc}$.
The global structure of the compact model has already been described in \cite{Cvetic:2023pgm}, and we will recast the results in terms of the product SymTFT.

The 0-form gauge symmetry on the disjoint union $X^\text{loc} \coprod X^\circ$ is $\mathfrak{su}(2)^{16} \oplus \mathfrak{u}(1)^6$.
The product SymTFT would have the symmetry group
\begin{align}
    [H_1(\partial X^\text{loc}) \oplus H_1(\partial X^\circ)]^{(1)} \oplus [H_1(\partial X^\text{loc}) \oplus H_1(\partial X^\circ)]^{(4)} = \bigoplus_{i=1}^{32} [{\Z_2}_{,i}^{(1)} \oplus {\Z_2}_{,i}^{(4)}] \, .
\end{align}
The linking pairing is only non-trivial between ${\Z_2}^{(1)}$ and ${\Z_2}^{(4)}$ factors with the same index $i$.

As explained in the previous section, the geometry on $X^\circ$ forces a $({\mathbb{Z}_2}^{(1)} \oplus {\mathbb{Z}_2}^{(4)})^{\oplus 5} \cong (H_1(X^\circ)^\vee)^{(1)} \oplus (H_1(X^\circ)^\vee)^{(4)}$ subgroup to have Dirichlet boundary conditions, and a $({\Z_2}^{(1)} \oplus {\Z_2}^{(4)})^{\oplus 5} \cong H_1(X^\circ)^{(1)} \oplus H_1(X^\circ)^{(4)}$ to have Neumann boundary conditions.
The charged defects arising from these factors are not charged under the center symmetries of the gauge theory, and, vice versa, the charged defects of the gauge theory are not charged under these higher-form symmetries.
On the $X^\text{loc}$ side, the Dirichlet boundary conditions are matched by a rank 16 lattice $M \subset H_2(X^\text{loc})^*$ such that $M/H_2(X^\text{loc}) \cong {\Z_2}^{\oplus 5}$, i.e., $M$ is generated by the generators $e_I$ of $H_2(X^\text{loc})$ plus five half-integer linear combinations of $e_I$ (see \cite{Cvetic:2023pgm} for details).
Because $M$ is an integer lattice, the linking of elements in $M/H_2(X^\text{loc}) \cong {\Z_2}^{\oplus 5}$ is trivial; but there is a Pontryagin-dual ${\Z_2}^{\oplus 5} \subset H_1(\partial X^\text{loc})$ which links with $M/H_2(X^\text{loc})$.
The former are given Dirichlet boundary conditions, as in the compact geometry M2- and M5-branes will wrap the 2-cycles in $M \subset H_2(X)$, and trivialize the symmetry generators of ${{\cal C}_\text{loc}}^{(1)} \oplus {{\cal C}_\text{loc}}^{(4)} \cong ({\Z_2}^{(1)} \oplus {\Z_2}^{(4)})^{\oplus 5}$;
the latter are given Neumann boundary conditions, and the symmetries generated by them will eventually be screened by the former in the gravitational theory.

For the remaining twelve factors $[{\Z_2}^{(1)} \oplus {\Z_2}^{(4)}]$ --- six of which come from $H_1(\partial X^\text{loc})^{(1)} \oplus H_1(\partial X^\text{loc})^{(4)}$, and the other six from $H_1(\partial X^\circ)^{(1)} \oplus H_1(\partial X^\circ)^{(4)}$ --- for which we have not specified the boundary conditions yet, there are six linearly independent compact 2-cycles $H_2(X) \ni F_\alpha \simeq (\tfrac12 T_\alpha) \cup_{\sigma_\alpha} (\sum_{I_\alpha} \tfrac12 e_{I_\alpha})$.
Here $T_\alpha$ are the generators of $H_2(X^\circ)$, and the relative cycles $\tfrac12 T_\alpha \in H_2(X^\circ)^* \subset H_2(X^\circ,\partial X^\circ)$ and $\sum_{I_\alpha} \tfrac12 e_{I_\alpha} \in H_2(X^\text{loc})^*$ are glued along $\sigma_\alpha = \partial_2 F_\alpha = D_2^\circ(\tfrac12 T_\alpha) = D_2^\text{loc}(\sum_{I_\alpha} \tfrac12 e_{I_\alpha}) \in H_1(\partial X^\circ) = H_1(X^\text{loc})$.
This specifies a diagonal $[{\Z_2}^{(1)} \oplus {\Z_2}^{(4)}]^{\oplus 6}$ inside the twelve factors above, generated by $(\sigma_\alpha, \sigma_\alpha) \in H_1(\partial X^\text{loc}) \oplus H_1(\partial X^\circ)$, with Dirichlet boundary conditions for both M2- and M5-branes.
This trivializes, or ``gauges'' the ${\cal C}_\text{extra}^{(1)} \oplus {\cal C}_\text{extra}^{(4)} \cong ({\Z_2}^{(1)} \oplus {\Z_2}^{(4)})^{\oplus 6}$, while screening/breaking the anti-diagonal ${\Z_2}^{\oplus 6}$ factors that are given Neumann boundary conditions.

In summary, this determines the global structure of the gauge group to be $[SU(2)^{16}/\Z_2^5 \times U(1)^6]/\Z_2^6 \cong [SU(2)^{16} \times U(1)^6]/[(\Z_2^\text{loc})^5 \times (\Z_2^\text{extra})^6]$.
With the additional information that the center symmetries of the $U(1)^6$ is screened to $({\Z_2}^{(1)} \oplus {\Z_2}^{(4)})^{\oplus 6}$, the second expression makes the ``maximal mixing'' manifest: there is a gauged ${\Z_2}^{\oplus 11}$ 1-form symmetry, and a global ${\Z_2}^{\oplus 22} / {\Z_2}^{\oplus 11} = {\Z_2}^{\oplus 11}$ 1-form symmetry; dual to that is a global ${\Z_2}^{\oplus 11}$ and a gauged ${\Z_2}^{\oplus 11}$ 4-form symmetry.

\subsection{Towards a bottom-up interpretation via lattices}\label{sec:lattice-interpretation_of-geometry}

One of the key ingredients in the geometric origin to the maximally mixed polarization is the lattice structure of (relative) 2-cycles in a complex surface.
The only non-trivial part not encoded in the intersection product is the interplay between relative 2-cycles in $\text{Ext}(H_1(X^\circ),\mathbb{Z}) = H_1(X^\circ)^\vee \subset H_2(X^\circ, \partial X^\circ)$ and $H_1(X^\circ)$, which resulted in a fixed boundary condition: Dirichlet for $\widetilde{D}_2^\circ[H_1(X^\circ)^\vee] \subset H_1(\partial X^\circ)$ and Neumann for $H_1(\partial X^\circ)/\text{im}(\widetilde{D}_2^\circ) = H_1(X^\circ)$.
As elaborated in detail above, when the dust settles, the SymTFT for the $\mathfrak{u}(1)^b$ gauge theory on $X^\circ$ has a ``left-over'' defect group
\begin{align}
    \left(\frac{H_2(X^\circ)^*}{H_2(X^\circ)}\right)^{(1)} \oplus \left(\frac{H_2(X^\circ)^*}{H_2(X^\circ)}\right)^{(4)} \subset H_1(\partial X^\circ)^{(1)} \oplus H_1(\partial X^\circ)^{(4)} \, ,
\end{align}
under which only the states from wrapping the relative 2-cycles in $\text{Hom}(H_2(X^\circ), \mathbb{Z}) = H_2(X^\circ)^*$ are charged.
This is the defect group for defects charged under the center symmetries of the $\mathfrak{u}(1)$'s, which has been screened by the dynamical states from $H_2(X^\circ)$ to the finite group $H_2(X^\circ)^*/H_2(X^\circ) \equiv Z$.
To obtain an absolute theory, one needs to specify boundary conditions for this defect group.

The global structure of the supergravity theory can be then fully specified in terms of a polarization for the defect group
\begin{align}\label{eq:field-theoretic_defect-group}
    \left[ \frac{H_2(X^\text{loc})^*}{H_2(X^\text{loc})} \oplus \frac{H_2(X^\circ)^*}{H_2(X^\circ)} \right]^{(1)} \oplus \left[ \frac{H_2(X^\text{loc})^*}{H_2(X^\text{loc})} \oplus \frac{H_2(X^\circ)^*}{H_2(X^\circ)} \right]^{(4)} = [{\cal Z}(\widetilde{G}) \oplus Z]^{(1)} \oplus [{\cal Z}(\widetilde{G}) \oplus Z]^{(4)},
\end{align}
where ${\cal Z}(\widetilde{G}) \cong H_1(\partial X^\text{loc})$ is the center of the simply-connected group $\widetilde{G}$ with algebra $\mathfrak{g}$.
The pairing on this group is entirely inherited from the lattice pairing on $H_2(X^\text{loc}) \subset H_2(X^\text{loc})^*$ and $H_2(X^\circ) \subset H_2(X^\circ)^*$.
Given an electric operator labelled by $(a_e,b_e) \in [{\cal Z}(\widetilde{G}) \oplus Z]^{(1)}$, and a magnetic operator $(a_m, b_m) \in [{\cal Z}(\widetilde{G}) \oplus Z]^{(4)}$, the pairing is given by
\begin{align}\label{eq:pairing-from-lattice}
    \langle (a_e, b_e), (a_m, b_m) \rangle = \vec{a}_e \cdot \vec{a}_m + \vec{b}_e \cdot \vec{b}_m \mod \mathbb{Z} \, ,
\end{align}
where $\vec{a} \in H_2(X^\text{loc})^*$ and $\vec{b} \in H_2(X^\circ)^*$ are the lattice representatives, and the product is the lattice pairing.

Note that the global gauge group structure dictated by the compact geometry --- which is to give Dirichlet boundary conditions to ${\cal C}_\text{loc} \oplus {\cal C}_\text{extra} \subset {\cal Z}(\widetilde{G}) \oplus Z$ for both the electric and magnetic defect groups --- is also a maximally mixed polarization in this defect group.
First, $\Lambda = [{\cal C}_\text{loc} \oplus {\cal C}_\text{extra}]^{(1)} \oplus [{\cal C}_\text{loc} \oplus {\cal C}_\text{extra}]^{(4)}$ is isotropic, because the product \eqref{eq:pairing-from-lattice} is simply the intersection product $\Sigma_2 \cdot \widetilde\Sigma_2 \mod \Z = 0$, where $\Sigma_2 = \Sigma_2^\text{loc} \cup_{\sigma_1} \Sigma_2^\circ$ translates into $H_2(X^\text{loc})^* \ni \Sigma_2^\text{loc} \equiv \vec{a}_e$, and $H_2(X^\circ)^* = H_2(X^\circ, \partial X^\circ)/H_1(X^\circ)^\vee \ni \Sigma_2^\circ \equiv \vec{b}_e$ (and analogously for $\widetilde\Sigma_2$ and $(\vec{a}_m, \vec{b}_m)$).
Because
\begin{align}
\begin{split}
    & \, |{\cal C}_\text{loc} \oplus {\cal C}_\text{extra}|^2 = \left| \frac{M}{H_2(X^\text{loc})} \right|^2 \times \left| \frac{H_2(X^\circ)^*}{H_2(X^\circ)} \right|^2 = \left| H_1(X^\circ)^\vee \right|^2 \times \left| \frac{H_2(X^\circ)^*}{H_2(X^\circ)} \right|^2 \\
    = & \, |H_1(X^\circ)^\vee| \times |H_1(X^\circ)| \times \left| \frac{H_2(X^\circ)^*}{H_2(X^\circ)} \right|^2 = |H_1(\partial X^\text{loc})| \times \left| \frac{H_2(X^\circ)^*}{H_2(X^\circ)} \right| = |{\cal Z}(\widetilde{G}) \oplus Z |\,,
\end{split}
\end{align}
where we have used the short-exact sequence \eqref{eq:les-to-ses-relative_Xo} in the second-to-last equality, $\Lambda$ is a maximal subgroup, i.e., a polarization.
Importantly, this subgroup is nothing but the cokernel of the injective map $\jmath_2: H_2(X^\circ) \oplus H_2(X^\text{loc}) \rightarrow H_2(X)$, which we now recognize as a \emph{lattice embedding} map.
We will return to this perspective in the next section.

Staying with the geometry for a moment longer, we emphasize that the part $H_1(X^\circ)^\vee$ that is not visible in the lattice $H_2(X^\circ)$ is guaranteed by geometric consistency to be glued to cycles in $M \subset H_2(X^\text{loc}, \partial X^\text{loc})$.
From the perspective of the $\mathfrak{g} \oplus \mathfrak{u}(1)^b$ gauge theory, the charged defects from wrapped branes on $H_1(X^\circ)^\vee \subset H_2(X^\circ,\partial X^\circ) $ are completely decoupled, in that they appear as uncharged objects in the charge lattice of $\mathfrak{g} \oplus \mathfrak{u}(1)^b$.
The geometry ensures that these charges are realized as dynamical states in the compact model, in order to screen the global symmetry from branes on $H_1(\partial X^\circ)/\text{im}(\widetilde{D}_2^\circ)$ with Neumann boundary conditions.
This makes the characterization in terms of the ``reduced'' defect group \eqref{eq:field-theoretic_defect-group} possible in the first place.

\subsubsection*{Quadratic refinement and relationship to 1-form anomaly}

The lattice description provides another bottom-up characterization of the global structure dictated by the geometry, which is related to a mixed anomaly between the 1-form symmetry and the instanton $U(1)_I^{(2)}$ symmetry with background gauge field $C_3$ \cite{Cvetic:2021sxm,Apruzzi:2021nmk}.
In geometric engineering on $Y$ a complex surface, with defect group $H_1(\partial Y)^{(1)} \oplus H_1(\partial Y)^{(4)}$, the anomaly is captured by the SymTFT term
\begin{align}\label{eq:mixed_anomaly_7d}
    \text{CS}(\partial Y, \sigma_1) \int_{M_8} dC_3 \cup B_2 \cup B_2 \, ,
\end{align}
where $\sigma_1 \in H_1(\partial Y)$ is the boundary 1-cycle Poincar\'{e}-dual to the torsion 2-cocycle on which the M-theory 4-form reduces to the 1-form symmetry background field $B_2$.
The anomaly coefficient $\text{CS}[\partial Y, \sigma_1] = \tfrac12 \int_{\partial Y} \breve\sigma \star \breve\sigma$, with $\breve\sigma$ the differential cohomology class that is Poincar\'e-dual to $\sigma_1$ on $\partial Y$, is the spin Chern--Simons invariant of the 3-manifold $\partial Y$, which is a quadratic refinement of the linking pairing on $H_1(\partial Y)$ \cite{Apruzzi:2021nmk,Witten:1996md,Hopkins:2002rd}.
As explained there, it can be evaluated as
\begin{align}
    \text{CS}(\partial Y, \sigma_1) = \tfrac12 \Sigma_2 \cdot \Sigma_2 \mod \Z \, ,
\end{align}
for $\Sigma_2 \in H_2(Y, \partial Y)$ with $D_2(\Sigma_2) = \sigma_1$.

In the lattice description above, the defect group takes the form $({\cal L}^*/{\cal L})^{(1)} \times ({\cal L}^*/{\cal L})^{(4)}$, where ${\cal L}^*/{\cal L}$ is also known as the \emph{discriminant group} of the lattice ${\cal L}$, see Appendix \ref{sec:latticeappendix}.
Crucially, ${\cal L} = H_2(X^\text{loc}) \oplus H_2(X^\circ)$ is an even lattice.
This means that for $\sigma_1 = [v \mod {\cal L}] \in {\cal L}^*/{\cal L}$, the anomaly coefficient is given by the natural quadratic refinement
\begin{align}
    q([v]) := \tfrac12 v \cdot v \mod \mathbb{Z}\, ,
\end{align}
which satisfies
\begin{align}\label{eq:rel-q-ell}
    \ell(\sigma_1, \widetilde\sigma_1) = q([v] + [\widetilde{v}]) - q([v]) - q([\widetilde{v}]) \mod \Z \, .
\end{align}
Since ${\cal L} = H_2(X^\text{loc}) \oplus H_2(X^\circ)$ and ${\cal L}^* = H_2(X^\text{loc})^* \oplus H_2(X^\circ)^*$ are orthogonal decompositions, the quadratic refinement can also be decomposed into
\begin{align}
    q([v]) = q([v^\text{loc} + v^\circ]) = q^\text{loc}([v^\text{loc}]) + q^\circ([v^\circ])
\end{align}
in the (hopefully) obvious notation.

It is then not hard to see that the subgroup ${\cal C}_\text{loc} \oplus {\cal C}_\text{extra} = \text{coker}(\jmath_2)$ is isotropic with respect to the quadratic refinement $q$, i.e., $q|_{\cal C} = 0$ (we will refer to this as $q$-isotropic).
This simply follows from the fact that for $\sigma_1 \in \text{coker}(\jmath_2)$ there is a corresponding compact 2-cycle $\Sigma_2 = \Sigma_2^\text{loc} \cup_{\sigma_1} \Sigma_2^\circ$ such that 
\begin{align}
\begin{split}
    q(\sigma_1) & \equiv q^\text{loc}([\Sigma_2^\text{loc}]) + q^\circ([\Sigma_2^\circ]) \\
    & = \tfrac12 \Sigma_2^\text{loc} \cdot \Sigma_2^\text{loc} + \tfrac12 \Sigma_2^\circ \cdot \Sigma_2^\circ \mod \Z = \tfrac12 \Sigma_2 \cdot \Sigma_2 \mod \mathbb{Z} = 0 \mod \mathbb{Z} \, ,
\end{split}
\end{align}
because the intersection product on $H_2(X)$ is also even.
This means in particular that the ``gauged'' 1-form symmetry $[{\cal C}_\text{loc} \oplus {\cal C}_\text{extra}]^{(1)}$ is free of the mixed anomaly, which must hold in supergravity theories, where the instanton $U(1)_I$ is also a gauge symmetry.
Note that demanding the absence of this anomaly for a subgroup ${\cal C}$ of ${\cal Z}(\widetilde{G}) \oplus Z$ automatically implies isotropy of ${\cal C}^{(1)} \oplus {\cal C}^{(4)}$ with respect to the defect group pairing because of the relation \eqref{eq:rel-q-ell}.

Finally, because $|{\cal Z}(\widetilde{G}) \oplus Z| = |\text{coker}(\jmath_2)|^2$, there cannot be a proper subgroup larger than $\text{coker}(\jmath_2)$.
So, unless the full group is $q$-isotropic, in which case there is no interesting defect group to begin with (since $q$-isotropy implies a trivial linking pairing), the global structure $[\widetilde{G} \times U(1)^b]/{\cal C}$ of the supergravity theory with center symmetry ${\cal Z}(\widetilde{G}) \times Z \supset {\cal C}$ is characterized as follows:
the subgroup ${\cal C}$ is a maximal proper subgroup which must also be $q$-isotropic, which is equivalent to say that the gauged 1-form symmetry group is the largest one free of the anomaly \eqref{eq:mixed_anomaly_7d}.
This then implies that the corresponding polarization of the defect group SymTFT is maximally mixed.

\subsection{Delineating the Landscape from the Swampland}\label{sec:land_v_swamp}

As we have emphasized before, imposing the maximally mixed polarization structure is a very restricting condition on the set of field-theoretically viable choices.
However, it requires the information about the dynamical Abelian charges, arising from branes wrapping cycles in $H_2(X^\circ)$ that screen the center of the $\mathfrak{u}(1)$ gauge factors to the finite discrete group $Z$.
Just specifying the gauge algebra $\mathfrak{g} \oplus \mathfrak{u}(1)^b$ does not provide this information.
Instead, from the perspective of the Swampland program, a more natural question to ask is: what are the global structures of 7d ${\cal N}=1$ supergravity models (of total rank 22) with a given non-Abelian ADE gauge algebra $\mathfrak{g}$ (with root lattice $L_\mathfrak{g}$ of rank $\leq 19$, which in the M-theory setting is given by $L_\mathfrak{g} = H_2(X^\text{loc})$) that appear in the Landscape?

From the discussion above, it is clear that there is a necessary condition that can be formulated just in terms of the lattice $T$ of the massive electric and magnetic states charged under the $\mathfrak{u}(1)$'s.
Namely, the lattice must be even, and the discriminant group $T^*/T \equiv Z$ must be a subgroup of the discriminant group $L_\mathfrak{g}^* / L_\mathfrak{g} \equiv {\cal Z}(\widetilde{G})$, such that ${\cal Z}(\widetilde{G}) \oplus Z$ has a maximal subgroup ${\cal C}$ which is isotropic with respect to the quadratic refinement $q = q_\mathfrak{g} \oplus q_Z$.
In particular, maximally mixed polarization would require $|{\cal Z}(\widetilde{G}) \oplus Z|$ to be a square number.

This is in general not a sufficient criterion, because geometrically there may not exist a suitable local patch $X^\circ$ with $H_2(X^\circ) = T$, which furthermore also has the necessary $H_1(X^\circ)$.
However, as we will elaborate in the next section, this criterion is sufficient when rank$(\mathfrak{g}) = 19$ is maximal.
In fact, the lattice description that we distilled from the geometry applies also to the Landscape of ${\cal N}=1$ theories of maximal non-Abelian gauge rank in other spacetime dimensions.

\section{Maximally mixed polarizations in \texorpdfstring{${\cal N}=1 , d > 7$}{N = 1, d > 7} SUGRA}
\label{S:MMP_d>7}

Recasting the global structure of M-theory models into properties of lattice embeddings allows a derivation of the necessity of a maximally mixed polarization also in untwisted heterotic compactifications on tori, giving rise to supergravity theories with 16 supercharges in dimensions other than seven.
We will focus on $d\geq 7$, in view of the recent classifications and explicit computations of the global structure of such theories \cite{Font:2020rsk,Fraiman:2021soq,Cvetic:2022uuu}.
In these cases, 16 supercharges gives minimal supersymmetry, ${\cal N}=1$.

\subsection{Heterotic vacua from lattice embedding}

We begin with a review of some of the relevant features of heterotic compactifications on $T^{\widetilde{d}} = (S^1)^{\widetilde{d}}$ where $\widetilde{d} = 10-d$; see \cite{Font:2020rsk} and references therein for a broader background.

The lattice that arises in the heterotic setting is the lattice of winding and momentum states for the heterotic string on the $\widetilde{d}$-dimension torus.
For untwisted compactifications (meaning no holonomies for the $\Z_2$ outer automorphism of the $E_8 \times E_8$, or for the center of the $Spin(32)$ heterotic strings), this lattice is an even self-dual lattice of signature $(\widetilde{d}, 16+\widetilde{d})$, denoted by ${\rm II}_{\widetilde{d}, 16+\widetilde{d}} \equiv {\rm II}$; it is also referred to as the \emph{Narain lattice} of signature $(\widetilde{d}, 16+\widetilde{d})$.\footnote{In our convention we have $\widetilde{d}$ negative eigenvalues and $16+\widetilde{d}$ positive eigenvalues for the symmetric pairing matrix.}
This lattice corresponds to the charge lattice of the rank $16+2\widetilde{d}$ gauge symmetry; its rank increases by 2 with every additional circle dimension, corresponding to the appearance of a KK-$U(1)$ and another $U(1)$ from the reduction of the 2-form field inside the gravity multiplet.

Because of the self-duality of ${\rm II}$, one can think of it as encoding both the electric and magnetic states, with the Dirac pairing between these states given by the integer lattice pairing.
For $\widetilde{d}=3$, the lattice ${\rm II}_{3,19}$ is precisely the same as the homology lattice $H_2(X)$ of a K3 surface $X$.

At a generic position of the moduli space (parametrized by the vevs of the Wilson lines of the 10d heterotic gauge symmetry on the torus), these states are all massive, and so the generic gauge algebra is $\mathfrak{u}(1)^{16+2\widetilde{d}}$.
For a supergravity with gauge algebra $\mathfrak{g} \oplus \mathfrak{u}(1)^b$ to be realized, one must be able to tune the moduli, such that the roots of $\mathfrak{g}$ --- which span the root lattice $L_\mathfrak{g} \equiv L \subset {\rm II}$ with signature $(0, \text{rank}(\mathfrak{g}))$ --- become massless.
Whether this is possible, and what the global structure of the gauge group of the resulting heterotic model is, depends on the embedding $L \hookrightarrow {\rm II}$. This is analogous to how the gluing of local geometry $X^\text{loc}$ from union of ADE singularities into a compact K3 manifold $X$ determines the consistency and the global structure of the M-theory model.
The geometric counterpart to the Wilson line moduli are the complex structure moduli of the K3.

In order to address the question of what possible global structures are realized given a fixed algebra $\mathfrak{g} \oplus \mathfrak{u}(1)^b$, one therefore needs to find all possible lattice embeddings $L \hookrightarrow {\rm II}$, and determine whether they are consistent with certain criteria that guarantees the masslessness of W-bosons in the heterotic model.
The analogous geometric task is to find all $X^\circ$ for a given $X^\text{loc}$ such that there is a consistent gluing $X = X^\text{loc} \cup X^\circ$ into a K3.

There is a necessary and sufficient criterion to find all such embeddings for $\text{rank}(\mathfrak{g}) = 16 + \widetilde{d}$ \cite{Font:2020rsk}, related to the classification of elliptic K3 surfaces \cite{shimada_k3}.
As we will see momentarily, this criterion translates directly into the polarization choice of the supergravity model.
From these maximally gauge-enhanced cases, consistent models with $\text{rank}(\mathfrak{g}) < 16 + \widetilde{d}$ can be obtained from adjoint Higgsing corresponding to moving to more general positions on the Wilson line / complex structure moduli space. The behavior of the SymTFT under these manipulations has been recently discussed in \cite{Baume:2023kkf}.

The criterion for a given maximal rank $\mathfrak{g}$ to be realized in $d$ dimensions utilizes heavily the discriminant group ${\cal L}^*/{\cal L}$ and the quadratic refinement $q_{\cal L}$ on it, and goes as follows.\footnote{We have modified the conditions slightly in order to streamline the comparison with the previous section. See Appendix A in \cite{Font:2020rsk} and also \cite{Braun:2013yya}.}
\begin{enumerate}
    \item Find an even, rank $16+\widetilde{d}$ overlattice $M \supset L_\mathfrak{g} \equiv L$ of the root lattice $L$, which embeds primitively into ${\rm II}_{\widetilde{d},16+\widetilde{d}}$.
    $M$ is necessarily a positive signature lattice, which is generated by $L$ and some additional vectors $w_a \in L^*$ that satisfies $q_L([w_a]) = 0$ for $[w_a] \equiv w_a \mod L \in L^*/L$, i.e., $M/L \subset L^*/L$ is $q_L$-isotropic.
    \item For such an $M$, find an even, \emph{negative signature}, rank $\widetilde{d}$ lattice $T$ such that the discriminant groups $T^*/T$ and $M^*/M$ are isomorphic, and their quadratic refinements $-q_T$ and $q_{M}$ agree.
    This is equivalent to the existence of generators $v_I^T$ of $T^*$ and $v_I^M$ of $M^* \subset L^*$ such that 
    \begin{align}\label{eq:qL+qT_condition}
        q_{L}([v_I^M]) + q_T([v_I^T]) = 0 \mod \Z \, .
    \end{align}
    Note that the subscript $L$ in the first term is not a typo:
    Because the pairing on $M^* \subset L^*$ is the natural restriction of the pairing on $L^*$, and $M$ is an integer even lattice, we have
    \begin{align}
        q_M([v \mod M]) = \tfrac12 v \cdot v \mod \Z = q_L([v \mod L]) \qquad \text{for} \quad v \in M^* \subset L^*.
    \end{align}
\end{enumerate}
If such a pair $(T, M)$ exists, the full lattice ${\rm II}_{\widetilde{d}, 16+\widetilde{d}} \equiv {\rm II}$ can be constructed as an overlattice of $T \oplus L$ spanned in addition by $w_a \equiv 0 + w_a \in T^* \oplus L^*$ and $v_I = v_I^T + v_I^M \in T^* \oplus L^*$:
\begin{align}\label{eq:II-as-overlattice}
    {\rm II} = T \oplus L + \{ w_a, v_I \}_\Z \subset T^* \oplus L^*\, ,
\end{align}
where $\{ w_a , v_I\}_\Z$ denotes the set of any integer linear combinations of $w_a$ and $v_I$.

To appreciate the restrictive nature of this criterion, we point out that there are, e.g., 1599 semi-simple ADE algebras $\mathfrak{g}$ of rank 18, but only 325 are realized in the landscape of 8d vacua \cite{shimada_k3, Font:2020rsk}.
Furthermore, for a given $\mathfrak{g}$ with root lattice $L$ there can in general be several different overlattices $M$ (including the option $M=L$), of which again only a subset may be realized.
The restricting nature comes from the existence of the lattice $T$ and the condition on the quadratic refinement on its discriminant group, see \cite{Font:2020rsk} for illustrative examples.

Physically, the lattice $T$ corresponds to the charge lattice of the Abelian gauge factors.
In the 7d case, we would have $T = H_2(X^\circ)$.
For $\text{rank}(\mathfrak{g}) = 16+\widetilde{d}$, these $U(1)$ factors are all associated with the vector fields in the gravity multiplet, hence, they are often referred to as gravitational $U(1)$s.
By supersymmetry, this charge lattice has the opposite signature.
If the non-Abelian gauge symmetry is broken to lower rank, the resulting $U(1)$s will retain the positive signature, so that the analogous lattice $T$ ($\cong H_2(X^\circ)$ in the geometric setting) would have indefinite signature.
The lattice points of ${\rm II}$ which do not lie in the sublattice $T \oplus L$ correspond to massive states, so can be viewed as charged defects in the supergravity EFT.

From this data, the gauge group $[(\widetilde{G}/{\cal C}_\text{loc}) \times U(1)^{\widetilde{d}}]/{\cal C}_\text{extra}$ is determined as follows \cite{Cvetic:2021sjm} (see also \cite{Font:2020rsk, Font:2021uyw} for the derivation of ${\cal C}_\text{loc}$):
\begin{align}\label{eq:gauge-group_from_lattice}
    {\cal C}_\text{loc} \cong \frac{M}{L} \, , \quad {\cal C}_\text{extra} \cong \frac{\pi_L({\rm II})}{L} \cong \frac{T^*}{T} \, .
\end{align}
Here, $\pi_{L}: {\rm II} \rightarrow L^*$ be the orthogonal projection, associated to the orthogonal sum $T \oplus L \subset {\rm II}$.
Note that this map always lands on a vector in $L^*$, hence making the identification of ${\cal C}_\text{extra}$ sensible in the first place, because ${\rm II}$ is spanned by vectors in $M \subset L^*$, vectors in $T$ which project to $0$ under $\pi_L$, and the gluing vectors $v_I$ which project to $v_I^M \in M^* \subset L^*$.\footnote{Obviously there is also an orthogonal projection $\pi_T: {\rm II} \rightarrow T^*$, which by similar arguments also maps onto lattice points in $T^*$.\label{footnote:projection}}

The argument in \cite{Font:2020rsk, Font:2021uyw, Cvetic:2021sjm} for this structure is essentially showing that the massive electric/magnetic states are all invariant under these subgroups of the center.
In what follows, we will rephrase this in terms of the polarization choice. 

\subsection{Polarization choice from lattice data}

Treating the supergravity EFT as a $\mathfrak{g} \oplus \mathfrak{u}(1)^{\widetilde{d}}$ gauge theory with Abelian charge lattice $T$, there is the natural defect group $[Z \oplus {\cal Z}(\widetilde{G})]^{(1)} \oplus [Z \oplus {\cal Z}(\widetilde{G})]^{(d-3)}$ with ${\cal Z}(\widetilde{G}) = L^*/L$ and $Z = T^*/T$.
The linking pairing is non-trivial between the 1-form and $(d-3)$-form symmetry defects, and is the natural one induced by the lattice,
\begin{align}\label{eq:linking_pairing-lattice}
    \langle ([s_T], [s_L])^{(1)} \, , \, ([t_T], [t_L])^{(4)} \rangle = s_T \cdot t_T + s_L \cdot t_L \mod \Z \, ,
\end{align}
where $s_L \in L^*, s_T \in T^*$.

Analogously to the geometric setting, the vectors $s$ in the self-dual lattice ${\rm II}$ specify charged objects via the orthogonal projections:
\begin{align}
    {\rm II} \ni s \mapsto (\pi_T(s), \pi_L(s)) \in T^* \oplus L^* \, .
\end{align}
As these vector label both electric and magnetic charges which are present in the absolute supergravity theory, we must impose Dirichlet boundary conditions on $([\pi_T(s)], [\pi_L(s)]) \in Z \oplus {\cal Z}(\widetilde{G})$ for both the 1-form and $(d-3)$-form defects of the SymTFT.
This mirrors the geometric M-theory setting where M2- and M5-branes wrap the same 2-cycles in the compact geometry, and leads to maximally mixed boundary conditions.

Next we show from the lattice description that this also defines a polarization in the defect group $[Z \oplus {\cal Z}(\widetilde{G})]^{(1)} \oplus [Z \oplus {\cal Z}(\widetilde{G})]^{(d-3)}$.
Given the explicit form \eqref{eq:II-as-overlattice}, the subgroup $\{ ([\pi_T(s)], [\pi_L(s)]) \, | \, s \in {\rm II} \}$ can be equivalently described as the cokernel of the embedding $T\oplus L \stackrel{\jmath_2}{\hookrightarrow} {\rm II}$ in \eqref{eq:II-as-overlattice}, which we have given the suggestive label $\jmath_2$:
\begin{align}\label{eq:coker_from_lattice}
    \{ ([\pi_T(s)], [\pi_L(s)]) \in Z \oplus {\cal Z}(\widetilde{G}) \, | \, s \in {\rm II} \} = \text{coker}(\jmath_2) = \frac{\rm II}{T \oplus L} \subset \frac{T^*}{T} \oplus \frac{L^*}{L} \, .
\end{align}
Given the presentation \eqref{eq:II-as-overlattice}, $\text{coker}(\jmath_2)$ is generated by $[w_a] = (0, [w_a]) \in T^*/T$ and $[v_I] = ([v_I^T], [v_I^M]) \in L^*/L$, and is clearly $q$-isotropic with respect to the quadratic refinement $q_{T \oplus L} = q_{T} \oplus q_L$ on $T^*/T \oplus L^*/L$, given the definitions of $w_a$ and $(v_I^M, v_I^T)$ which satisfies \eqref{eq:qL+qT_condition}.
Therefore, $\text{coker}(\jmath_2)^{(1)} \oplus \text{coker}(\jmath_2)^{(d-3)}$ is isotropic also with respect to the linking pairing \eqref{eq:linking_pairing-lattice}.
To see the maximality of this subgroup, we use the general fact that for a sublattice ${\cal M} \subset {\cal L}$ of the same rank, one has $|{\cal M}/{\cal L}|^2 = d({\cal L})/d({\cal M})$, where $d({\cal L}) = |{\cal L}^*/{\cal L}|$.
Then, we have
\begin{align}\label{eq:square_center-order_lattice}
    |\text{coker}(\jmath_2)|^2 = \left| \frac{\rm II}{T \oplus L}  \right|^2 = \frac{d(T \oplus L)}{d({\rm II})} = d(T \oplus L) = \left| \frac{T^*}{T} \right| \times \left| \frac{L^*}{L} \right| = | Z \oplus {\cal Z}(\widetilde{G}) |\, ,
\end{align}
because $d({\rm II}) = |{\rm II}^*/{\rm II}| = 1$ due to self-duality of ${\rm II}$.
This confirms that $\Lambda_{\text{maximally mixed}} = [\text{coker}(\jmath_2)]^{(1)} \oplus [\text{coker}(\jmath_2)]^{(d-3)}$ is indeed a polarization.

\subsection{Examples in \texorpdfstring{$d>7$}{d > 7}}

Let us exemplify the maximally mixed polarization structure in $d>7$ models which do not have a geometric M-theory description.

\subsubsection*{10d heterotic vacua}

In ten dimensions, the lattice ${\rm II}_{\widetilde{d}, 16+\widetilde{d}}$ with $\widetilde{d}=0$ is a rank 16 lattice with definite signature, and it is well-known that there are two such self-dual lattices given by the root lattice ${\rm II}_\mathfrak{e}$ of $\mathfrak{e}_8 \oplus \mathfrak{e}_8$, and the lattice ${\rm II}_{\mathfrak{so}}$ spanned by the roots and one of the spinorial representations of $\mathfrak{so}(32)$.
These give rise to the only two consistent ${\cal N}=1$ supergravity theories in 10 with gauge algebra $\mathfrak{e}_8 \oplus \mathfrak{e}_8$ or $\mathfrak{so}(32)$, consistent with the fact that there are no moduli in 10d that can be activated to break the gauge symmetry.

As there are no Abelian gauge factors, the maximally mixed polarization pertains to the non-Abelian gauge group alone; this is irrelevant for $\mathfrak{g} = \mathfrak{e}_8 \oplus \mathfrak{e}_8$ because it has trivial electric and magnetic defect charges.
For $\mathfrak{g}=\mathfrak{so}(32)$, the defect group is $[\Z_2 \oplus \Z_2]^{(1)} \oplus [\Z_2 \oplus \Z_2]^{(4)}$; the linking pairing is induced by the quadratic refinement $q: \Z_2 \oplus \Z_2 \rightarrow \mathbb{Q}/\Z$ given by
\begin{align}
    q((1,0)) =  q((0,1)) = 0 \, , \quad q((1,1)) = \tfrac12 \, .
\end{align}
The maximal $q$-isotropic subgroup is therefore either of the $\Z_2$ subgroups generated by $(1,0)$ or $(0,1)$ (the ``spinorial'' $\Z_2$s), but not their diagonal (the ``vector'' $\Z_2$).
The maximally mixed polarization defined by this subgroup then corresponds to the gauge group $Spin(32)/\Z_2$.
Obviously, this corresponds to the global structure of the gauge group of the $\mathfrak{so}(32)$ heterotic string.

\subsubsection*{9d examples}

T-duality between the $\mathfrak{e}_8 \times \mathfrak{e}_8$ and the $\mathfrak{so}(32)$ heterotic strings manifests itself at the level of the lattices in
\begin{align}
    {\rm II}_{\mathfrak{e}} \oplus U = {\rm II}_{1,17} = {\rm II}_{\mathfrak{so}} \oplus U \, ,
\end{align}
where $U$ is the rank 2 hyperbolic lattice with pairing lattice $\left(\begin{smallmatrix}
    0 & 1 \\ 1 & 0
\end{smallmatrix}\right)$, and corresponds to the momentum and winding modes along the $S^1$ that screen the center symmetries of the KK-$U(1)$ and the vector from reduction of the 2-form field in 10d.

By tuning the moduli, these $U(1)$s allow an enhancement of the non-Abelian algebra $\mathfrak{g}$ to rank 17.
One such example is $\mathfrak{g} = \mathfrak{su}(18)$, with $L^*/L = \Z_{18}$, and $q_L(k) = \tfrac{17}{36}  k^2 \mod \Z$.
The $q_L$-isotropic subgroups are the trivial subgroup and $\Z_3 =\{0, 6, 12\} \subset \Z_{18}$; these correspond to the lattices $M = L$, and $M = L + \{w_6\}_\Z$, where here and in the following $w_i$ is a fundamental weight vector of $\mathfrak{su}(n>i)$ algebra which in the Dynkin basis has a ``1'' in the $i$-th entry.

For $M = L$, criterion 2 would require a rank one even lattice $T$ with $T^*/T = \Z_{18}$.
Since any even rank one lattice is just $T = Z$, with negative signature pairing given by $x \cdot y := -2m x y$ ($m \in \Z)$ and dual lattice $T^* = \tfrac{1}{2m}\Z$, one would need $m=9$ to satisfy the condition $T^*/T = \Z_{2m} \stackrel{!}{\cong} M^*/M$.
Then it can be straightforwardly verified that there is no generator $k$ of $T^*/T$ for which the quadratic form, given by $q_T(k) = -\tfrac{k^2}{4m} = -\frac{k^2}{36}$, satisfies $q_T(k) = -\tfrac{17}{36}$, which is necessary to satisfy the rest of criterion 2.

For $M=L+ \{w_6\}_\Z$, we can find $d(M) = |M^*/M| = \frac{d(L)}{|M/L|^2} = \frac{18}{|\Z_3|^2} = 2$, so $M^*/M = \Z_2$.
The quadratic refinement $q_M$ can be computed by finding the generator of $M^*$, i.e., a vector $w \in L^*$ such that $w \cdot w_6 = 1$.
One such vector is the fundamental weight $w_9$, as can be verified using $w_{i} \cdot w_j = (C^{-1})_{ij}$, with $C^{-1}$ the inverse Cartan matrix of $\mathfrak{su}(18)$.
Then, we have $q_M(1) = \tfrac12 w_9 \cdot w_9 \mod \Z = \tfrac14$.
This is matched by the quadratic refinement on the rank one lattice defined above with $m=1$; the resulting charges screen the center of the $\mathfrak{u}(1)$ gauge factor to $Z = \Z_2$.
Notice how this is necessary to obtain a center ${\cal Z}(SU(18)) \oplus Z \cong \Z_{9} \oplus \Z_2 \oplus \Z_2$ whose order is a square number. 
The gluing vector $v = (v^T, v^M) \in T^* \oplus L^*$ which satisfies \eqref{eq:qL+qT_condition} is then simply $v = (\tfrac12 , w_9)$.
So the only consistent model in 9d with gauge algebra $\mathfrak{su}(18)$ has gauge group $[(SU(18)/\Z_3) \times U(1)]/\Z_2$.

Another rank 17 gauge algebra would be $\mathfrak{g} = \mathfrak{su}(7) \oplus \mathfrak{so}(22)$.
The center is $\Z_7 \times \Z_4 \cong \Z_{28}$,\footnote{The natural isomorphism by the Chinese remainder theorem is $k \! \! \mod 28 \mapsto (k \! \! \mod 7, k \! \! \mod 4)$.} with quadratic refinement given by
\begin{align}
    q_L(k) = k^2 \left( \tfrac37 + \tfrac{11}{4} \right) \mod \Z = \tfrac{5}{28} \, k^2 \mod \Z\, .
\end{align}
One can check that there are no non-trivial $q$-isotropic subgroups, so $M=L$ is the only choice.
However, the only even rank one lattice with discriminant group $T^*/T \cong \Z_{28}$ has quadratic refinement $q_T(l) = -\tfrac{l^2}{56}$, and there is no generator $l \in \Z_{28}$ with $q_T(l) = 5/28$.
This means that there is no 9d model with non-Abelian gauge algebra $\mathfrak{g} = \mathfrak{su}(7) \oplus \mathfrak{so}(22)$ regardless of the global structure (see also \cite{Font:2020rsk}).

\subsubsection*{8d examples}

The maximal non-Abelian rank in 8d is 18, which is saturated by the example $\mathfrak{g} = \mathfrak{su}(10)^{\oplus 2}$.
The quadratic refinement on the center $\Z_{10} \oplus \Z_{10} = L^*/L$, where $L = (L_{\mathfrak{su}(10)})^{\oplus 2}$, is 
\begin{align}
    q_L((k_1, k_2)) = \tfrac12 (k_1 w_1, k_2 w_1) \cdot (k_1 w_1, k_2 w_1) \mod \Z = \tfrac{9}{20}(k_1^2 + k_2^2) \mod \Z \, ,
\end{align}
where $w_1$ is the first fundamental weight of $\mathfrak{su}(10)$, and $(w, \widetilde{w}) \in (L^*_{\mathfrak{su}(10)})^{\oplus 2} = L^*$.
It is easy to check that, besides the trivial subgroup, there is only one $\Z_5$ subgroup (up to permuting the two $\Z_{10}$ factors) which is $q$-isotropic:
\begin{align}
    \Z_5^\text{iso} =  \{(0,0) , (2,4) , (4,8), (8,6), (6,2) \} \, .
\end{align}
The corresponding overlattices are $M = L$ and $M = L + \{ (w_2, w_4) \}_\Z$, respectively.

For $M = L$, one must find an even rank 2 lattice $T$ with discriminant group $\Z_{10}^{\oplus 2}$ with two suitable generators $v_1^T, v_2^T$ such that the quadratic refinement $q_T$ evaluates on them to $q_T(v_I^T) = -\tfrac{9}{20}$ for $I=1,2$.
Such a lattice is given by $T \cong \Z^2$ with pairing
\begin{align}
    (x_1,x_2) \cdot (y_1, y_2) := (x_1, x_2) \begin{pmatrix}
        -10 & 0 \\ 0 & -10
    \end{pmatrix}
    \begin{pmatrix}
        y_1 \\ y_2
    \end{pmatrix} \, ,
\end{align}
with $v_1^T = (3,0)$ and $v_2^T = (0,3)$.
The gluing vectors are $v_1 = (v_1^T, (w_1,0))$ and $v_2 = (v_2^T, (0,w_1)) \in T^* \oplus [(L_{\mathfrak{su}(10)})^* \oplus (L_{\mathfrak{su}(10)})^*]$.
The resulting gauge group is $[(SU(10) \times U(1))/\Z_{10}]^2$.

We can also find a suitable rank 2 lattice for $M = L +\{(w_2, w_4)\}_\Z$.
This has $M^*/M = \Z_2 \oplus \Z_2$, generated by $v_1^M = (w_5,0) \in M^* \subset (L_{\mathfrak{su}(10)})^* \oplus (L_{\mathfrak{su}(10)})^*$ and $v_2^M = (0, w_5) \in M^*$.
We have $q_L(v_I^M) = \tfrac12 w_5^2 \mod \Z = \tfrac14$ for $I=1,2$.
It is easy to see that there is a suitable rank 2 lattice $T = \Z^{\oplus 2}$ with pairing
\begin{align}
    (x_1,x_2) \cdot (y_1, y_2) := (x_1, x_2) \begin{pmatrix}
        -2 & 0 \\ 0 & -2
    \end{pmatrix}
    \begin{pmatrix}
        y_1 \\ y_2
    \end{pmatrix} \, .
\end{align}
The gluing vectors are $v_1 = ( (\tfrac12, 0), (w_5,0))$ and $v_2 = ((0, \tfrac12), (0, w_5))$, resulting in the gauge group $[SU(10)^2/{\Z_5^\text{iso}} \times U(1)^2]/\Z_2^2$.

\subsection{Maximally mixed polarization and 1-form symmetry anomalies}

As we have emphasized in Section \ref{sec:land_v_swamp}, the maximally mixed polarization puts constraints on the possible string-/M-theory model that accommodates a given non-Abelian gauge symmetry $\mathfrak{g}$.
Using the lattice description, we can formulate the constraint as a genuine field theoretic statement, without explicit reference to a top-down construction.

A key ingredient for this bottom-up formulation is the mixed anomaly between the 1-form symmetry and the instanton symmetry.
In previous works, this anomaly was mainly discussed only for the center of non-Abelian symmetries.
However, in the gravitational setting, the contributions of the $U(1)$s are equally important.

\subsubsection*{Mixed anomalies involving the gravitational $U(1)$s}

As already eluded to in the previous section, the mixed 1-form anomaly, which in the geometric setting is computed as in \eqref{eq:mixed_anomaly_7d}, arises from a coupling between the instanton density and the gauge field $C_{d-4}$.
This field is the Hodge dual of the 2-form $B_2$ inside the gravity multiplet, and are both dynamical in supergravity.
Hence their gauge transformations must be non-anomalous.

$C_{d-4}$ couples democratically to the instanton densities of non-Abelian and Abelian gauge factors.
Indeed, the topological part of the ${\cal N}=1$ supergravity action, written using $C_{d-4}$ instead of $B_2$, contains the terms \cite{Townsend:1983kk,Gates:1984kr,Salam:1985ns,Awada:1985ag}\footnote{In the dual frame with $B_2$, these terms translate into the Bianchi identity for $H_3 = d B_2$.}
\begin{align}\label{eq:u1-instanton-coupling}
    \frac12 C_{d-4} \wedge \left(\sum_g \text{tr}({\cal F}_g^2) + \sum_{i,j} T_{ij} F_i \wedge F_j \right) = C_{d-4} \wedge \left( \sum_g c_2(\cF_g) - \sum_{i,j} \frac{T_{ij}}{2} c_1(F_i) \wedge c_1(F_j) \right) .
\end{align}
Here, $c_2(\cF_g) = \tfrac12 \text{tr}(\cF_g^2)$ is the second Chern-class of the $g$-th factor of the non-Abelian algebra $\mathfrak{g} = \bigoplus_g \mathfrak{g}_g$, and $c_1(F_i) = \sqrt{-1} F_i$ the first Chern-class of the $i$-th $\mathfrak{u}(1)$ factor.\footnote{
At generic points of the moduli space, the non-Abelian gauge symmetry is broken to its Cartan subalgebra, and $c_2(\cF_g) \rightarrow \tfrac12 \sum_{a,b \leq \text{rk}(\mathfrak{g}_g)} K_{ab}^{(g)} \cF_{g,a} \wedge \cF_{g,b}$, with $K_{ab}^{(g)}$ the Cartan matrix of $\mathfrak{g}_g$, which is the lattice pairing on $L_{\mathfrak{g}_g}$.
}
$T_{ij}$ denotes the pairing matrix of the $T$ lattice.

Turning on a 1-form background leads to fractional shifts of the instanton term which breaks the large gauge transformations of $C_{d-4}$, thus manifesting the anomaly.
In a consistent ${\cal N}=1$ supergravity theory, this anomaly must be absent.
In the non-Abelian sector, the fractionalization of $\sum_g c_2(\cF_g)$ in the presence of a 1-form background $B$ labelled by $[w_a] \in {\cal C}_\text{loc} \subset L^*/L = {\cal Z}(\widetilde{G})$ is given by\footnote{See Appendix \ref{app:fractionalizations} for details.}
\begin{align}
    \sum_g c_2(\cF_g) \equiv -q_L([w_a]) \, B \cup B \mod \Z \, . 
\end{align}
Therefore, the first condition which requires $q_L$ to be trivial on $M/L = {\cal C}_\text{loc}$ is just the requirement that the anomaly is absent for 1-form symmetries allowed in a non-simply-connected non-Abelian gauge group $\widetilde{G} / {\cal C}_\text{loc}$ \cite{Cvetic:2020kuw}.

Now, by including the $U(1)$ factors, one can potentially cancel more fractional configurations of $c_2(\cF_g)$ with fractional $U(1)$ Chern-classes $c_1(F_i)$.\footnote{This is analogous to the criterion used in \cite{Heckman:2022suy} to determine the global structure of the flavor symmetry, including $U(1)$ factors, for 6d SCFTs.}
The corresponding 1-form symmetry background $B$ is then labelled by $[v_I] = ([v_I^T], [v_I^M]) \in T^* / T \oplus L^*/L = Z \oplus {\cal Z}(\widetilde{G})$ with
\begin{align}
    \sum_g c_2(\cF_g) - \sum_{i,j} \frac{T_{ij}}{2} c_1(F_i) \wedge c_1(F_j) \equiv (-q_L([v_I^M]) - q_T([v_I^T])) \, B \cup B \mod \Z \, ,
\end{align}
which now vanishes thanks to \eqref{eq:qL+qT_condition}.

\subsubsection*{Towards String Universality for gauge group topologies}

Combining the mixed anomaly with the maximally mixed polarization structure, we arrive at the following bottom-up (``Swampland'') constraint for an ${\cal N}=1$ supergravity theory in $d \geq 7$ dimensions with rank $16+d$ non-Abelian gauge algebra $\mathfrak{g}$ to be realized.

First, identify a subgroup ${\cal C}_\text{loc} \subset {\cal Z}(\widetilde{G})$ which is free of the mixed 1-form--instanton anomaly.
Then, find an instanton coupling matrix $T_{ij}$ for the gravitational $U(1)^{10-d}$ factors which (when interpreted as a lattice pairing) defines a discriminant group $T^*/T = {\cal C}_\text{extra} \subset {\cal Z}(U(1)^{10-d})$ that is also a subgroup of ${\cal C}(\widetilde{G})$; the mixed 1-form anomalies for the ${\cal C}_\text{extra} \subset {\cal Z}(\widetilde{G})$ 1-form background must cancel that of the ${\cal C}_\text{extra} \subset {\cal Z}(U(1)^{10-d})$.
The existence of a maximally mixed polarization then requires $|Z \oplus {\cal Z}(\widetilde{G})|$ to be a square, and ${\cal C}_\text{loc} \oplus {\cal C}_\text{extra}$ to be a maximal subgroup.
Then, the global structure of the supergravity EFT is given by imposing the maximally mixed polarization.

This constraint is a natural extension of the bottom-up constraints on the non-Abelian group topology $\widetilde{G}/{\cal C}_\text{loc}$ \cite{Cvetic:2020kuw}.
However, just as in that case, the absence of the mixed 1-form anomaly including the $U(1)$ factors is only a necessary but not sufficient condition to have a gauged 1-form symmetry, i.e., a non-trivial subgroup ${\cal C} \subset [{\cal Z}(\widetilde{G}) \times {\cal Z}(U(1)^{10-d})]^{(1)}$ of the defect group with Dirichlet boundary condition.
Indeed, this anomaly argument, and arguments based on other Swampland principles \cite{Montero:2020icj}, can never rule out gauge groups $\widetilde{G} \times U(1)^{10-d}$ that have no center-quotient.
This gap is closed by imposing the maximally mixed polarization, as the 9d examples have demonstrated.
That is, any gauge group topology which does not arise from giving Dirichlet boundary condition to the largest anomaly-free subgroup of the 1-form symmetry in the corresponding SymTFT is ruled out.
Furthermore, the relationship to the lattice description of heterotic compactifications demonstrates that any supergravity model that has a maximally mixed polarization with an anomaly-free 1-form symmetry is also realized from string theory.

Unfortunately, while this a simple-to-state condition using the SymTFT framework, we currently do not have a field theoretic argument why it must be true.
Moreover, we have also not studied the SymTFT structure in twisted heterotic compactifications, which in 7d would be dual to M-theory on frozen singularities \cite{Witten:1997bs, deBoer:2001wca, Tachikawa:2015wka}.
Finding a unified condition that applies to all such vacua, and then providing a purely field theoretic reason for it would show String Universality for consistent gauge groups in ${\cal N}=1$ $d\geq 7$ supergravity theories.

\section{The correlated boundary conditions requirement}
\label{S:CBCreq}
In this section we will generalize the discussion of Section \ref{S:MMP-K3} by considering placing $D$-dimensional string/M-theory on a compact internal space $X_{D-d}$. We will similarly split the internal space into two parts: The ``local'' part $X^{\text{loc}}=\bigcup_i X_i$, with $X_i$ corresponding to local neighborhoods of the singular loci of $X$, and $X^\circ$ the completion of this part in $X$, which has no singularities. Thus, we can write the compact internal space as $X=X^{\text{loc}}\cup X^\circ$. The two parts are glued along their boundaries $\partial X^{\text{loc}}= \partial X^\circ$ to generate the full compact space $X$. As in the $K3$ case we argue that this gluing specifies the boundary conditions for the SymTFT defined on $\partial X^{\text{loc}} \coprod \partial X^\circ$. 

The aim of this section is to be more general; thus, it will only include string/M-theory construction evidence, and won't include lattice interpretations as in Section \ref{S:MMP-K3}, as these are only relevant for theories with eight real supercharges.

We will start by considering the underlying geometry. The covering $X=X^{\text{loc}}\cup X^\circ$ leads to a truncation of the Mayer--Vietoris sequence as follows
\be
\label{E:MVS_gen}
0 \to H_{p-n+1}(X^\circ) \oplus H_{p-n+1}(X^\text{loc}) \xrightarrow{\jmath_{p-n+1}} H_{p-n+1}(X) \xrightarrow{\partial_{p-n+1}} H_{p-n}(\underbrace{\partial X^\text{loc}}_{=\partial X^{\circ}}) \xrightarrow{\imath_{p-n}} H_{p-n}(X^\circ) \rightarrow 0\,.
\ee
In this general case the geometric picture of possible completions of relative cycles from the local perspective is the same as discussed and shown in Figure \ref{F:CycleCompK3} and in Subsection \ref{SS: Gluing}, but with cycle dimensions matching the sequence above. Thus, from such a geometric construction one can immediately see that the discrete gauge symmetry of the full $d$-dimensional theory including gravity depends only on the homological structure of the internal space $X$. This means that in cases where we can wrap more than one type of brane on both the relative cycle $\Sigma_{p-n+1}^\text{loc}\in H_{p-n+1}(X^\text{loc}, \partial X^\text{loc})$ and a cycle $\gamma_{D-d-(p-n+2)}\in H_{D-d-(p-n+2)}(\partial X^\text{loc})$ linked with $D^\text{loc} \Sigma_{p-n+1}^\text{loc}= \sigma_{p-n}\in H_{p-n}(\partial X^\text{loc})$ in $\partial X^\text{loc}$, the resulting higher form gauge symmetries will be correlated. For example, assume there are both $p$-branes and $(p+q)$-branes in the spectrum of the $D$-dimensional theory, with $H_{p-n+1}(X)=\Z_N$. The resulting gauge symmetries will be correlated, i.e. both $\Z_N$ $n$-form and $\Z_N$ $(n-q)$-form gauge symmetries, assuming $n\ge q$.

Note that in the general case, unlike the K3 internal space case, the boundary cycles $H_{p-n}(\partial X^*)$ will not be self linked. We instead expect them to be linked with cycles in $H_{D-d-(p-n+2)}(\partial X^*)$. 
Therefore, the operators generated by wrapping $p$-branes on cycles in $H_{p-n}(\partial X^*)$ and wrapping $(D-p-4)$-branes on cycles in $H_{D-d-(p-n+2)}(\partial X^*)$ will be paired in the SymTFT. This means that in a consistent coupling to gravity the above Mayer--Vietoris sequence will be correlated with a similar sequence where $(p-n)$ is replaced with $(D-d-(p-n+2))$ and the resulting Dirichlet and Neumann boundary conditions are flipped. Nevertheless, this will not affect the possible completions of relative cycles described above.

As in the K3 case we can write down a SymTFT for the above general case by using the geometry on $\partial X^\text{loc} \coprod \partial X^\circ$. Considering the resulting SymTFT as associated with some spacetime theory, one would need to choose boundary conditions for the symmetries of the SymTFT in order to make the resulting theory absolute. The boundary conditions will be determined completely by the geometry of $X^\text{loc}$ and $X^\circ$ and the identification of their boundaries as we have shown above. The question we will try to answer is what are the options for the global structure of the gauge symmetry of the full gravitational theory given just the boundaries and their associated SymTFT. One might naively expect that the only requirements will be that the resulting theory has mutually local operators charged under a global symmetry that is screened by dynamical operators due to the no global symmetries conjecture, and in addition that the resulting gauge symmetry has no mixed gauge gravity anomalies, that are inconsistent in a gravity coupled theory. In what comes next we will claim the options are more restricted.

Using the understandings above about different branes wrapping the same cycle, we can constrain the boundary conditions to be those where all the branes wrapping a specific cycle get the same boundary conditions.

\subsection{Implications}

\paragraph{Correlated boundary conditions requirement:}
Consider a $(d+1)$-dimensional SymTFT that can be constructed geometrically from a $D$-dimensional string/M-theory. Furthermore we assume there are electric BPS branes of spacial dimensions $p,(p+p_1),...,(p+p_k)$ with $(p+p_i)\le (D-4)/2$ for $i=1,..,k$ and their magnetic dual branes wrapped on the combined boundary space $\partial X^\text{loc} \coprod \partial X^\circ$. Considering SymTFT operators created by wrapping $p$-branes and their magnetic dual $(D-p-4)$-branes on boundary $(p-n)$- and $(D-d-(p-n+2))$-cycles, respectively, one can choose any boundary conditions for them as long as the charged objects under the resulting global symmetry are mutually local, with the constraint that the resulting gauge symmetry has no mixed gauge gravity anomalies. These boundary conditions can be translated to boundary conditions for background gauge fields for spacetime $n$-form and $(d-(n+2))$-form symmetries, respectively. After this choice, the boundary conditions for the $(n-p_i)$-form and $(d-(n-p_i+2))$-form symmetries, arising from wrapping the other branes on the same $(p-n)$- and $(D-d-(p-n+2))$-cycles for all $i=1,...,k$, should be the same as the boundary conditions for the $n$-form and $(d-(n+2))$-form symmetries, respectively. These boundary conditions are required for consistency with spaces $X^\text{loc}$ and $X^\circ$ that can be glued to a space $X$ that generates a consistent $d$-dimensional gravitational theory.

If we restrict ourselves to branes wrapping cycles $\sigma_{p-n}$ and $\gamma_{D-d-(p-n+2)}$ that have the same dimension, i.e.
\be
D-d = 2(p-n+1)\,.
\ee
We find that the correlated boundary conditions constraint now limits the types of boundary conditions to polarizations that are maximally mixed, meaning they are ``half" magnetic and ``half" electric. We will denote this constraint as \emph{the maximally mixed polarization conjecture}. When considering only supersymmetric preserving setups we find that this is the general case due to the dimensions of the magnetic dual branes in the different string and M-theories and the homology of the possible supersymmetry-preserving internal geometries.

\paragraph{Maximally mixed polarization requirement:}
Consider the same $(d+1)$-dimensional SymTFT, now only with electric BPS $p$-branes and their magnetic dual $(D-(p+4))$-branes placed on the combined boundary space $\partial X^\text{loc} \coprod \partial X^\circ$, with non-trivial $H_{p-n}(\partial X^\text{loc})=H_{p-n}(\partial X^\circ)$ and $D-d=2(p-n+1)$. We construct SymTFT operators  by wrapping $p$-branes and their magnetic dual $(D-p-4)$-branes on boundary $(p-n)$-cycles. There is no free choice of boundary conditions for these operators as long as the charged objects under the resulting global symmetry are mutually local, and with no mixed gauge gravity anomalies. Instead one is required to choose a maximally mixed polarization, this choice will amount to having the same gauge group for the $n$-form symmetry and the $(d-(n+2))$-form symmetry. These boundary conditions are required for consistency with spaces $X^\text{loc}$ and $X^\circ$ that can be glued to a space $X$ that generates a consistent $d$-dimensional gravitational theory.

One can then restrict the above constraints even further by setting also the dimension of the two magnetic dual branes to be the same, meaning
\be
D=2p+4\,.
\ee
In this case the boundary conditions will be limited to those choosing the maximally isotropic subgroup of the SymTFT defect group as the gauge symmetry. For such cases the choice of boundary conditions can be understood fully from the combined SymTFT as the maximally mixed choice is inherently required in order to get an absolute theory.

\subsection{Examples}

Examples for the general case of the \emph{correlated boundary conditions requirement} with $D-d \ne 2(p-n+1)$ are less common, as we need two cycles of the internal geometry with linked boundary cycles that can be wrapped by a two pairs of magnetic dual branes. One example for this case is found by engineering 6d $(2,0)$ in type IIB string theory placed on a K3 manifold with local geometries of the type $\mathbb{C}^2/\Gamma$. We consider both 2-form symmetries and the pair of Poincare dual 0-form and 4-form symmetries, which are generated by operators created by placing D3-branes, D5/NS5-branes and D1/F1-branes on boundary 1-cycles of the internal space $\mathbb{C}^2/\Gamma$, respectively. In this case the dimensional parameters are $D=10,\,d=6,\,p=1,\,p_1=2,\,n=1$. In this case fixing the boundary conditions for the 0- and 4-form symmetries will fix the boundary conditions for the 2-form symmetries and vice versa. We will review this example in more detail bellow. Finding other supersymmetric examples is hard due to dimensional constraints.\footnote{One can perhaps engineer some non-supersymmetric examples, but this is beyond the scope of this paper.}

Examples for the \emph{maximally mixed polarization requirement}, i.e. $D-d = 2(p-n+1)$ are actually the more common case in supersymmetric cases and there are many such examples:
\begin{itemize}
    \item 7d $\cN=1$ SYM can be engineered from M-theory by compactifying on a K3 manifold with local geometries of the type $\mathbb{C}^2/\Gamma$. For the purposes of this discussion we are interested in 1-form and 4-form symmetries. Thus, we choose our dimensional parameters to take values $D=11,\,d=7,\,p=2,\,n=1$. In this case wrapping M2- and M5-branes on the boundary cycles will generate a pair of defect group symmetries coupled by a ``BF'' term in the SymTFT. The polarization consistent with the full gravity theory is the maximally mixed one. We discussed this example extensively in Section \ref{S:MMP-K3}.
    \item We can also consider the 0- and 4- form symmetries of 6d (2,0) SCFTs, constructed in type IIB string theory by compactifying on a K3 manifold with local geometries of the type $\mathbb{C}^2/\Gamma$. Thus, we choose our dimensional parameters to take values $D=10,\,d=6,\,p=1,\,n=0$. In this case wrapping D1/F1- and D5/NS5-branes on the boundary cycles will generate a pair of defect group symmetries coupled by a ``BF'' term in the SymTFT. The polarization consistent with the full gravity theory is the maximally mixed one.
    \item Lastly we can contemplate the 0- and 3-form symmetries in 5d $\cN=1$ SCFTs engineered in M-theory by compactifying on a CY3 manifold if $\text{Tor}H_2(\partial X^\text{loc})$ is non-trivial.\footnote{We expect such a case to have an isolated terminal singularity, which in compact models is associated with discrete 0-form gauge symmetries \cite{Grimm:2010ez,Camara:2011jg,Grimm:2011tb,Braun:2014oya,Mayrhofer:2014laa,Arras:2016evy,Grassi:2018rva}, and analyzing such cases will be reserved for future works.} These symmetries would be generated by wrapping M2- and M5-branes on 2-cycles on the boundary of the CY3, and are coupled by a ``BF'' term in the SymTFT. In this case the parameters are $D=11,\,d=5,\,p=2,\,n=0$, and the polarization is required to be maximally mixed.
\end{itemize}

Finally, we end with several examples for \emph{the maximally mixed polarization requirement}, with the additional constraint $D=2p+4$. Such examples can be found when considering type IIB string theory, where D3-branes are magnetic dual to themselves. Some examples are:
\begin{itemize}
    \item Consider the 2-form symmetries in 6d (2,0) SCFTs engineered in type IIB string theory by compactifying on a K3 manifold with local geometries of the type $\mathbb{C}^2/\Gamma$. These are generated by wrapping D3-branes on 1-cycles on the boundary of $\mathbb{C}^2/\Gamma$. The parameters of this case are $D=10,\,d=6,\,p=3,\,n=2$ and the 2-form defect group give ``CS'' terms in the SymTFT. The polarization is determined fully by field theory considerations of mutual locality and no mixed gauge gravity anomalies.
    \item We can also consider the 1-form symmetries of 4d $\cN=2$ SCFTs engineered in type IIB string theory by compactifying on a CY3 manifold with non-trivial $\text{Tor}H_2(\partial X^\text{loc})$. These are generated by wrapping D3-branes on 2-cycles on the boundary of the CY3. The parameters of this case are $D=10,\,d=4,\,p=3,\,n=1$. From the geometric analysis above one might suspect such symmetries will have a ``CS'' SymTFT, but we will reserve such analysis to future work.
    \item Engineering 2d $(2,0)$ or $(1,0)$ theories from type IIB string theory by compactifying on a CY4 or Spin(7)-holonomy manifolds with non-trivial $H_3(\partial X^\text{loc})$,\footnote{The existence of such manifolds is not guaranteed. We reserve analysing such a case for future work.} we can look at the 0-form symmetries generated by wrapping D3-branes on 3-cycles on the boundary of the CY3. The parameters of this case are $D=10,\,d=2,\,p=3,\,n=0$. Such theories are expected to have a ``CS'' SymTFT terms similar to the 6d case.
\end{itemize}

\subsection{6d (2,0) SCFTs}

In this section we focus on 6d $\cN=(2,0)$ SCFTs and their associated 7d bulk SymTFT description \cite{Gukov:2020btk}, which describes their allowed global structures.\footnote{6d $(2,0)$ models were studied in the context of higher form symmetries in \cite{Heckman:2022muc}, also see \cite{Braun:2021sex} that discusses coupling such models to gravity.} These theories can be engineered from type IIB string theory by placing it on
\be
\mathbb{R}^{1,5}\times X_4^\text{loc}\,,
\ee
where $X_4^\text{loc}=\bigcup_i \mathbb{C}^2/\Gamma_i$, with $\Gamma_i\subset SU(2)$ a finite group of ADE type. In this specific case it is hard to derive the SymTFT directly from a reduction of the IIB string theory on $\partial X_4^\text{loc}$, following the differential cohomology procedure of \cite{Apruzzi:2021nmk}, due to the self duality of the $C_4$ background gauge field. Thus, here we will discuss the 2-form symmetries and their ``CS'' term contributions to the SymTFT, and the 0- and 4-form symmetries and their ``BF'' term contributions to the SymTFT, separately. Finally we will show there is an additional mixed gauge gravity anomaly term in the SymTFT that can be derived for the IIB string theory action that restricts the choice of boundary conditions for the 0- and 2-form symmetries to be correlated.

\paragraph{7d SymTFT for 0- and 4-form symmetries. } The 0- and 4-form symmetries of these theories are generated by operators created by wrapping D5(NS5)-branes and D1(F1)-branes on 1-cycles of $\partial X_4^\text{loc}$, respectively. Accordingly, the charged operators under these symmetries are created by wrapping D1(F1)-branes and D5(NS5)-branes on relative 2-cycles of $X_4^\text{loc}$ that their boundary 1-cycles are linked with the 1-cycles wrapped to create the symmetry generators. In this case the SymTFT terms are the regular ``BF''-terms of the form:
\be
\label{E:SymTFT0-4F}
S_{\rm SymTFT}^{(0-4)} = \frac{1}{4\pi}\sum_{i,j}K_{ij} \int_{M_{7}}(B_5^i dA_1^j + \widetilde{B}_5^i d\widetilde{A}_1^j)\,,
\ee
where $K_{ij}$ is the Cartan matrix associated with the singularities $\mathbb{C}^2/\Gamma_i$,\footnote{When there is more than on singularity the matrix $K_{ij}$ is block diagonal containing the Cartan matrix associated with each singularity in each block.} $B_5^i(\widetilde{B}_5^i)$ and $A_1^i(\widetilde{A}_1^i)$ are the background gauge fields associated with the symmetries generated by wrapping D5(NS5)-branes and D1(F1)-branes on 1-cycles of $\partial X_4^\text{loc}$, respectively.

As in the 7d $\cN=1$ case discussed before we must assign boundary conditions to the fields in the SymTFT above in order to get an absolute 6d theory.

\paragraph{7d SymTFT for 2-form symmetries. } The 2-form symmetries are somewhat special in this case along with their 7d SymTFT terms. This is due to the fact that they are generated by operators engineered by wrapping self magnetic dual D3-branes on 1-cycles of $\partial X_4^\text{loc}$. Similarly the background gauge fields for the 2-form symmetries come from KK reduction of the self-dual 3-form gauge fields of IIb string theory. One physical consequence of this is that there is no natural distinction between ``electric'' and ``magnetic'' contributions.

In this case the ``BF''-terms of interest above are now abelian Chern-Simons terms of the form:
\be
\label{E:SymTFT2F}
S_{\rm SymTFT}^{(2)} = \frac{1}{4\pi}\sum_{i,j}K_{ij} \int_{M_{7}}C_3^i dC_3^j \,.
\ee
Here $K_{ij} \in \Z$ and $C_3^i$ are $U(1)$-valued gauge fields. Specifically for 6d SCFTs, based on a lie algebra $\mathfrak{g}$, the matrix $K$ can be identified with the associated Cartan matrix. This bulk term describes a non-invertible TQFT: different boundary conditions potentially encode 6d absolute theories with different global structures. However, it is a non-trivial step to ascertain whether or not such boundary conditions exist for these theories.

A general 3d topological operator in this bulk theory takes the form
\be
U_\lambda(M_3) = e^{ i \int_{M_3} \lambda_i C_3^i} \,.
\ee
The group of such topological operators under \textit{fusion} is exactly the \textit{defect group} of the associated SCFT. The existence of boundary condition corresponding to an absolute 6d theory is related to the linking of these bulk operators. In particular, the correlation of two such operators obeys \cite{Gukov:2020btk}
\be
\langle U_\lambda(M_3) U_{\lambda'}(M_3') \rangle= e^{-2\pi i \, \lambda' T (K^{-1}) \lambda \cdot \rm{Link}_{M_7}(M_3, M_3')}\,,
\ee
where Link$_{M_7}(\cdot, \cdot)$ denotes the linking number inside $M_7$. This correlation defines a bilinear pairing for the 2-form part of the defect group $D^{(2)}$. 
\be
\langle \cdot,\cdot \rangle\ :\ D^{(2)} \times D^{(2)} \to U(1)\,.
\ee
Concretely, if two such topological operators with non-trivial braiding were given the same boundary condition this would lead to a SCFT with mutually non-local extended operators.

The bilinear pairing, emanating from a quadratic form
\be
q: \lambda \to \half \lambda^T K^{-1} \lambda \,,
\ee
with the relation
\be
\langle \lambda,\lambda' \rangle = \text{exp} [-2\pi i (q(\lambda + \lambda') - q(\lambda)- q(\lambda')) ]\,.
\ee
We denote a choice of subgroup generated by $\{\lambda_i\}$ that gives $\langle \lambda_i,\lambda_j \rangle=1$ for all $i,j$, as $\ell$-isotropic,\footnote{The pairing $\langle\cdot, \cdot \rangle$ is obviously the ``exponentiated'' version of the linking pairing on lattices, which we called $\ell$, see Appendix \ref{sec:latticeappendix}.} while integer $q$ choice is denoted as $q$-isotropic.
In Table \ref{tab:qvaluesforADE} we write the quadratic form for some A,D,E type algebras.

\begin{table}[h]
\setlength\extrarowheight{5pt}
    \centering
    \begin{tabular}{|c|c|}
        \hline
         ADE Algebra & $q$  \\
         \hline\hline
         $A_{n-1}$& $\frac{n-1}{2n}$ \\
         \hline
         $D_{2n}$ & $q(v) =\half$, $q(s) = q(c) = \frac{2n}{8}$ \\
         \hline
         $D_{2n+1}$ & $\frac{2n+1}{8}$ \\
         \hline
         $E_6$ & $\frac{2}{3}$ \\
         \hline
         $E_7$ & $\frac{3}{4}$ \\
         \hline
    \end{tabular}
    \caption{Quadratic form/ spin evaluated for the generators of $A$,$D$ and $E$ type algebras \cite{Gukov:2020btk} which give non-trivial fusion algebras as 6d theories. For $D_{2n}$, $v$ denotes the vector, $s$ the spinor, and $c$ the cospinor generators.}
    \label{tab:qvaluesforADE}
\end{table}
Consider a basic example: a 6d SCFT based on algebra $\mathfrak{g} = A_{N-1}$. The 2-form part of the defect group is given by the center, $D^{(2)}=\Z_N$. We can obtain an absolute theory if there is a subgroup of the defect group chosen by the generator $a$ which gives a $\ell$-isotropic group or equivalently:
\be
q(a) = \frac{N-1}{2N} a^2 \in \half \Z \,.
\ee
This is only possible for $N=n^2$. In which case the only solution is $\frac{a}{n}\in \Z$, which generates a $\Z_n$ subgroup of $\Z_N=\Z_{n^2}$: this is the 2-form symmetry of the associated absolute 6d SCFT. These considerations lead to the classification of the absolute 6d $\cN=(2,0)$ SCFTs of \cite{Gukov:2020btk}.

Applying boundary conditions at the level of the SymTFT Lagrangian can be instructive. We focus concretely on a simple example to illustrate this point:
\be
S_{\rm SymTFT} = \frac{N}{4\pi} \int C_3 dC_3 \,.
\ee
The most general boundary conditions we can choose involve setting $C_3$ equal to some $B_3$, a fixed $\Z_n$-valued field on the boundary. Such a condition can be imposed explicitly via an extra boundary term
\be
S_{\rm SymTFT} + S_\delta = \frac{N}{4\pi} \int C_3 dC_3 + \frac{n}{4\pi} \int_{\partial M_7} (C_3 - B_3) Y_3 \,.
\ee
Attempting to retain gauge invariance enforces two features: the first that $N=n^2$ and secondly that there is a residual 't Hooft anomaly
\be
\frac{n^2}{4\pi} \int B_3 dB_3 \,,
\ee
of the 2-form symmetry in the absolute theory.

\paragraph{$q$-isotropy requirement. } From a field theory perspective, the $q$-isotropic requirement is well-motivated. If one selects a $\ell$-isotropic subgroup, but strictly not $q$-isotropic, we call the resulting three-dimensional operators ``fermionic'' when they have half-integer spin $q$. With such operators in a theory, one must equip spacetime with a 4th Wu class to make the theory independent of a boundary manifold, or equivalently to specify how these fermionic operators depend on an attached bounding 4-manifold \cite{Gukov:2020btk}. Once we couple gravity we cannot fix this additional structure: we therefore must require $q$-isotropy. This could equally be interpreted as a condition to avoid a mixed gravity anomaly. 

\paragraph{The relation between 0-, 2- and 4-form symmetries. } From the topological terms of type IIB string theory\footnote{In general the kinetic terms of this action are not well defined due to self duality of the 4-form RR $C_4$ field strength.} one can derive a term which correlates the 0-, 2- and 4- form symmetries \cite{Lawrie:2023tdz}. Specifically we begin with the term
\be \label{E:IIBtop}
S_\text{top}\supset 2\pi \int_{M_{10}} (-\half  F_5 \wedge B_2 \wedge F_3)\,, 
\ee
where $M_{10}$ is the 10d spacetime, and $F_5$ and $F_3$ are the 4-form $C_4$ and 2-form $C_2$ field strength in the RR sector, while $B_2$ is the 2-form field in the NSNS sector. 

In topologically non-trivial situations where the above fields are not globally well defined this term needs to be slightly modified \cite{Apruzzi:2021nmk}. Specifically, one will need to model the $B_2$ field as a class $\breve{H}_3 \in \breve{H}^3(M_{10})$ in (ordinary) differential cohomology. Under this refinement the term in \eqref{E:IIBtop} is interpreted as the $\mathbb{R}/\Z$-valued secondary invariant of a differential cohomology class $\breve{I}_{11}\in \breve{H}^{11}(M_{10})$,\footnote{Note that deriving the full SymTFT action from the type IIB supergravity action requires to move one dimension higher to a manifold $M_{11}$ such that $\partial M_{11}=M_{10}$ \cite{Yu:2023nyn}. This is required in order to deal with the self duality that makes the kinetic term of $F_5$ not well defined. In our case since we are not interested in the kinetic term this will not have any effect.}
\be
\label{E:StopCohom}
\frac{S_\text{top}}{2\pi}\supset \int_{M_{10}} \breve{I}_{11} \mod 1=  \int_{M_{10}} (-\half  \breve{F}_5 * \breve{H}_3 * \breve{F}_3) \mod 1 \,.
\ee
Now, we can consider engineering 6d $(2,0)$ by reducing type IIB string theory on $\mathbb{C}^2/\Gamma$. Note that the $S^3/\Gamma$ cohomology torsion is localized in degree two shown in Table \ref{T:ADE_H1_Link},
\be
H^0(S^3/\Gamma,\Z)=H^3(S^3/\Gamma,\Z)=\Z\,,\quad H^1(S^3/\Gamma,\Z)=0\,,\quad H^2(S^3/\Gamma,\Z)=\Gamma_{ab}\,.
\ee
Using the matching cohomology generators we can expand
\be
\ba
\breve{F}_5 = \breve{f}_5 * \breve{1} + \sum_i \breve{c}_3^{(i)}  * \breve{t}_{2(i)}  + \breve{f}_2  * \breve{v}_3\,,\\ 
\breve{F}_3 = \breve{\widetilde{h}}_3 * \breve{1} + \sum_i \breve{a}_1^{(i)}  * \breve{t}_{2(i)}  + \breve{f}_0  * \breve{v}_3\,,\\ 
\breve{H}_3 = \breve{h}_3 * \breve{1} + \sum_i \breve{\widetilde{a}}_1^{(i)}  * \breve{t}_{2(i)}  + \breve{h}_0  * \breve{v}_3\,,
\ea
\ee
where the above generators are associated with the following cohomologies
\be
\breve{1} \leftrightarrow H^0(S^3/\Gamma,\Z)=\Z\,,\quad \breve{t}_{2(i)} \leftrightarrow H^2(S^3/\Gamma,\Z)=\Gamma_{ab}\,,\quad \breve{v} \leftrightarrow H^3(S^3/\Gamma,\Z)=\Z\,.
\ee
Note that fields coming from the torsional part of the cohomology are associated with discrete symmetries while the rest are associated with continuous $U(1)$ symmetries. Focusing only on the torsional part of the above expansion and substituting it to \eqref{E:StopCohom} we find the following SymTFT terms,
\be
\label{E:SymTFTmix}
\frac{S_\text{SymTFT}}{2\pi} \supset - \sum_{i,j} \text{CS}[S^3/\Gamma]_{ij}  \int_{M_{7}} \breve{c}_3^{(i)} * (\breve{h}_3 * \breve{a}_1^{(j)} + \breve{\widetilde{h}}_3 * \breve{\widetilde{a}}_1^{(j)})\,,
\ee
where 
\be
\text{CS}[S^3/\Gamma]_{ij} =\half \int_{S^3/\Gamma} \breve{t}_{2(i)} \breve{t}_{2(j)} \mod 1 \,.
\ee
In this SymTFT action the discrete fields $\breve{c}_3^{(i)}$, $\breve{a}_1^{(j)}$( $\breve{\widetilde{a}}_1^{(j)}$) are associated with the 2- and 0-form symmetries background gauge fields $C_3^i$ and $A_1^i(\widetilde{A}_1^i$) appearing in \eqref{E:SymTFT2F} and \eqref{E:SymTFT0-4F}, respectively, in the case $X_4^\text{loc}$ is built from a single singularity. This can be trivially generalized for the case of multiple singularities. In addition, the ordinary cohomology representatives $h_3, \widetilde{h}_3$ of the fields $\breve{h}_3$ and $\breve{\widetilde{h}}_3$ are field strengths for a self dual $U(1)$ 2-form symmetry, i.e. $h_3 = db_2$ and $\widetilde{h_3}=\star_6 db_2$. This 2-form field is associated with the $B_2$ gravitational 2-form; thus, it will not appear in the 6d $(2,0)$ SCFT theory, but it will become relevant once we couple gravity.

The terms appearing in \eqref{E:SymTFTmix} will in general generate a mixed gauge gravity anomaly; thus, one will need to choose boundary conditions for the 2-form and 0-form fields in a way that will trivialize these terms when we couple the theory to gravity. One can notice that the coefficient of these terms $-\text{CS}[S^3/\Gamma]_{ij}$ in Table \ref{T:ADE_H1_Link} are exactly the same as the $q$ quadratic form in Table \ref{tab:qvaluesforADE}.\footnote{Note that for $D_{2n}$ there is a choice of basis for the $\Z_2\times \Z_2$ such that $\Z_2^{(s)}=\Z_2^{(1)}$ and $\Z_2^{(c)}=\Z_2^{(2)}$, and $\Z_2^{(v)}$ is the diagonal $\Z_2$; this shows the results in both tables are equivalent for the $D_{2n}$ case as well.} This exactly means that the boundary conditions for the 2- and 0-form symmetries will be correlated. Taking this constraint together with the BF coupling of the 0- and 4-form symmetries will fully constrain the 4-form symmetries boundary conditions; this will correspond to \emph{the correlated boundary conditions requirement} presented earlier.

\begin{table}[t]
\setlength\extrarowheight{5pt}
    \centering
    \renewcommand{\arraystretch}{1.2}
    \begin{tabular}{|c|c|c|}
        \hline
         ADE Algebra & $\Gamma_{ab}$ & $-\text{CS}[S^3/\Gamma]_{ij}$  \\
         \hline\hline
         $A_{n-1}$ & $\Z_n$ & $\frac{n-1}{2n}$  \\
         \hline
         $D_{2n}$ & $\Z_2 \oplus \Z_2$  & $\frac{1}{4} \begin{pmatrix}
           n & n-1 \\
           n-1 & n \\
         \end{pmatrix}$ \\
         \hline
         $D_{2n+1}$ & $\Z_4$ & $\frac{2n+1}{8}$ \\
         \hline
         $E_6$ & $\Z_3$ & $\frac{2}{3}$ \\
         \hline
         $E_7$ & $\Z_2$ & $\frac{3}{4}$ \\
         \hline
         $E_8$ & $0$ & $0$ \\
         \hline
    \end{tabular}
    \caption{The group $H^2(S^3/\Gamma)$ and the coefficients $-\text{CS}[S^3/\Gamma]_{ij}$ for the different ADE algebras associated with $\Gamma$ \cite{Apruzzi:2021nmk,GarciaEtxebarria:2019caf}.}
    \label{T:ADE_H1_Link}
\end{table}

\subsection*{Acknowledgements}

We thank Andreas P.~Braun, Markus Dierigl, Antonella Grassi, Sakura Sch\"afer-Nameki, Ethan Torres, and Jingxiang Wu for useful discussions. ES is supported by the European Union’s Horizon 2020 Framework: ERC grant 682608 and the ``Simons Collaboration on Special Holonomy in Geometry, Analysis and Physics''.

\appendix

\section{Some lattice facts}
\label{sec:latticeappendix}
Here, we collect some useful information about lattices.
\begin{itemize}
    \item \textbf{Lattice}: a free, finitely generated Abelian group ${\cal L} \cong \Z^n$ with a non-degenerate symmetric pairing ${\cal L} \times {\cal L} \rightarrow \mathbb{Q}$, denoted by $x \cdot y$.
    One can always embed ${\cal L} \hookrightarrow \mathbb{R}^n$ such that the pairing is induced by a suitable symmetric bilinear form $\langle \cdot, \cdot \rangle$ on $\mathbb{R}^n$ via restriction:
    \begin{align}
        \forall x,y \in \cL \subset \bbR^n : \quad x \cdot y = \langle x, y \rangle \, .
    \end{align}
    
    \item \textbf{Gram matrix}: Given a basis $\{e_a\}_{a=1,..,n}$ of $\cL$, the Gram matrix is $G_{ab} = e_a \cdot e_b = G_{ba}$.

    \item \textbf{Signature:} Let $n_\pm$ be the number of positive/negative eigenvalues of $G_{ab}$ (including multiplicities).
    Then $(n_-, n_+)$ is the signature of the lattice $\cL$.
    This does not depend on the choice of basis $\{e_a\}$.
    
    \item \textbf{Integral lattice}: A lattice $\cL$ such that $x \cdot y \in \Z$ for all $x,y \in \cL$.
    \item \textbf{Even lattice}: An integer lattice $\cL$ such that $x \cdot x \in 2\Z$ for all $x \in \cL$.
    
    \item \textbf{Dual lattice}: Given a lattice $\cL \hookrightarrow \mathbb{R}^n$ with associated bilinear form $\langle \cdot, \cdot \rangle$, the dual lattice is the rank $n$ lattice
    \begin{align}
        \bbR^n \supset \cL^* := \{v \in \bbR^n \, \, | \, \, \forall x \in \cL \subset \bbR^n : \, \langle v, x\rangle \in \Z \} \, .
    \end{align}
    One has $(\cL^*)^* = \cL$.
    For an integer lattice $\cL$ with basis $e_a$ of $\cL$, any vector $v \in \cL^*$ can be written as a \emph{fractional} linear combination of $e_a$.
    So $\cL \subset \cL^*$

    \item \textbf{Discriminant group}: The discriminant group of an integer lattice $\cL$ is the finite Abelian group $\cL^*/\cL$.
    The order $|\cL^*/\cL|$ is equal to $\det(G_{ab})$.

    \item \textbf{Unimodular / self-dual}: a lattice $\cL$ is unimodular/self-dual if $\cL = \cL^\ast$.
    Equivalently, it has a trivial discriminant group.

    \item \textbf{``Linking'' pairing}: Given an integer lattice $\cL$, there is a natural non-degenerate, or perfect, ``linking'' pairing on $\cL^*/\cL$, given by
    \begin{align}
        \ell: \frac{\cL^*}{\cL} \times \frac{\cL^*}{\cL} \rightarrow \mathbb{Q}/\bbZ \, , \qquad \ell([v \! \! \mod \cL], [w \! \! \mod \cL]) := v \cdot w \mod \Z \, .
    \end{align}
    A subgroup ${\cal C} \subset {\cal L}^*/{\cal L}$ is isotropic, or $\ell$-isotropic, if $\ell|_{\cal C} = 0$.

    \item \textbf{Quadratic refinement}: Given a finite Abelian group $Z$ with a linking pairing $\ell$, a quadratic refinement (of $\ell$) is a quadratic form $q: Z \rightarrow \mathbb{Q}/\Z$ (i.e., $q(z+\ldots +z) = q(n \, z) = n^2 q(z)$ and $q(0) = 0$) such that
    \begin{align}
        \ell(z, z') = q(z+z') - q(z) - q(z') \, .
    \end{align}
    In general, there can be multiple inequivalent quadratic refinements for one linking pairing.

    For ${\cal L}$ an even lattice, there is a natural choice of quadratic refinement induced by the lattice pairing,
    \begin{align}
        q_{\cal L}: \frac{\cL^*}{\cL} \rightarrow \mathbb{Q}/\Z \, , \quad q_{\cL}([v \! \! \mod \cL]) := \tfrac12 v^2 \mod Z \, .
    \end{align}

    We call a subgroup ${\cal C} \subset {\cal L}^*/\cL$ $q$-isotropic if $(q_\cL)|_{\cal C} = 0$.
    Notice that a $q$-isotropic subgroup must necessarily be $\ell$-isotropic, but not vice versa.

    \item \textbf{Properties of ADE root lattices}:
    The root lattice $L_\mathfrak{g} \equiv L$ of a semi-simple non-Abelian ADE algebra $\mathfrak{g}$ is an even lattice with signature $(0, \text{rank}(\mathfrak{g}))$.
    For a decomposition $\mathfrak{g} = \bigoplus_g \mathfrak{g}_g$ into simple factors, $L_\mathfrak{g} = \bigoplus_g L_{\mathfrak{g}_g}$ is the orthogonal sum of lattices.
    \begin{itemize}
        \item The simple roots of $\mathfrak{g}$ define a basis $\{e_a\}$ of $L$.
        The Gram matrix $G_{ab}$ in this basis is nothing but the Cartan matrix of $\mathfrak{g}$.
    
        \item The discriminant group $L^*/L = {\cal Z}(\widetilde{G})$ is the center of the simply-connected cover $\widetilde{G}$ with algebra $\mathfrak{g}$.

        \item The dual lattice $L^*$ is generated by the \emph{weights} of $\mathfrak{g}$, hence, it is also called the weight lattice.
        Let $\{w_a\}$ be the dual basis, i.e., $w_a \cdot e_b = \delta_{ab}$, then $w_a \cdot w_b = (G^{-1})_{ab}$ is the inverse Cartan matrix.

        \item For a simple ADE algebra $\mathfrak{g}$, the quadratic refinement on $L^*/L$ takes the values listed in Table \ref{tab:qvaluesforADE} for the generators.
    \end{itemize}
    
\end{itemize}

\section{Instanton fractionalization in the lattice description}
\label{app:fractionalizations}

For a simple non-Abelian factor $\mathfrak{g}$, the fractionalization of $c_2({\cal F})$ in the presence of a generic ${\cal Z}(\widetilde{G})$ 1-form symmetry background $B$ has been computed before \cite{Kapustin:2014gua,Gaiotto:2014kfa,Gaiotto:2017yup,Cordova:2019jnf}, and takes the form $c_2(\cF) \equiv -\alpha B \cup B$, where $\alpha$ are fractional coefficients that are listed in Table \ref{tab:qvaluesforADE} as the values of $q$.
This is because it is equivalent to the quadratic refinement $q([v])$ for the generator $[v]$ of $L_\mathfrak{g}^* / L_\mathfrak{g} = {\cal Z}(\widetilde{G})$.
Turning on a generic background in a subgroup ${\cal C}\subset {\cal Z}(\widetilde{G})$, generated by another element $[w] \in L^*/L$, the fractional coefficient would then be $q([w])$.

For a semi-simple non-Abelian ADE algebra $\mathfrak{g} = \bigoplus_g \mathfrak{g}$, any background field $B$ is labelled by $[w] = ([w_g]) \in L^*/L = \bigoplus_g (L_{\mathfrak{g}_g}^*/L_{\mathfrak{g}_g})$.
Since the root lattice $L$ is just the orthogonal sum of $L_{\mathfrak{g}_g}$, the pairings on the lattices and the discriminant groups decompose.
Therefore, the fractionalization of the instanton densities sum is
\begin{align}
    \sum_g c_2(\cF_g) \equiv -\sum_g q_{g}([w_g]) \, B \cup B \mod \Z = -q_L([w]) \, B \cup B \mod \Z \, .
\end{align}

The fractionalization of the Abelian instanton density $\frac{T_{ij}}{2} c_1(F_i) \wedge c_1(F_j)$ in \eqref{eq:u1-instanton-coupling} can also be expressed in terms of the lattice pairing $T_{ij}$.
First, we can think of this expression as the self-pairing inside the $T$ lattice for a vector $V = \sum_i c_1(F_i) t_i$, where $t_i$ are the basis vectors of $T$ which leads to the Gram matrix $T_{ij}$.
Now, for a single $U(1)$ factor with $ T_{ij} = T_{11} = N$, turning on a $\Z_N = T^*/T$ 1-form symmetry background $B$ labelled by the generator $[v]$ of the $\Z_N$ would fractionalize as $c_1(F) \equiv \tfrac{1}{N} B \mod \Z$; the coefficient here is nothing but the coefficient of the dual basis vector $v = \tfrac{1}{N} t$ when expressed as a fractional linear combination of the generators of $T$.
This generalizes straightforwardly to a higher rank $T$ lattice, meaning that the fractionalization $c_1(F_i) \equiv \lambda_i B \mod \Z$ in the presence of a 1-form background $B$ labelled by an element $[v] \in T^*/T$ is encoded in the fractional expression $v = \sum_i \lambda_i \, t_i \in T^*$.
Note that $c_1(F_i)$ still contains an integer part $p_i \, A$, with $A$ an integer 2-cocycle, which, similar to $c_1(F_i)$ without any 1-form symmetry backgrounds, can be regarded as a vector in $T$.
Then,
\begin{align}
\begin{split}
    \tfrac{T_{ij}}{2} c_1 (F_i) \wedge c_1(F_j) & = \tfrac{T_{ij}}{2} (p_i\,A + \lambda_i \,B) \cup (p_j\,A + \lambda_j\,B ) \\
    & = \tfrac{T_{ij}}{2} p_i p_j \, A \cup A+ T_{ij} p_i \lambda_j \, A \cup B+ \tfrac{T_{ij}}{2} \lambda_i \lambda_j \, B\cup B\\
    & \equiv (\tfrac12 v \cdot v) \, B \cup B \quad \text{+ integer 4-cocycles } = q_T([v]) \, B\cup B \mod \Z \, ,
\end{split}
\end{align}
because $T_{ij}/2 p_i p_j = \tfrac12 p \cdot p \in \Z$ since $T$ is even, and $T_{ij} p_i \lambda_j = p \cdot v \in \Z$ since $v$ is a vector in the dual lattice.

\bibliographystyle{JHEP}
\bibliography{refs.bib}

\end{document}